\documentstyle[prd,aps,epsf]{revtex}

\def\simlt{\rlap{\lower 3.5 pt\hbox{$\mathchar \sim$}}\raise 1pt \hbox {$<$}}
\def\simgt{\rlap{\lower 3.5 pt\hbox{$\mathchar \sim$}}\raise 1pt \hbox {$>$}}
\newcommand{\lw}[1]{\smash{\lower2.ex\hbox{#1}}}
\newcommand{\be} {\begin{equation}}
\newcommand{\ee} {\end{equation}}
\newcommand{\bea} {\begin{eqnarray}}
\newcommand{\eea} {\end{eqnarray}}

\begin{document}
\draft

\title{
\begin{flushright}
\normalsize
UTCCP-P-122, UTHEP-456  \\
\end{flushright}
Light Hadron Spectrum and Quark Masses from Quenched Lattice QCD}

\author{CP-PACS Collaboration\\[3mm]
S.~Aoki$^1$,
G.~Boyd$^2$\thanks{present address : 
            Packaging and Device Development, Procter and Gamble,
            Temselaan 100, B-1853 Strombeek-Bever, Belgium},
R.~Burkhalter$^{1,2}$\thanks{present address : 
            KPMG Consulting AG, Badenerstrasse 172, 
            8804 Zurich, Switzerland},
S.~Ejiri$^2$\thanks{present address :
            Department of Physics, University of Wales, 
            Swansea SA2 8PP, U.K.},
M.~Fukugita$^3$,
S.~Hashimoto$^4$,
Y.~Iwasaki$^{1,2}$,
K.~Kanaya$^{1,2}$,
T.~Kaneko$^2$\thanks{present address :
             High Energy Accelerator Research Organization (KEK), 
             Tsukuba, Ibaraki 305-0801, Japan},
Y.~Kuramashi$^4$,
K.~Nagai$^2$\thanks{present address :
            CERN, Theory Division, CH--1211 Geneva 23, Switzerland},
M.~Okawa$^4$\thanks{present address :
            Department of Physics, Hiroshima University, 
            Higashi-Hiroshima, Hiroshima 739-8526, Japan}, 
H.~P.~Shanahan$^2$\thanks{present address :
            European Bioinformatics Institute,
            Wellcome Trust Genome Campus,
            Hinxton, Cambridge, CB10 1SD, England, U.K.},
A.~Ukawa$^{1,2}$,
T.~Yoshi\'e$^{1,2}$\\[2mm]
}

\address{
$^1$Institute of Physics, University of Tsukuba, 
Tsukuba, Ibaraki 305-8571, Japan \\
$^2$Center for Computational Physics, University of Tsukuba, 
Tsukuba, Ibaraki 305-8577, Japan \\
$^3$Institute for Cosmic Ray Research, University of Tokyo, 
Tanashi, Tokyo 188-8502, Japan \\
$^4$High Energy Accelerator Research Organization (KEK), 
Tsukuba, Ibaraki 305-0801, Japan
}

\date{June 13, 2002}

\maketitle


\tightenlines

\begin{abstract}
We present details of simulations for the light hadron spectrum
in quenched QCD carried out on the CP-PACS parallel computer.
Simulations are made with the Wilson quark action and the plaquette gauge 
action on lattices of size $32^3\times 56-64^3\times 112$
at four values of lattice spacings in the range $a \approx 0.1$--0.05~fm 
and the spatial extent $L_sa \approx$ 3~fm. Hadronic observables 
are calculated at five quark masses corresponding to 
$m_{PS}/m_V \approx 0.75$--0.4, assuming the $u$ and $d$ quarks
being degenerate but treating the $s$ quark separately.
We find that the presence of quenched chiral singularities is
supported from an analysis of the pseudoscalar meson data.
Physical values of hadron masses are determined using
$m_\pi$, $m_\rho$ and $m_K$ (or $m_\phi$) as input
to fix the physical scale of lattice spacing and the $u$, $d$ and $s$ 
quark masses.
After chiral and continuum extrapolations, 
the agreement of the calculated mass spectrum 
with experiment is at a 10\% level.
In comparison with the statistical accuracy of 1--3\% and 
systematic errors of at most 1.7\% we have achieved, 
this demonstrates a failure of the quenched
approximation for the hadron spectrum: 
the hyperfine splitting in the meson sector is too small, 
and in the baryon sector the octet masses and 
the mass splitting of the decuplet are both smaller than experiment.  
Light quark masses are calculated using two definitions:
the conventional one and
the one based on the axial-vector Ward identity. The two results 
converge toward the continuum limit, yielding 
$m_{ud}=4.29(14)^{+0.51}_{-0.79}$~MeV where the first error 
is statistical and the second one is systematic due to chiral extrapolation.
The $s$ quark mass depends on the strange hadron mass chosen for input: 
$m_s = 113.8(2.3)^{+5.8}_{-2.9}$~MeV from $m_K$ and
$m_s = 142.3(5.8)^{+22.0}_{-0}$~MeV from $m_\phi$,
indicating again a failure of the quenched approximation.
We obtain  the scale of QCD, 
$\Lambda_{\overline {\rm MS}}^{(0)}=$ 219.5(5.4) MeV
with $m_\rho$ used as input. An $O(10\%)$ deviation from experiment is 
observed in the pseudoscalar meson decay constants.
\end{abstract}



\twocolumn

\section{Introduction}

Theoretical derivation of the light hadron spectrum from 
the first principles of Quantum Chromodynamics (QCD) is a fundamental issue 
in our understanding of the strong interactions.  
The binding of quarks due to gluons cannot be treated perturbatively,
and numerical simulations based on the lattice formulation
of QCD, therefore, provide a unique means to approach this problem. 

The calculation of the hadron spectrum is made for given
quark masses and hence it in turn enables us to determine the light
quark masses, which are the fundamental parameters of QCD.
The dynamical scale $\Lambda$ of QCD is determined by
measurements of lattice spacing $a$ as a function of the bare coupling 
constant. 
Lattice QCD also provides us with a method to explore the chiral
structure which is approximately realized in the real world.
A further subsidiary verification of QCD may include the examination of 
the decay matrix elements against experiment.

Lattice QCD simulations, however, are computationally demanding, 
particularly when effects of dynamical quarks are to be included. 
Therefore, since the pioneering attempts in 
1981 \cite{ref:HamberParisi,ref:Weingarten}, the majority of lattice QCD 
simulations have been made within 
the quenched approximation in which pair 
creation and annihilation of sea quarks are ignored.  
In fact such calculations have given hadron spectrum in a gross
agreement with experiment, but clear understanding has not been
achieved yet as to where this approximation would break down.
In order to study this point, 
a calculation with a much higher precision is needed.
Such a high precision study requires accurate controls of a number of
systematic errors, which is not an easy task even within the quenched
approximation. 
The origins of systematic errors include finiteness of lattice size,
coarseness of lattice spacing and extrapolations in quark masses 
from relatively large values.

The work of GF11 Collaboration carried out 
in 1991--1993 \cite{ref:GF11mass} has advanced
the control of systematic errors from a finite lattice
spacing and a finite lattice size. Taking an advantage of large
computing power, the GF11 Collaboration calculated the light hadron spectrum
with three sets of the coupling constant and three different lattice
sizes at one coupling constant, which is used to take continuum limit
and estimate finite lattice effects. 
They claimed that the resulting spectrum is in agreement with
experiment
within 6\%, the difference for each hadron being within their errors.  

We feel that their results need a further verification by an
independent analysis, since we consider that their conclusions depend
crucially on the error estimate at simulation points, and on a rather 
long chiral extrapolation from the region of the pseudoscalar to vector
meson mass ratio $m_{PS}/m_V=0.9$ -- 0.5. 
Another issue is that GF11 simulations were made
only for degenerate quarks. Masses of strange mesons and decuplet
baryons were estimated using mass formulae, while strange octet
baryons were not calculated.

We have embarked on the program to push the calculation of the
quenched light hadron spectrum beyond that of the GF11 Collaboration
to answer the posed problems. We have aimed to achieve a precision of
a few percent for statistical errors and to reduce systematic errors to be
comparable to or smaller than statistical errors.
Taking the Wilson quark action and the plaquette gluon action,
simulations are made with lattices of physical spatial size 
$L_sa\approx 3$~fm for the range of $a\approx 0.1$ -- 0.05 fm.
The smallest value of $m_{PS}/m_V$ is lowered to $\approx 0.4$. 
We take an advantage of the recent development of quenched chiral
perturbation theory (Q$\chi$PT) \cite{ref:QChPTSharpe,ref:QChPTBGmass}, 
which suggests us the form of chiral extrapolations.
We assume that the light $u$ and $d$ quarks are degenerate, but the heavier
$s$ quark is treated separately giving a different quark mass.

During the time, the MILC Collaboration carried out studies in a
similar spirit \cite{ref:MILC-KS} using the Kogut-Susskind quark
action.  Because of complications with the
spin-flavor content of this action, they reported only the 
nucleon mass taking $m_\pi$ and $m_\rho$ as input,
leaving aside all other hadrons.

Our calculation was made by the CP-PACS computer, 
a massively parallel computer developed at the 
University of Tsukuba completed in September 1996 \cite{ref:CPPACS}.
With 2048 processing nodes,   the peak speed of the CP-PACS is 614 GFLOPS 
(614 billion double precision floating-point operations per second).   
Our optimized program achieves a sustained
speed of 237.5 GFLOPS for the heat-bath update of gluon variables and 
264.6 GFLOPS for the over-relaxation update,  
and 325.3 GFLOPS for the quark propagator solver for which the core 
part is written in the assembly language
\cite{ref:CPPACSPerformance}. 
The simulations were executed from the summer of 1996 to 
the fall of 1997. 

A brief description of our results has been published in 
Ref.~\cite{ref:CPPACSletter}, and preliminary reports
have appeared in Refs.~\cite{ref:CPPACSreports}.  
In this article we present full details of analyses and results. 

The organization of the paper is as follows:
in Sec.~\ref{sec:model} the lattice action and simulation parameters
are explained.  
In Sec.~\ref{sec:summary} 
we present a summary of results for the light hadron spectrum,
quark masses and meson decay constants.
Subsequent sections describe details of our analyses.  
In Sec.~\ref{sec:measurement} measurements of hadron masses and quark
masses at simulation points are discussed. 
We then examine in Sec.~\ref{sec:qChPT} the prediction of Q$\chi$PT
for light hadron masses against our data. 
In Sec.~\ref{sec:spectrum} we describe the extrapolation procedure of 
hadron masses to the chiral and continuum limits.  
Comparisons with other studies are given in this section.
In Sec.~\ref{sec:Mq} we discuss determinations of  
the light quark masses, 
and in Sec.~\ref{sec:AlphaLambda}, the QCD $\Lambda$ parameter.
In Sec.~\ref{sec:decay} results for meson decay constants are presented.
Finally, Sec.~\ref{sec:methodB} presents an alternative analysis 
in which the order of the chiral and continuum extrapolations are reversed.
Our conclusions are given in Sec.~\ref{sec:conclusions}. 
Technical details are relegated to 
Appendices~\ref{appendix:gfix} to \ref{appendix:mqlocal}. 

\section{Lattice action and simulation parameters}
\label{sec:model}

\subsection{Lattice action}
\label{sec:action}

We generate gauge configurations using 
the one-plaquette gluon action,
\be
S_g = \frac{\beta}{3} \sum_P {\rm Re\,Tr} (U_P),
\label{eq:gaugeaction}
\ee
where $\beta=6/g^2$ with $g$ the bare gauge coupling constant.
On gauge configurations, we evaluate
quark propagators using the Wilson fermion action, 
\be
S_q = - \sum_{n,m} \bar\psi(n) D(\kappa;n,m)\psi(m), 
\label{eq:quarkaction}
\ee
\bea
D(\kappa;n,m) = \delta_{n,m} -
\kappa \sum_\mu\{&&(I-\gamma_\mu)U_{n,\mu}\delta_{n+\hat\mu,m} 
\nonumber \\
       +   &&(I+\gamma_\mu)U^\dagger_{m,\mu}\delta_{m+\hat\mu,n}\},
\label{eq:quarkmatrix}
\eea
where the hopping parameter $\kappa$ 
controls the quark mass.

\subsection{Simulation parameters}
\label{sec:parameters}

Simulation parameters are summarized in Tables~\ref{tab:param1} and \ref{tab:param2}.
Four values of $\beta$ are chosen so as to cover the range of 
$a\approx 0.1$ -- 0.05 fm ($a^{-1}\approx 2$--4 GeV). 

We employ lattices with the physical extent of $L_s a \approx 3$ fm 
in the spatial directions.  
In a previous study, no significant finite lattice size effect was 
observed for $L_s a \geq 2$ fm beyond a statistical error of about 
2\% \cite{ref:MILC-W}.
For a large lattice, the dominant size effect comes from 
spatial wrappings of pions whose magnitude decreases as 
$m_{L_s} - m_\infty \propto \exp(-c\,m_\pi L_s)$\cite{ref:Luescher}.
For smaller lattices squeezing of hadron wave functions enhances 
the finite size effect, leading to a power low behavior,
$m_{L_s} - m_\infty \propto c/L_s^3$\cite{ref:fukugitaFS,ref:aokiFS}.
Assuming the latter behavior, we expect the finite-size effects on 
lattices with $L_sa \approx 3$ fm to be about 0.6\%, which is 
sufficiently small compared with our statistical errors.
This requires us to use a $64^3$ lattice for simulations at $a\approx$ 
0.05 fm.

For the temporal extent of the lattices, we adopt $L_t=(7/4) \cdot L_s$. 
This gives the maximal physical time separation of $L_ta/2 \approx 2.5$ fm.
With our smearing method described below, we find 
that this temporal extent is sufficient to extract ground state 
signals in hadron propagators, 
suppressing contaminations from excited states. 

For the quark mass, 
we select five values of $\kappa$, so that they give
$m_{PS}/m_V \approx 0.75$, 0.7, 0.6, 0.5, and 0.4.
The two heaviest values, which we denote as $s_1$ and $s_2$, are chosen
to interpolate hadron mass data to the physical point of the $s$ quark.
The three lighter quarks denoted as $u_1$, $u_2$ and $u_3$ are used to
extrapolate to the physical point of the light $u$ and $d$ quarks, 
$m_{PS}/m_V = m_\pi/m_\rho = 0.176$.

The quark mass at the smallest value of $m_{PS}/m_V \approx 0.4$ is 
closer to the chiral limit than that in any previous studies with the
Wilson quark action, in which calculations were limited to $m_{PS}/m_V
\simgt 0.5$. Reducing the quark mass further is not easy.  Test runs we carried
out for $m_{PS}/m_V\approx 0.3$ at $\beta=5.9$ show that fluctuations
become too large and the computer time for this point alone 
exceeds the sum of those for the
five $\kappa$ down to $m_{PS}/m_V\approx$ 0.4.

Gauge configurations are generated by the 5-hit pseudo-heat-bath
algorithm\cite{ref:HeatBath} and an over-relaxation 
algorithm\cite{ref:OverRelaxation}, mixed in the ratio of 1:4. 
We call the combination of one pseudo-heat-bath update sweep 
followed by four over-relaxation sweeps ``iteration''. 
The periodic boundary conditions are imposed in all four directions.
The acceptance rate in the pseudo-heat-bath step is 82--85\% 
as listed in Table~\ref{tab:param1}.
For vectorization and parallelization of the computer program,
we adopt an even-odd algorithm.

After 2000--20000 thermalization iterations, 
we calculate quark propagators and measure 
hadronic observable on configurations separated by 200 to 2000 
iterations depending on $\beta$, 
while we measure the gluonic observable, such as the plaquette 
expectation value, at every iteration. 
The total number of configurations and their separation
are summarized in Table~\ref{tab:param1}.

We estimate errors by the jackknife method except otherwise stated.
Tests on the bin size 
dependence do not show the presence of correlations between successive 
configurations, and hence we use the unit bin size for error analyses. 

Table~\ref{tab:simultime} shows the number of employed processors of the
CP-PACS and the execution time required for generating 
and analyzing one configuration. 
Simulations at $\beta=$ 5.9, 6.1 and 6.25 are carried out
on subpartitions of the CP-PACS computer
while at $\beta=6.47$ the whole system with 2048 processing units is used.

\section{Summary of results}
\label{sec:summary}

\subsection{Quenched chiral singularity}
\label{sec:summary:chiral_extrap}

Quenched chiral perturbation theory (Q$\chi$PT)~\cite{ref:QChPTSharpe,ref:QChPTBGmass}
predicts that hadron mass as a function of quark mass $m_q$ exhibits
a characteristic singularity in the chiral limit.
Data for $m_{PS}^2$
strongly support the existence of an expected singular term $\delta m_q \ln m_q$ 
with $\delta\approx 0.1$. 
For vector mesons and baryons, the accuracy of mass data and 
the covered range of $m_q$ are not sufficient to establish the
presence of quenched singularities. 

In Figs.~\ref{fig:chiralFit590} to~\ref{fig:chiralFit647}, 
the Q$\chi$PT fit is shown by solid lines for (a) pseudoscalar meson, 
(b) vector meson, (c) octet baryon and (d) decuplet baryon. 
The data are consistent with the theoretical expectations 
from Q$\chi$PT, not only for pseudoscalar mesons but also for
vector mesons~\cite{ref:Booth} and baryons~\cite{ref:LabrenzSharpe}.
We therefore adopt functional forms based on Q$\chi$PT for chiral
extrapolations for all cases. 

\subsection{Quenched light hadron spectrum}
\label{sec:summary:spectrum}

We take experimental values of $m_\pi= 0.1350$ GeV and 
$m_\rho= 0.7684$ GeV as input for the mean $u,d$ quark mass
$m_{u,d}$ and the lattice spacing $a$. We use either   
$m_K = 0.4977$ GeV or $m_\phi = 1.0194$ GeV for the strange quark mass 
$m_s$.
As shown by solid lines in Fig.~\ref{fig:ContinuumFP}, hadron masses 
determined at each $\beta$ are well described by a linear function of $a$.   

The quenched hadron spectrum in the continuum limit is compared 
in Fig.~\ref{fig:MassFinal} with experiment shown by horizontal bars,
with the numerical values given in Table~\ref{tab:MassFinal}. 
Filled symbols use $m_K$ as input, and open ones employ $m_\phi$.
The two error bars show both statistical error and the sum of statistical and 
systematic errors (see Sec.~\ref{sec:spectrum}).
Statistical errors are 1--2\% for mesons and 2--3\% for baryons.
Estimated systematic errors are at worst 1.8$\sigma$ of statistical
ones, which add only extra 1.7\% to statistical ones.

Figure~\ref{fig:MassFinal} shows that quenched QCD reproduces 
the global pattern of the light hadron spectrum reasonably well, 
but at the same time systematic deviations exist between the quenched
spectrum and experiment.  
An important manifestation of this discrepancy is that the quenched prediction 
depends sizable on the choice of particle ($K$ or $\phi$) 
to fix $m_s$. While an overall agreement in the baryon 
sector is better if $m_\phi$ is employed as input, $m_K$ disagrees by
11\% (6$\sigma$), which is the largest difference 
between our result and experiment.    

In the meson sector, the discrepancy is seen in the hyperfine splitting, 
that is too small compared to experiment. 
If one uses the $m_K$ as input, the vector meson masses 
$m_{K^*}$ and $m_\phi$ are smaller by 4\% (4$\sigma$) and 6\% (5$\sigma$). 
If $m_\phi$ is employed instead, 
$m_{K^*}$ agrees with experiment within 0.8\% (2$\sigma$) but 
$m_K$ is larger by 11\% (6$\sigma$). 

The smallness of the hyperfine splitting is observed in a different way 
in Fig.~\ref{fig:hyperfine}, which plots $m_V^2-m_{PS}^2$ 
as a function of $m_{PS}^2$.  The figure shows an approximate scaling over 
the four values of $\beta$.  The convergence of data toward the experimental 
point corresponding to ($m_\pi$,$m_\rho$) is due to our choice of 
these particles as input. 
Toward heavier quark masses, the mass square difference decreases faster than 
experiment, and is about 10\% smaller at the point corresponding to 
$(m_K,m_{K^*})$ mesons.
 
A faster decrease of $m_V^2-m_{PS}^2$ can be quantified through the $J$ 
parameter\cite{ref:Jparameter} defined by 
\be
J = m_V \frac{d m_V}{d m_{PS}^2}.  
\label{eq:J}
\ee 
A large negative value of the slope seen in Fig.~\ref{fig:hyperfine} 
translates into a small $J$ as shown in Fig.~\ref{fig:valJ}; 
we obtain 
\begin{equation}
J=0.346(23) 
\label{eq:Jvalue}
\end{equation}
in the continuum limit, to be compared with the experimental value 
$\sim0.48$ at $m_V/m_{PS} = 1.8$. 

In the octet baryon sector, the masses are all smaller compared to experiment.  
The nucleon mass is lower than experiment by 7\% ($2.5\sigma$). 
The strange octet baryons are lighter by 6--9\% with $m_K$ as input, and 
by 2--5\% even with $m_\phi$ as input. 
The $\Sigma$-$\Lambda$ hyperfine splitting is larger by 30\% (50\%)
with $m_K$ ($m_\phi$) input, though the deviation of $0.8\sigma$ ($2.3\sigma$)
is statistically marginal.
The Gell-Mann-Okubo (GMO) relation 
\begin{equation} 
\frac{1}{2}(m_N+m_\Xi) = \frac{1}{4}(3m_\Lambda+m_\Sigma)
\end{equation}
based on first-order flavor $SU(3)$ breaking 
is well satisfied, at 1\% in both $m_K$ and $m_\phi$ input,
though the two sides take values (1.04(2) GeV for the $m_K$-input 
and 1.09(1) GeV for the $m_\phi$-input) smaller than experiment (1.13 GeV).

For decuplet baryons, the mass of $\Delta$ turns out to be consistent
with experiment within statistical error of 2.0\% (0.7$\sigma$). 
An equal spacing rule is well satisfied, 
the three spacings mutually agreeing within statistical errors.
However, the mass splitting is smaller by 30\% on average 
compared to experiment for $m_K$ input, and by 10\% for $m_\phi$ input. 

The results discussed above are based on Q$\chi$PT chiral fits. In order to see
effects of choosing different chiral fit functions, we repeat the
procedure using low-order polynomials in $m_q$,
as was done in traditional analyses. 
Chiral fits and continuum extrapolations for this case are illustrated 
in Figs.~\ref{fig:chiralFit590} to~\ref{fig:chiralFit647} by dashed
lines and in Fig.~\ref{fig:ContinuumFP} by open symbols and dashed
lines, respectively.
Q$\chi$PT and polynomial fits lead to masses which agree within 1.5\%
or $1.6\sigma$. The pattern of the quenched spectrum
remains the same even if one adopts the polynomial
chiral fits. 

\subsection{Reversibility of order of the chiral and continuum extrapolations}
\label{sec:summary:methodB}

In order to obtain the physical hadron mass, one conventionally
carries out chiral extrapolation first and then takes the continuum  
extrapolation (we refer to this as method A). 
These two limiting operations can in principle be reversed, and the
resulting spectrum should be unchanged.
An advantage with the reversed limiting procedure (method B) is that
one need not worry about possible $O(a)$ terms that are present in 
Q$\chi$PT formulae at finite lattice spacings.

The light hadron spectrum from the two methods 
are compared in Fig.~\ref{fig:ReverseK} for the case of the $m_K$ input.
The prediction from the method B denoted by open symbols 
is in good agreement with that of the method A plotted 
with filled symbols within $1.5\sigma$. 

An additional advantage of method B is that the hadron mass formula
can be obtained as a function of an arbitrary quark mass, 
as shown in the Edinburgh plot in Fig.~\ref{fig:EDcont}.

\subsection{Fundamental parameters of QCD}
\label{sec:summary:alphaS-Mq}
The scale parameter $\Lambda$ is the fundamental parameter of QCD. 
We evaluate it in the $\overline {\rm MS}$ scheme to be
\be
\Lambda_{\overline {\rm MS}}^{(0)} = 219.5(5.4) {\rm \ MeV},  \label{eq:Flmd}
\ee
when the scale is fixed by $m_\rho$.

The definition of quark mass for the Wilson quark action is not
unique, because chiral symmetry is broken by terms of $O(a)$. 
We analyze quark masses from two definitions, 
the conventional one through the hopping parameter,
which we call the VWI quark mass (see Sec.~~\ref{sec:measurement:Mq}),
and another defined in terms of the
Ward identity for axial-vector currents (AWI).  

Figures~\ref{fig:Mnormal-MeV} and \ref{fig:Mstrange-MeV} show
$m_{ud}$ and $m_s$ renormalized in the $\overline{\rm MS}$
scheme at $\mu=2$~GeV as functions of $a$.
The VWI and AWI quark masses, differing at finite $a$,  
extrapolate to a universal value in the continuum limit, in accordance
with a theoretical expectation. 

A combined linear extrapolation assuming a unique value in
the continuum limit yields 
\begin{eqnarray}
m_{ud} &=& 4.29(14)^{+0.51}_{-0.79} \;\; {\rm MeV} \label{eq:Fmud}\\
m_s&=& 113.8(2.3)^{+5.8}_{-2.9} \;\; {\rm MeV}\ \ \ m_K{\rm -input},\label{eq:FmsK}\\
   &=& 142.3(5.8)^{+22.0}_{-0}  \;\; {\rm MeV}\ \ \ m_\phi{\rm -input}.\label{eq:FmsP}
\end{eqnarray} 
We indicate systematic error arising mainly from chiral extrapolations.
The value of $m_s$ differs by about 20\% depending on $m_K$ or $m_\phi$ 
used as input. The difference arises from the small value of meson hyperfine 
	splitting in the simulation.

\subsection{Meson decay constants}
\label{sec:summary:f}

The pseudoscalar meson decay constant $f_{PS}$ is defined by
\be
\langle 0 | A_\mu | {\rm PS} \rangle = i p_\mu f_{PS},
\label{eq:fps}
\ee
in the continuum notation, where $A_\mu$ is the axial-vector current. 
The experimental value for $\pi$ is $f_\pi = 132$ MeV.
Data for $f_{PS}$ are shown in Fig.~\ref{fig:fPSCont}
as a function of $a$. We obtain for physical values
\begin{eqnarray}
f_\pi&=&120.0(5.7) \text{\ MeV}, \label{eq:fpifnl}\\
f_K&=&138.8(4.4) \text{\ MeV}\ \ \ m_K{\rm -input}. \label{eq:fKfnl}
\end{eqnarray}
These values are smaller than experiment by 9\% (2$\sigma$) and 
13\% (5$\sigma$), respectively.
Q$\chi$PT predicts that the ratio $f_K/f_\pi-1$ in the quenched QCD is
smaller than experiment by about 30\%. 
This quantity is shown in Fig.~\ref{fig:fPSRatCont}. 
We obtain $f_K/f_\pi-1 = 0.156(29)$ which is smaller than experiment by
26\% ($1.9\sigma$) as Q$\chi$PT predicts. 

The vector meson decay constant $F_V$ in the continuum theory is defined by
\be
\langle 0|V_i|V\rangle = \epsilon_i F_V m_V, 
\label{eq:fv}
\ee
where $\epsilon_i$ and $m_V$ are the polarization vector and the mass of the 
vector meson $V$.
This is related with another conventional definition by $f^{-1}_V=F_V/m_V$.
The experimental value of $F_\rho$ is 220(5) MeV, 
where the charge factor is removed. 
Figure~\ref{fig:FVcont} summarizes the vector meson decay constants. 
We obtain 
\begin{eqnarray}
F_\rho&=&205.7(6.6) \text{\ MeV}, \label{eq:frhofnl}\\
F_\phi&=&229.4(5.7) \text{\ MeV}\ \ \ m_\phi{\rm -input}. \label{eq:fphifnl}
\end{eqnarray}
These values are slightly smaller than experiment; 
by 6.7\% (2.2$\sigma$) for $F_\rho$ and by 3.8\% (1.6$\sigma$) 
for $F_\phi$.

We summarize meson decay constants in Table~\ref{tab:fFinal}.

\section{Measurements of hadron masses and quark masses}
\label{sec:measurement}

\subsection{Quark propagators}
\label{sec:measurement:quark_prop}

We calculate the quark propagator $G(m)$ at a value of $\kappa$ 
by solving
\be
 \sum_m D(\kappa,n,m) G(m) = S(n), 
\label{eq:qprop}
\ee 
where $D(\kappa,n,m)$ is the quark matrix defined in 
Eq.~(\ref{eq:quarkmatrix}), and $S(n)$ is the quark source. 
In order to enhance the ground state signal in the hadronic measurements, 
we use smeared quark sources. 
For this purpose, we fix gauge configurations to the Coulomb gauge
as described in Appendix~\ref{appendix:gfix}.

For the smeared source, we employ an exponential form given by 
\be
S(n) =  \left\{  \begin{array}{ll}
                  A \exp( -B|n| ) & \quad\mbox{for}\quad   n \ne 0 \\
                  1.0             & \quad\mbox{for}\quad   n = 0
                  \end{array}
         \right.,  
\label{eq:expsmear}
\ee
as motivated by the pion wave function measured by 
the JLQCD Collaboration\cite{ref:JLQCD-smear}. 
The smearing radius is approximately constant, $a/B \approx 0.33$ fm,
over the range of $\beta$ we simulate. 
The quark propagator solver and the smearing function
are discussed in Appendix~\ref{appendix:qprop}.

\subsection{Hadron masses}
\label{sec:measurement:Mhad}

From quark propagators, we construct hadron propagators 
corresponding to degenerate combinations,
$ff$ and $fff$ ($f=s_1, s_2, u_1, u_2, u_3$), 
as well as non-degenerate combinations of the type $s_i u_j$ for
mesons, and $s_i s_i u_j$ and $s_i u_j u_j$ for baryons; 
two quarks in baryons are taken to be degenerate.
We study pseudoscalar and vector mesons, and 
spin $1/2$ octet and spin $3/2$ decuplet baryons. 
The hadron operators are summarized in Appendix~\ref{appendix:hadron}. 

Hadron propagators are calculated for all possible combinations of 
point and smeared sources. At the sink we use only point operators. 
Effective masses $m_{eff}(t)$ for various combination of quark sources
are compared in Fig.~\ref{fig:smear}. 
With our choice of smearing function, $m_{eff}(t)$, in almost all
cases reaches a plateau from above, suggesting that the smearing
radius is smaller than the actual spread of hadron wave functions. 
The onset of plateau is the earliest when the smeared
source is used for all quarks, and statistical error 
is the smallest for this case. 
In light of this advantage, we extract masses from hadron propagators with 
all quark sources smeared.

In order to illustrate the quality of data, typical effective masses are
shown in Figs.~\ref{fig:pro590}--\ref{fig:pro647}
for degenerate octet baryons at the four $\beta$ values. 
We extract the ground state masses using a single hyperbolic cosine 
fit for mesons and a single exponential fit for baryons,
taking account of correlations among different time slices.
In Figs.~\ref{fig:pro590}--\ref{fig:pro647}, 
the horizontal lines are the fit and error,
with the range of the lines representing the fit range. 

The fit range $[t_{min},t_{max}]$ is chosen based on the following 
observations:
(1) 
Value of $\chi^2/N_{df}$ decreases as $t_{min}$ increases and
becomes almost constant at a time slice which we denote as $t_\chi$.
$t_\chi$ in general depends on quark masses.
(2)
The effective mass 
shows a plateau for $t \simgt t_\chi$.
(3)
When $t_{max} \ge t_{\chi}+3$, $\chi^2/N_{df}$ is insensitive to the
choice of $t_{max}$. 
From these findings, we may use $t_\chi$ for $t_{min}$.
However, in order to avoid subjectivity in the identification
of $t_\chi$ and plateau, we adopt $t_{min}$ which satisfies
the following conditions:
\begin{itemize}
\item[(1)]
$t_{min}$ is larger than $t_\chi$.
\item[(2)]
For each kind of particle at each $\beta$, $t_{min}$ is common
to all quark masses.
\item[(3)]
For each particle, values of $t_{min}$ in physical units is
approximately constant for all $\beta$.
\end{itemize}
We find that these conditions 
are satisfied by $t_{min} \approx 1.0$ fm, 0.7 fm and 0.5 fm 
for pseudoscalar mesons, vector mesons and baryons, respectively.

The largest time slices $t_{max}$ for vector mesons and
baryons at $\beta=5.9$, 6.1 and 6.25 are chosen by the requirement
that the error of propagator at $t_{max}$ does not exceed 3\%.
We employ the same criterion for vector mesons at $\beta=6.47$.
For baryons at $\beta=6.47$, the fitting interval becomes too narrow
for light quark masses if we employ the cut at 3\%. 
We therefore adopt the cuts at 3.2\% for octet baryons and 4\% for
decuplet baryons, respectively.

Values of $t_{max}$ are determined with a different strategy
for pseudoscalar mesons, for which we make chiral extrapolations 
taking account of correlations among different quark masses.
(Correlation among different quark masses is ignored for other hadrons.)
A large value of $t_{max}$ results in full covariance 
matrices with too large dimensions.
Such matrices frequently have quite small eigenvalues due to statistical 
fluctuations, and lead to a failure of the convergence of the fit. 
In order to avoid instability of chiral extrapolations, we determine 
$t_{max}$ by trial and error.
We adopt $t_{max}=28$, 35, 25 and 35 for 
$\beta= 5.9$, 6.1, 6.25 and 6.47.

All hadron masses are stable under a variation of the fit
range. As an example, the fits with the range $[t_{min}+2,t_{max}]$ 
give results consistent within $1\sigma$.
Uncorrelated fits yield masses consistent with those from 
correlated fits within 1$\sigma$ for most cases, although the
differences are about 2$\sigma$ for some cases. 

Errors are determined by a unit increase of $\chi^2$. 
Jackknife errors are similar in magnitude, the difference being
at most about 25\%. 
The values of $\chi^2/N_{df}$ turn out to be consistent or smaller 
than unity within errors estimated by the jackknife method,
meaning that our one-mass fits reproduce the hadron propagator data well.
This suggests that the contamination of excited states are well 
suppressed in our fits. 
Our results for the hadron masses 
are reproduced in
Tables~\ref{tab:massmeson} for mesons and 
Tables~\ref{tab:massoct} and 
\ref{tab:massdec} for octet and decuplet baryons.

\subsection{Quark masses}
\label{sec:measurement:Mq}

The bare quark mass is conventionally defined by
\be
m_q^{VWI(0)} = \frac{1}{2} \left(\frac{1}{\kappa}-\frac{1}{\kappa_c}\right),
\label{eq:VWImq0}
\ee
where $\kappa_c$ is the critical hopping parameter at which the pion
mass vanishes.  
This quark mass is called the vector Ward identity(VWI) 
quark mass, since the divergence of the vector current is proportional to
the mass difference of quark flavors in the current,
which looks similar to Eq.~(\ref{eq:VWImq0}).

The definition of the AWI quark mass is based on the Ward identity for
axial-vector currents \cite{ref:Bochicchio}.  For the flavor combination 
$(f,g)$, it takes the form,   
\bea
\frac{1}{a} \langle\overline\nabla_\mu A_\mu(n){\cal O}\rangle
&=& (m_f + m_g) \langle P(n){\cal O}\rangle 
\nonumber \\
&& + \langle \delta {\cal O} \rangle+O(a),
\label{eq:AWI}
\eea
where
\be
A_\mu (n) = \frac{1}{2} \Big\{
  \bar f_{n+\hat\mu}U_{n,\mu}^\dagger i\gamma_\mu\gamma_5 g_n
+ \bar f_nU_{n,\mu} i\gamma_\mu\gamma_5 g_{n+\hat\mu} \Big\}
\label{eq:extendAmu}
\ee
is the axial-vector current  and 
\be
P(n) = \overline{f}_n \gamma_5 g_n
\ee
is the pseudoscalar density.
In Eq.(\ref{eq:AWI}), 
$\overline\nabla_\mu F(n) = F(n) - F(n-\hat\mu)$ is 
the backward lattice derivative, and 
$\delta {\cal O}$ is the response of the operator ${\cal O}$ under 
the chiral transformation. 
The $O(a)$ term is due to explicit violation of chiral
symmetry with the Wilson quark action.

In order to extract $m_q^{AWI(0)}$, we use the relation \cite{ref:Itoh86} 
\be
m_f^{AWI(0)} + m_g^{AWI(0)}  = \lim_{t\to\infty}
\frac{\langle\overline\nabla_4 A_4(t)P(0)\rangle}
     {\langle P(t)P(0)\rangle}, 
\label{eq:AWImq0}
\ee
where
\bea
A_4(t) = && \sum_{\vec n} A_4(n=({\vec n},t)), \\
P(t)   = && \sum_{\vec n} P(n=({\vec n},t)).
\eea
are projected to zero spatial momentum.
The right-hand side of Eq.(\ref{eq:AWImq0}) is evaluated by a constant fit to
the ratio  
\be
\frac{\langle\overline\nabla_4 A_4(t)P(0) \rangle_{\rm sym}}
{\langle P(t) P(0) \rangle} ,
\label{eq:etaE}
\ee
where the suffix $sym$ implies a symmetrization of the derivative
defined by 
\bea
&& \langle \overline\nabla_4 A_4(t)P(0) \rangle_{\rm sym} 
\equiv \nonumber \\
&& ( G(t) - G(t-1) + G(L_t-t) - G(L_t-t-1))/4 
\eea
with $G(t) = \langle A_4(t)P(0) \rangle$. 
For the pseudoscalar operator at the origin, 
we use the smeared source for two quarks.
Figures~\ref{fig:e-mq590} and \ref{fig:e-mq647} illustrate 
our data for Eq.~(\ref{eq:etaE}) at $\beta=5.9$ and 6.47.
The constant fit is carried out without taking account of
correlations between different time slices or between the 
two correlators. 
Fit ranges are determined from the plateau of the effective mass.
The results for $m_f^{AWI(0)}+m_g^{AWI(0)}$ are summarized in
Table~\ref{tab:AWImq}.

\section{Quenched chiral singularities}
\label{sec:qChPT}

The chiral extrapolation is conventionally carried out assuming 
a low-order polynomial in quark masses.
Chiral perturbation theory\cite{ref:CPT}, however, predicts   
a singular quark mass dependence in the chiral limit
due to the presence of massless pions.
The singularity is expected to be enhanced in quenched QCD 
\cite{ref:QChPTSharpe,ref:QChPTBGmass}
since the $\eta\prime$ meson is also massless in this approximation.
In order to choose the functional form for the chiral extrapolation, 
we examine whether hadron mass data are consistent with the predictions of 
Q$\chi$PT. 

\subsection{Mass ratio test for pseudoscalar meson}
\label{sec:qChPT:mPS}

For pseudoscalar mesons made of quarks with masses $m_1$ and $m_2$,
Q$\chi$PT predicts the mass formula\cite{ref:QChPTSharpe,ref:QChPTBGmass}
given by 
\bea
&& m_{PS,12}^2 = \nonumber \\
&& A \, (m_1+m_2)\Big\{ 1 - \delta \big[\ln\frac{2m_1A}{\Lambda_\chi^2}
  + \frac{m_2}{m_2-m_1} \ln\frac{m_2}{m_1} \big] \nonumber \\
&& + \frac{\alpha_\Phi A}{12\pi^2 f^2} 
 \big[m_1\ln\frac{2m_1A}{\Lambda_\chi^2} 
     +m_2\ln\frac{2m_2A}{\Lambda_\chi^2}  \nonumber \\
&&     +\frac{m_1 m_2}{m_2-m_1}\ln\frac{m_2}{m_1} \big] \Big\} \nonumber \\
&& + B \, (m_1+m_2)^2 + O(m^3,\delta^2), 
\label{eq:mps0}
\eea
where terms proportional to $(m_1-m_2)^2$ are absent\cite{ref:SharpePC}.
The logarithmic term proportional to $\delta$ represents the leading 
quenched singularity.
To the leading order in the $1/N_c$ expansion in terms of the number of 
colors $N_c$, $\delta$ is related to
the pseudoscalar meson mass and the pion decay constant $f$ by 
\be
\delta = \frac{m_{\eta'}^2 + m_\eta^2 - 2 m_K^2}{24\pi^2 f^2}.
\ee
Taking the experimental values of $f$ and pseudoscalar meson masses, 
one finds $\delta \approx 0.2$ as a phenomenological estimate.
The constant $\alpha_\Phi$ in Eq.(\ref{eq:mps0}) represents the
coefficient of the kinetic term of the flavor singlet meson field,
which is sub-leading in terms of $1/N_c$.  
The mass formula also contains a scale $\Lambda_\chi$ of $O(1)$ GeV. 
The parameters such as $\delta, \alpha_\Phi$ and $\Lambda_\chi$ may
differ in quenched QCD from those in the full theory. 

In order to see whether pseudoscalar mass data exhibit the presence 
of the logarithmic terms, 
we investigate the ratio $m_{PS,12}^2/(m_1+m_2)$ as a function
of $m_1+m_2$.
We use the AWI quark mass rather than the VWI mass to avoid uncertainties 
due to the necessity of choosing $\kappa_c$, which 
in turn depends on details of chiral extrapolations. 
Another important point with the use of the AWI quark mass is 
that it is free from chiral singularities which cancel 
between the numerator and the denominator 
in Eq.~(\ref{eq:AWImq0}) \cite{ref:SharpePC,ref:GoltermanPC}.

In Fig.~\ref{fig:PS2mq}, we plot $m_{PS,12}^2/(m_1+m_2)$
as a function of $m_1+m_2$.  The AWI quark mass  is converted 
to renormalized values in the ${\overline {\rm MS}}$ scheme at the scale 2 GeV 
(see Sec.~\ref{sec:Mq}) and the ratio is translated to physical units.
This enables us to compare the results at different values of $\beta$
with the same scale of the figure.
We find a clear increase of the ratio towards the chiral limit 
at all values of $\beta$ as expected from Eq.~(\ref{eq:mps0}).

In order to make a more quantitative analysis, 
we consider the ratio defined by 
\be
y = {\frac{2m_1}{m_1+m_2} \frac{m_{PS,12}^2}{m_{PS,11}^2}} \times 
   {\frac{2m_2} {m_1+m_2} \frac{m_{PS,12}^2}{m_{PS,22}^2}}. 
\label{eq:qchpty}
\ee
Assuming that $\delta$ and $\alpha_\Phi$ as well as quark masses are small, 
we expect
\be
y = 1 + \delta\, x + \alpha_X \, z + O(m^2,\delta^2), 
\label{eq:qchiptxy}
\ee
where 
\be
\alpha_X = \frac{\alpha_\Phi A^2}{12 \pi^2 f^2},
\ee
and the parameters 
\be
x = 2 + \frac{m_1+m_2}{m_1-m_2}\ln\Big(\frac{m_2}{m_1}\Big),
\label{eq:qchptx}
\ee
and 
\be
z = \frac{1}{A}\Big(\frac{2m_1m_2}{m_2-m_1}\ln\frac{m_2}{m_1}-m_1-m_2\Big),
\ee
represent terms of $O(m_q\ln m_q)$ and $O(m_q^2\ln m_q)$ in Eq.~(\ref{eq:mps0}). 

We plot $y$ as a function of $x$ in Fig.~\ref{fig:PSratio}, with 
numerical values listed in Table~\ref{tab:PSratio}.  
The points fall within a narrow ridge limited by two
lines $y \approx 1 + (0.08$--$0.12)\,x$. 
A one-parameter fit ignoring $\alpha_X z$ and higher order 
terms yields $\delta=0.10$--0.12 depending on $\beta$ as listed in 
Table~\ref{tab:delalpha}. 

A two-parameter fit keeping the $\alpha_X z$ term requires the 
value of $A$. We estimate $A\approx$ 3 GeV for all $\beta$ values
from data at $m_{PS}/m_V$=0.75 and 0.7, assuming 
$m_{PS}^2 = 2Am_q$. Setting $f=132$ MeV,  
we obtain $\delta$ and $\alpha_\Phi$ given in the right column 
of Table~\ref{tab:delalpha}.
For the two finer lattices at $\beta=6.25$ and 6.47, 
$\alpha_\Phi$ is consistent with zero and $\delta \approx 0.1$. 
At the coarser lattices the values of $\alpha_\Phi$ and $\delta$ are 
not stable. 

Further data are needed to pin down precise values 
of $\delta$ and $\alpha_\Phi$. We consider 
results at finer lattices, closer to the continuum limit,
are more reliable, and take $\delta=0.10(2)$ and $\alpha_\Phi=0$ 
as our best estimate.

The ratio test for the existence of quenched logarithm terms was
originally proposed in Ref.~\cite{ref:QChPTBGmass}, in which one
plots $y_0$ defined by
\be
y_0 = \frac{m_{PS,12}^2}{m_{PS,11}^2}\frac{2m_1}{m_1+m_2} 
\label{eq:qchpty0}
\ee
as a function of $x_0$ defined by
\be
x_0 = 1 + \frac{m_2}{m_1-m_2} \ln \Big(\frac{m_2}{m_1}\Big). 
\label{eq:qchptx0}
\ee
The relation $y_0 = 1 + \delta \cdot x_0$ follows if we ignore
$O(m_q^2)$ term in Eq.(\ref{eq:mps0}).
As shown in Fig.~\ref{fig:PSratioOrg}, however, $y_0$ systematically 
varies with quark masses, suggesting a contribution from the $O(m_q^2)$ term.
The double ratio defined by Eq.~(\ref{eq:qchpty}) is designed to 
cancel the $O(m_q^2)$ terms, and hence is more effective to observe
the quenched singularity. 

\subsection{Ratio test of pseudoscalar meson decay constant}

Quenched chiral singularities are also expected in meson 
decay constants \cite{ref:QChPTBGdecay}. Let $f_{fg}$ be 
the pseudoscalar meson decay constant for the flavor combination $(f,g)$.   
In Ref.~\cite{ref:QChPTBGdecay}, the ratio
\be
y_f = \frac{f_{12}^2}{f_{11} f_{22}}
\label{eq:decayratio}
\ee
was shown to satisfy the relation 
\be
y_f =1- \frac{\delta}{2} \, x
\ee
with $x$ defined by Eq.(\ref{eq:qchptx}), 
where $\alpha_\Phi$ is set to zero. 
Values of $y_f$ are listed in Table~\ref{tab:PSratio} and
the ratio test with Eq.(\ref{eq:decayratio}) is summarized 
in Fig.~\ref{fig:PSdecayRatio}.
We find $\delta \approx 0.08$--0.16, in agreement with the 
value from the mass ratio analysis.

\subsection{Pseudoscalar meson mass fit}
\label{sec:qChPT:PSfit}

The parameter $\delta$ is estimated also by fitting 
pseudoscalar meson mass data to Eq.~(\ref{eq:mps0}) 
assuming $\alpha_\Phi=0$ for all $\beta$.
The AWI quark mass introduces errors in the fit variable.
Therefore the VWI quark mass, $m_q^{VWI(0)} =
(1/\kappa-1/\kappa_c)/2$, is employed taking $\kappa_c$ as 
a parameter. We carry out fully-correlated fits described in
Appendix~\ref{appendix:CorrFits}, 
independently for degenerate and non-degenerate cases.

A noticeable property of the Q$\chi$PT formula Eq.~(\ref{eq:mps0}) is 
that $A$, $\delta$ and $\Lambda_\chi$ cannot be determined
simultaneously because the three conditions to minimize $\chi^2$ 
are not mathematically independent.
A possible method is to fix $f_\Lambda \equiv 2A^2/\Lambda_\chi^2$. 
Values of $A$ and $\delta$ depend on the choice of $f_\Lambda$,
while $\kappa_c$ and $B$, as well as $\chi^2$ and the 
fit curve are independent.
We consider $f_\Lambda =$ 4, 8, 16, and 32, which correspond to
$\Lambda_\chi \approx 1.32$, 1.00, 0.76, and 0.57 GeV, respectively.%
\footnote{
These values of $\Lambda_\chi$ in physical units are computed 
using $A$ determined by a degenerate fit at $\beta=$ 5.90.
We confirm that $\beta$ dependence of $A$ is very weak if translated
into physical units 
and that degenerate and non-degenerate fits lead to $A$ consistent 
with each other. See.Table \ref{tab:PSQchPTPrm}.}
This range of $\Lambda_\chi$ contains a natural scale for chiral 
perturbation theory, $\Lambda_\chi=m_\rho$, or 1 GeV.

The results are summarized in Table~\ref{tab:delta}.
The value of $\delta$ is stable against a variation of $f_\Lambda$
and $\beta$, and is consistent within $2\sigma$ with our estimate 0.10(2) 
from the mass ratio test.

\subsection{Comparison with other results for $\delta$}
The value of $\delta$ has recently been estimated by other groups. 
The FNAL group reported $\delta = 0.065(13)$~\cite{ref:QchPTFNAL} 
using the clover quark action, and the QCDSF Collaboration 
obtained $\delta \approx 0.14(2)$~\cite{ref:QCDSF} with a non-perturbatively
improved clover action.  
These estimates are consistent with ours. 

It has been pointed out in Ref.~\cite{ref:QchPTAlpha} that 
the $x$--$y$ correlation seen in Fig.~\ref{fig:PSratio} may be reproduced 
with a small $\delta\approx$ 0.03 -- 0.07, if $\alpha_\Phi\approx$
0.5. Our data for large $\beta$, however, do not seem to be
compatible with such a large $\alpha_\Phi$.

\subsection{Vector meson and baryon masses}

Q$\chi$PT predicts singularities of the form $O(m_{PS}) \sim O(\sqrt{m_q})$
for vector mesons and baryons \cite{ref:Booth,ref:LabrenzSharpe}. 
The ratio tests similar to those for the pseudoscalar mesons indicate 
that the coefficient of the $O(m_{PS})$ term is non-vanishing both for 
vector mesons and baryons.  
It is difficult, however, to reliably estimate the coefficients from the 
ratios because of large errors.
Direct fits of mass data to Q$\chi$PT formula are also difficult as
they are not very stable.  
While our data are consistent with Q$\chi$PT, statistics and the range
of the quark mass in our study do not allow conclusive results.
Our tests of the Q$\chi$PT mass formulae for these cases are 
described in Appendix~\ref{appendix:QChPT}.

\section{Hadron mass spectrum}
\label{sec:spectrum}

\subsection{Chiral fits}
\label{sec:spectrum:chiral}
\begin{figure}[htb]
\centerline{\epsfxsize=8.5cm \epsfbox{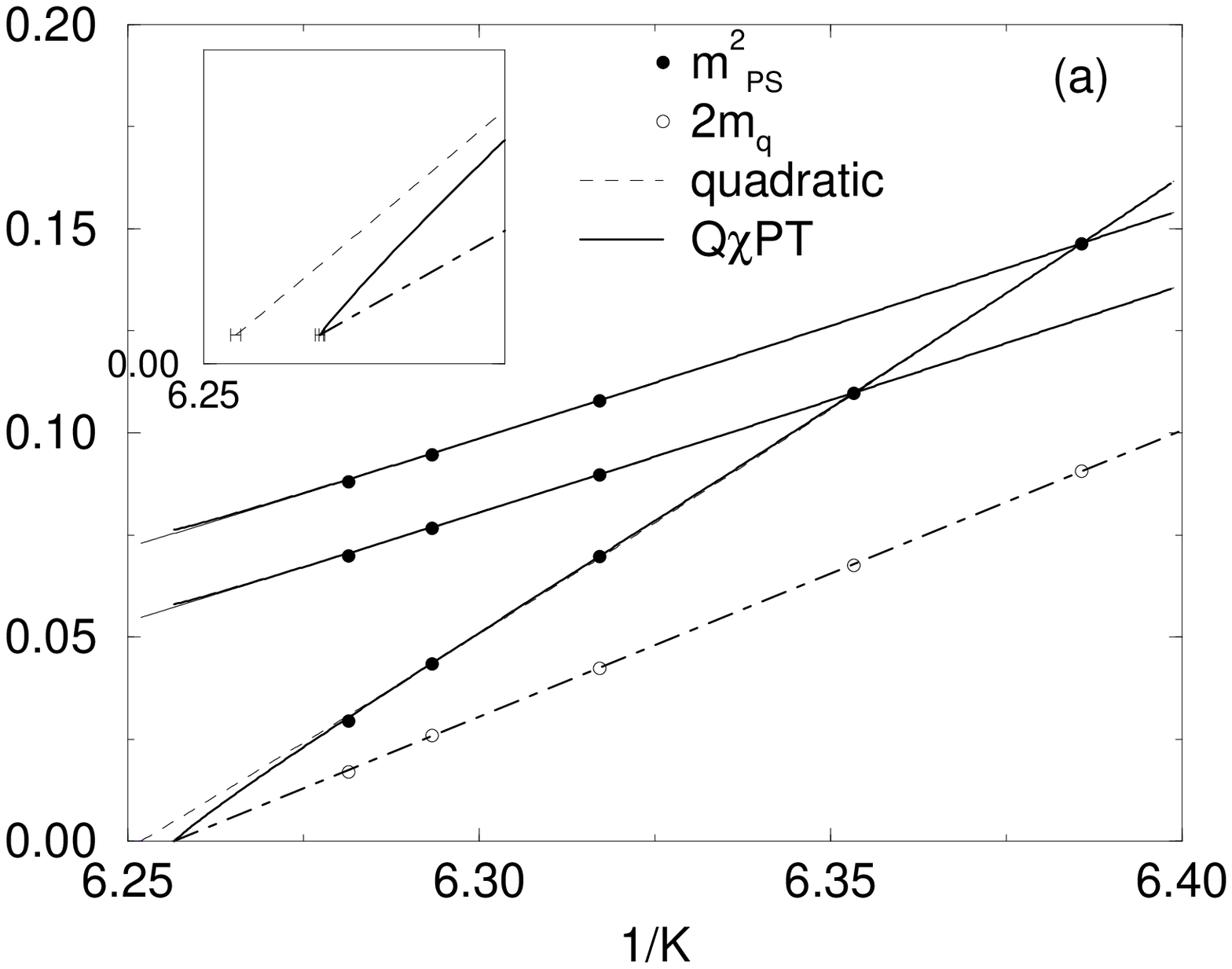}}
\begin{minipage}{8.5cm}
\centerline{\epsfxsize=8.5cm \epsfbox{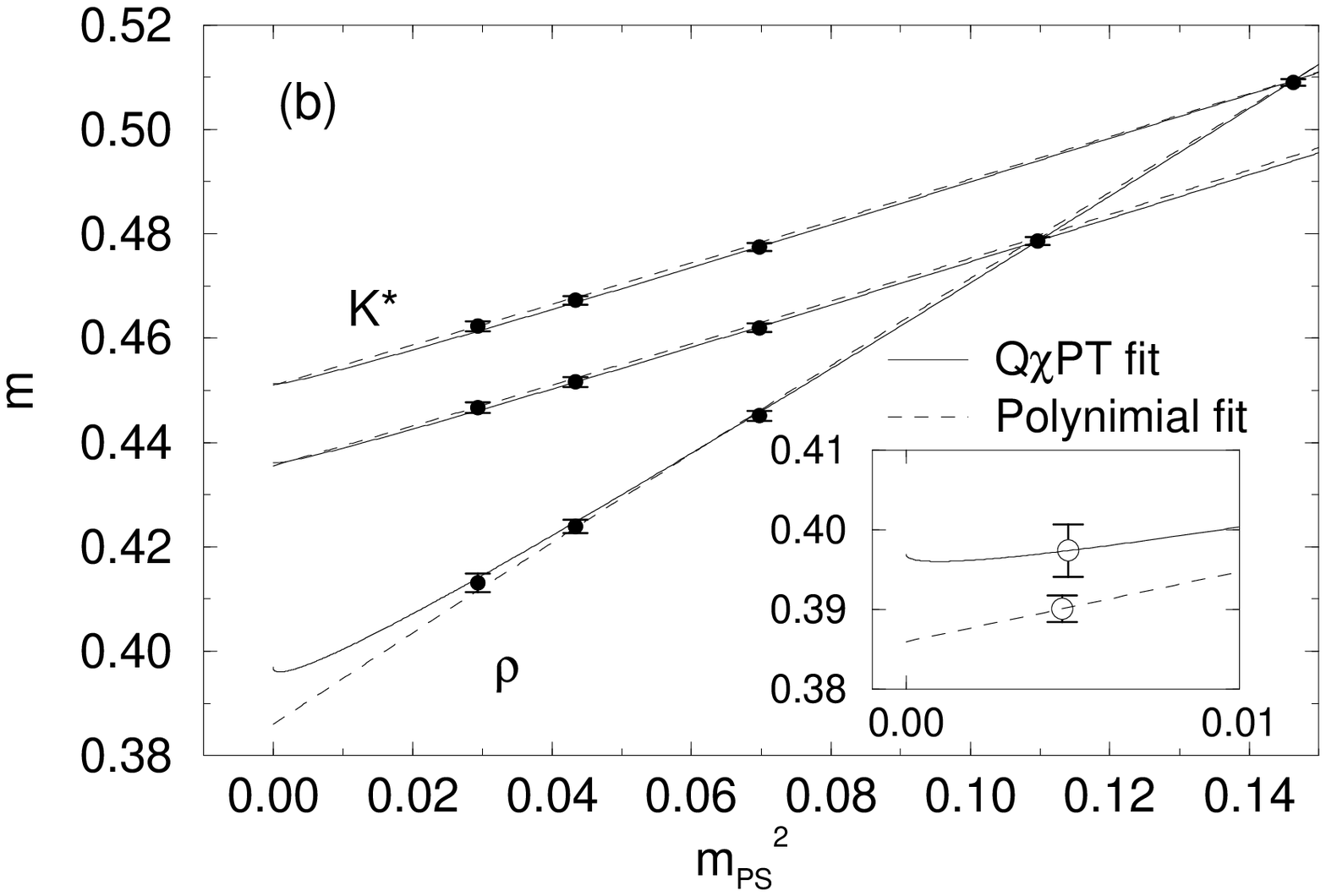}}
\end{minipage}
\begin{minipage}{8.5cm}
\centerline{\epsfxsize=8.5cm \epsfbox{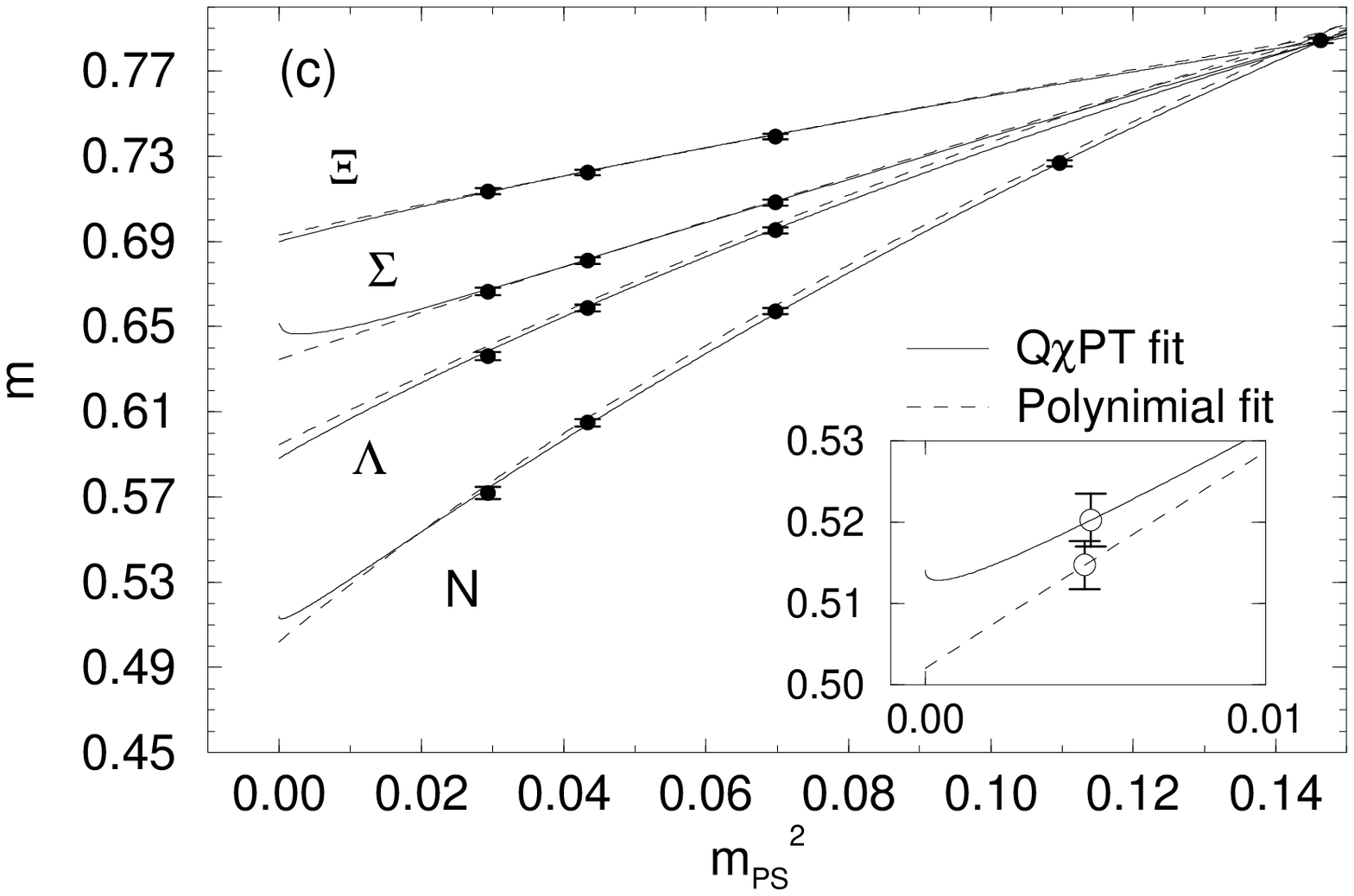}}
\end{minipage}
\centerline{\epsfxsize=8.5cm \epsfbox{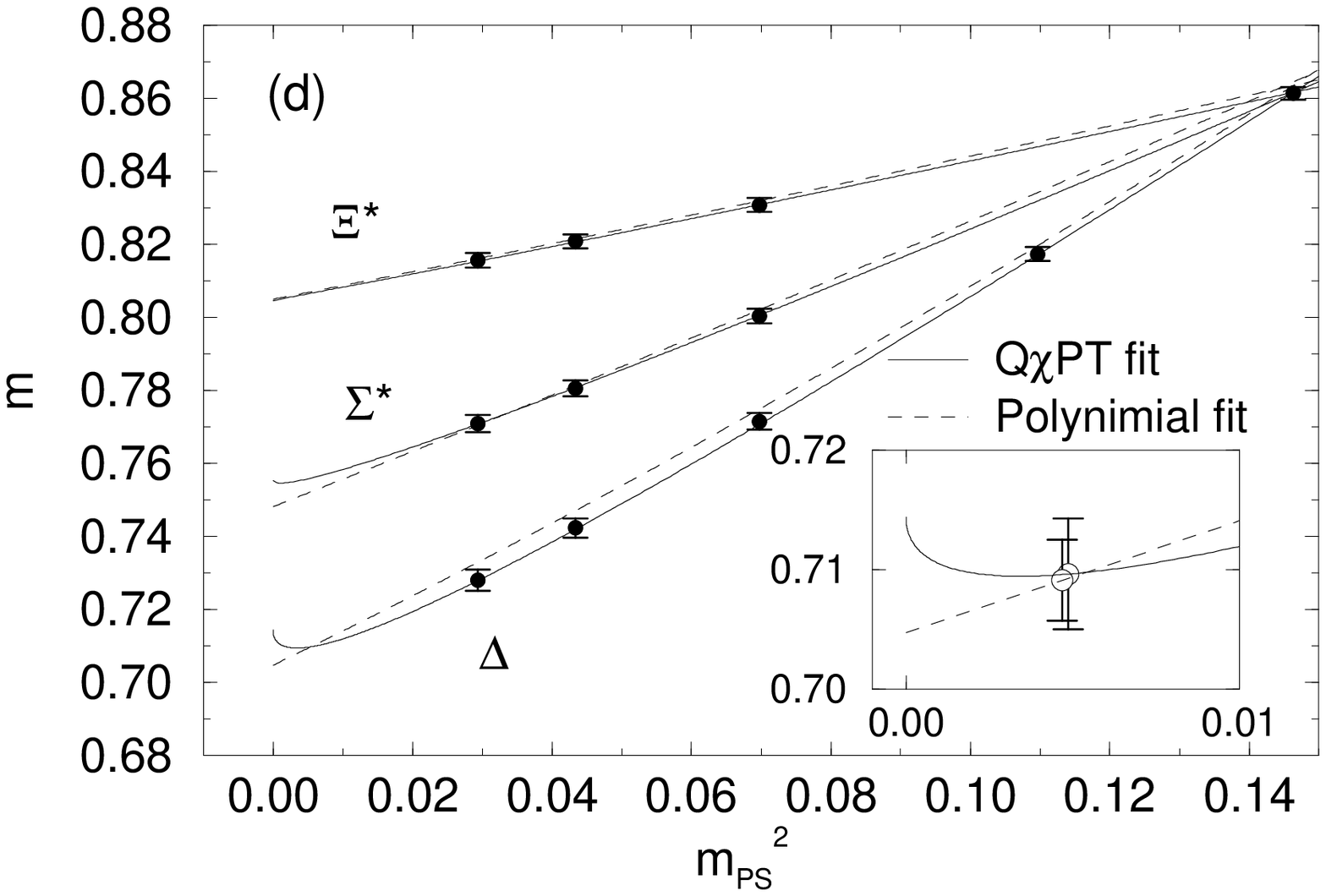}}
\caption{Chiral extrapolations for 
(a) pseudoscalar meson, 
(b) vector meson, (c) octet baryon and (d) decuplet baryon at $\beta=5.9$. 
The Q$\chi$PT and polynomial chiral fits are shown by 
solid and dashed lines. 
The insets are expanded displays for degenerate cases. 
In panel (a) AWI quark mass, $m_q^{AWI(0)}$ 
and linear chiral extrapolation are given (discussed in
Sec.~\protect\ref{sec:Mq}).
In panels (c) and (d), we give data only for combinations of
$(s_1,u_i,u_i)$ and $(u_i,u_i,u_i)$.}
\label{fig:chiralFit590}
\end{figure}

\begin{figure}[htb]
\centerline{\epsfxsize=8.5cm \epsfbox{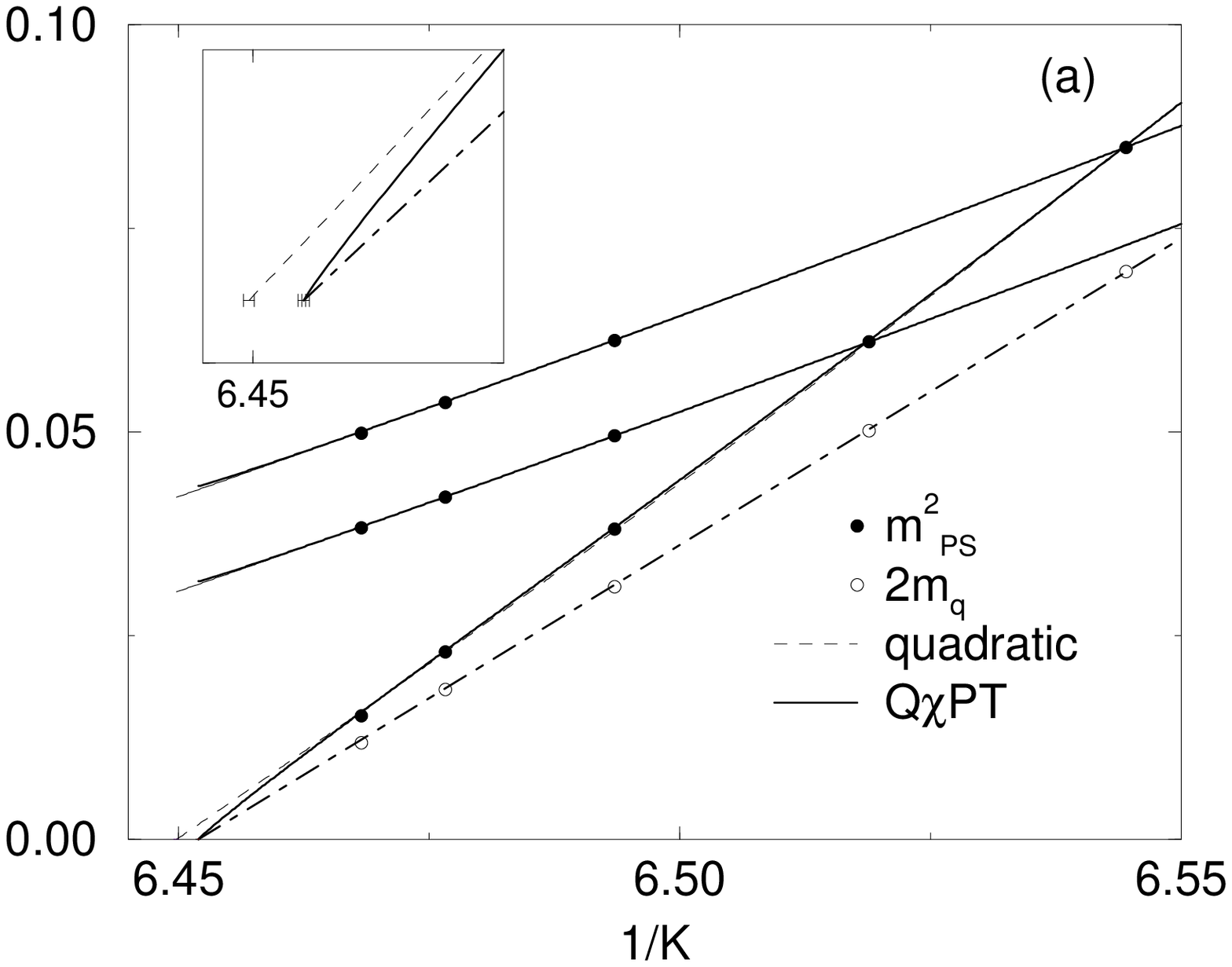}}
\begin{minipage}{8.5cm}
\centerline{\epsfxsize=8.5cm \epsfbox{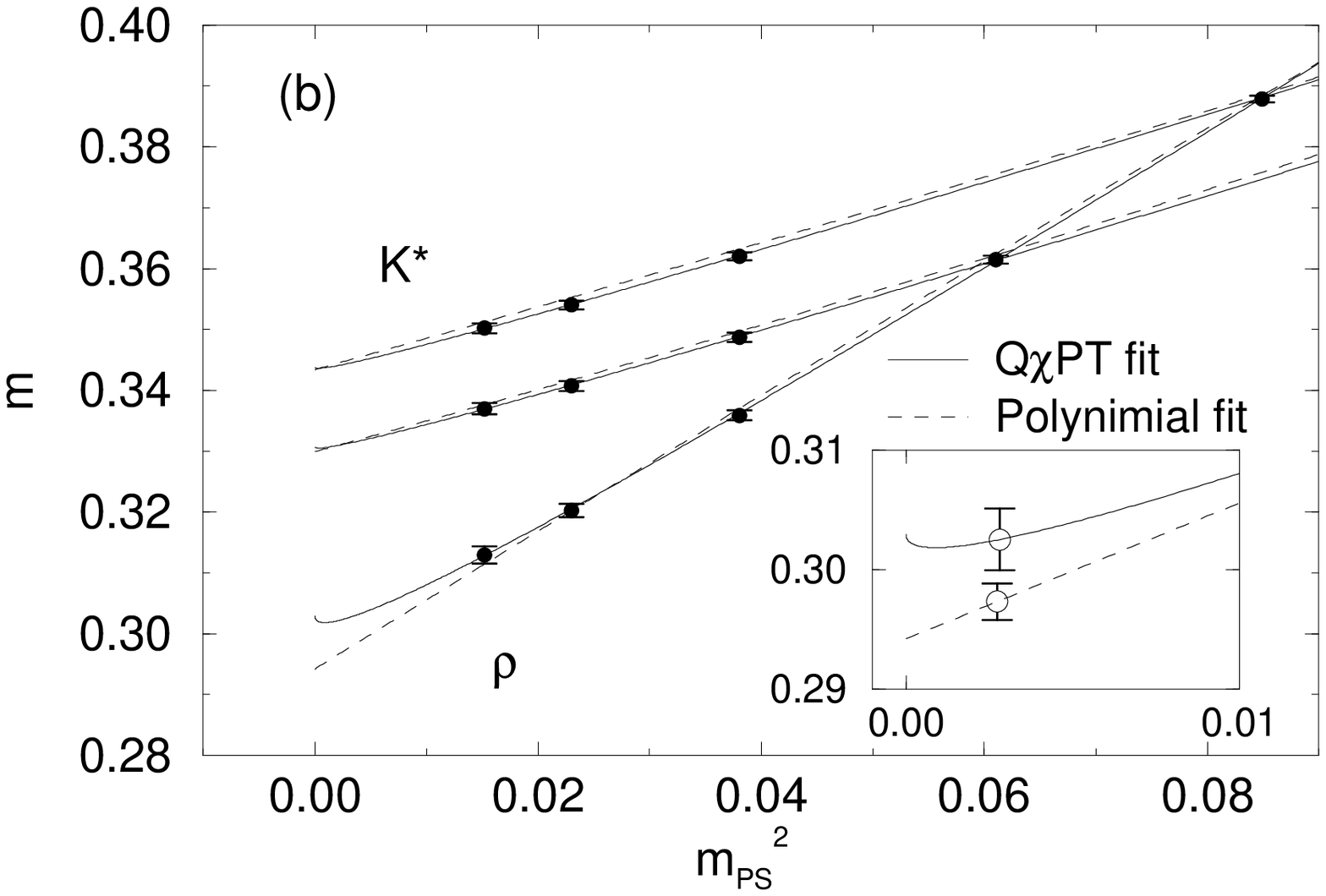}}
\end{minipage}
\begin{minipage}{8.5cm}
\centerline{\epsfxsize=8.5cm \epsfbox{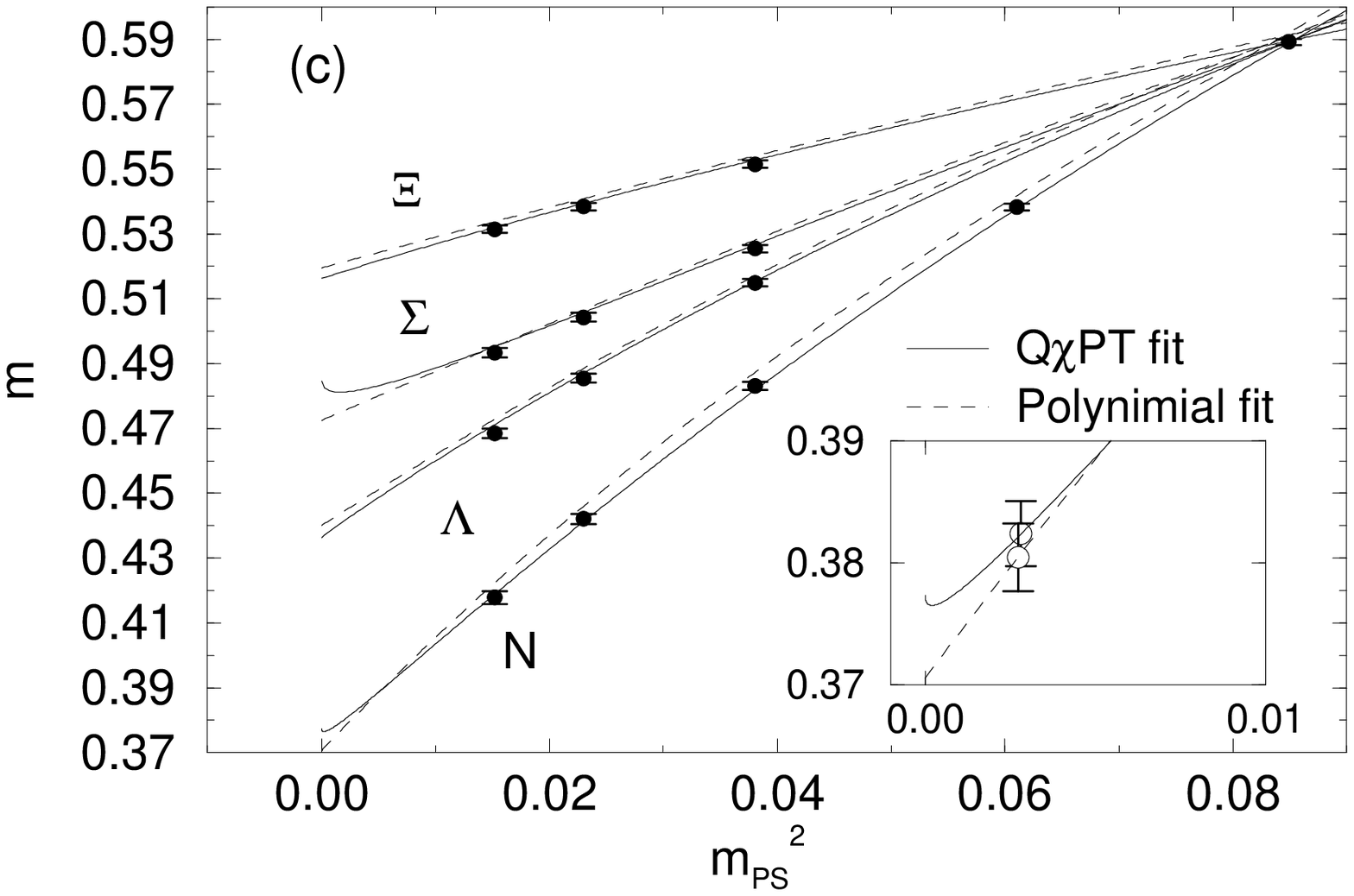}}
\end{minipage}
\centerline{\epsfxsize=8.5cm \epsfbox{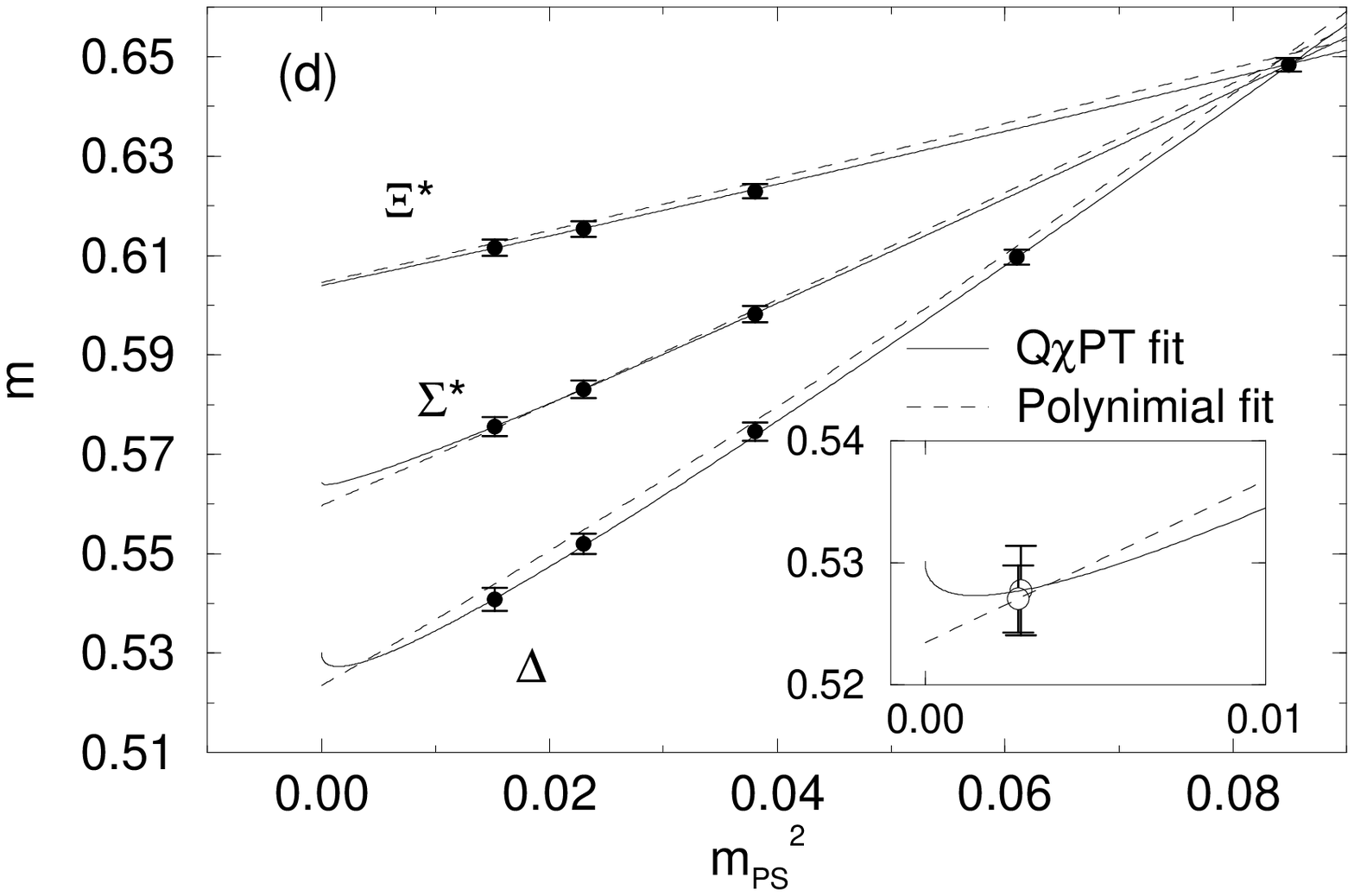}}
\caption{The same for Fig.~\protect\ref{fig:chiralFit590}
at $\beta=$ 6.10.}
\label{fig:chiralFit610}
\end{figure}

\begin{figure}[htb]
\centerline{\epsfxsize=8.5cm \epsfbox{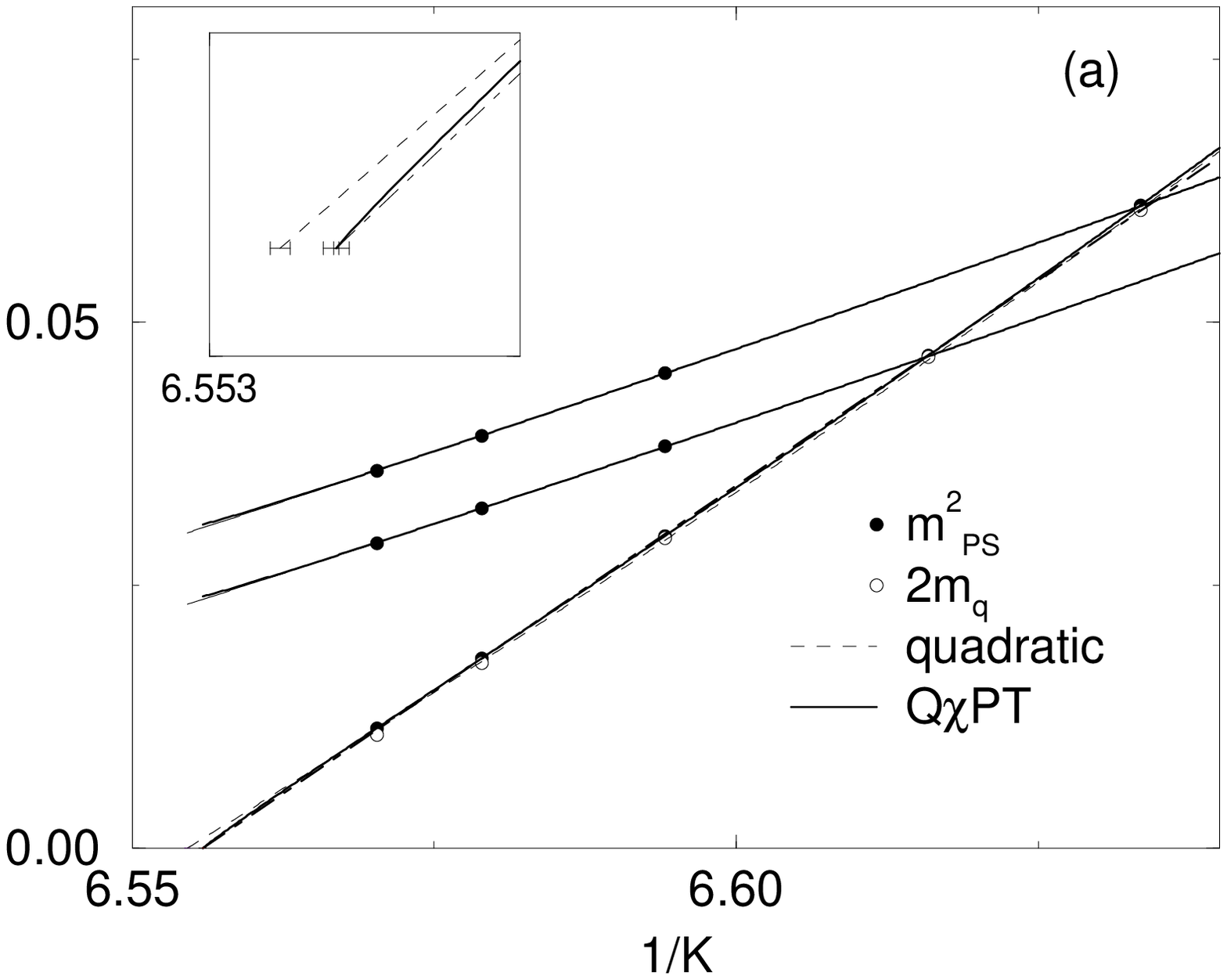}}
\begin{minipage}{8.5cm}
\centerline{\epsfxsize=8.5cm \epsfbox{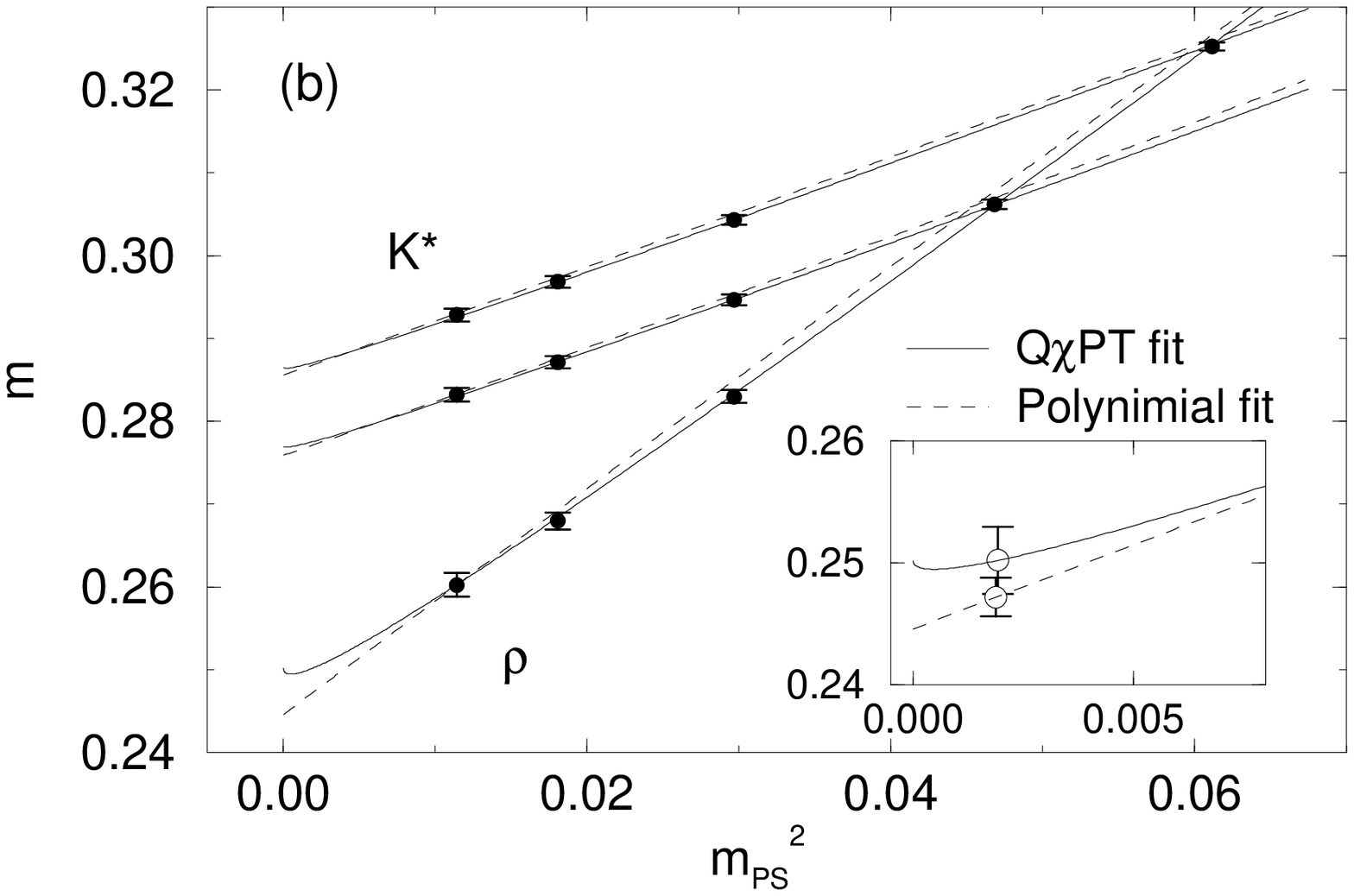}}
\end{minipage}
\begin{minipage}{8.5cm}
\centerline{\epsfxsize=8.5cm \epsfbox{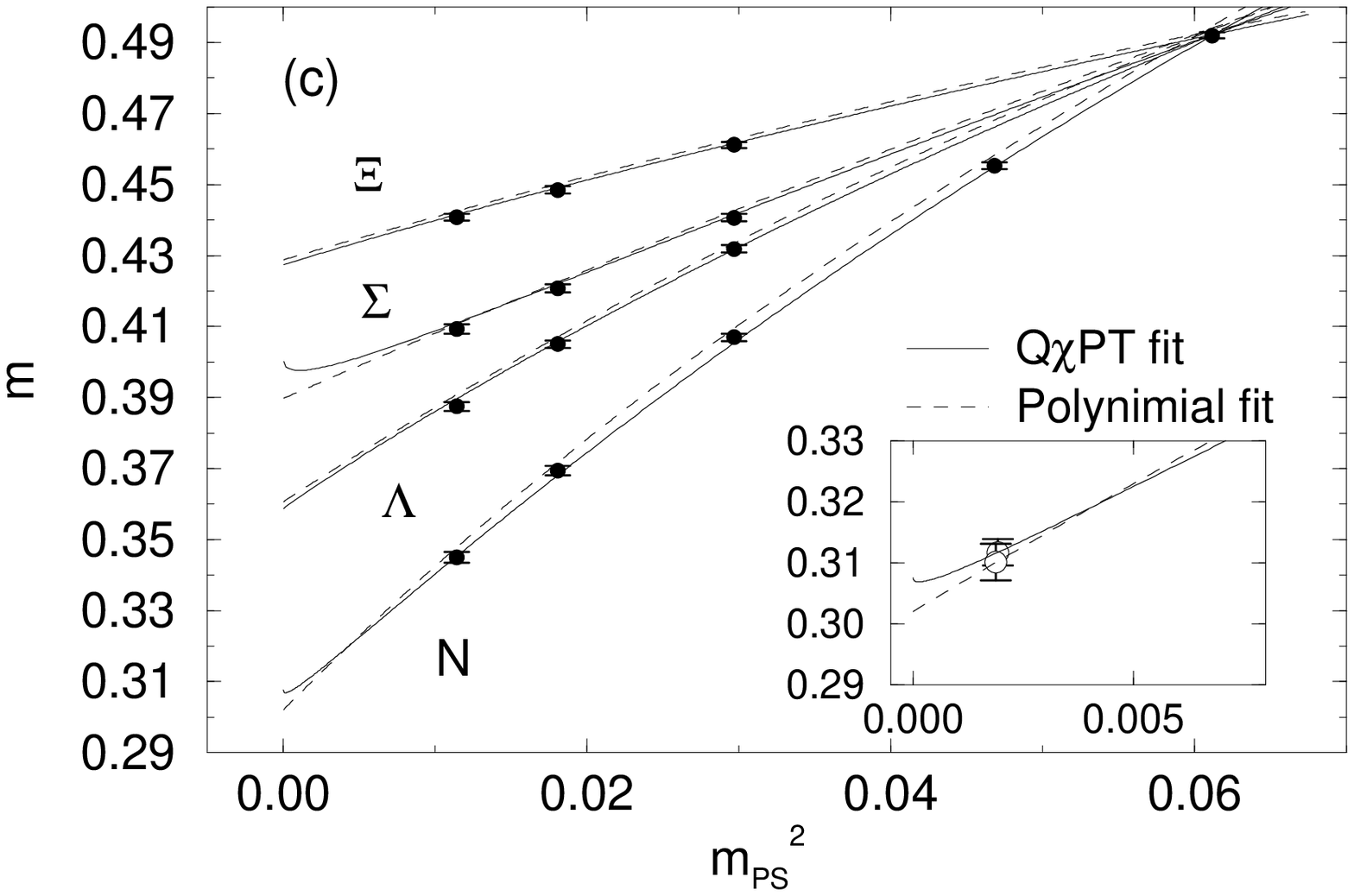}}
\end{minipage}
\centerline{\epsfxsize=8.5cm \epsfbox{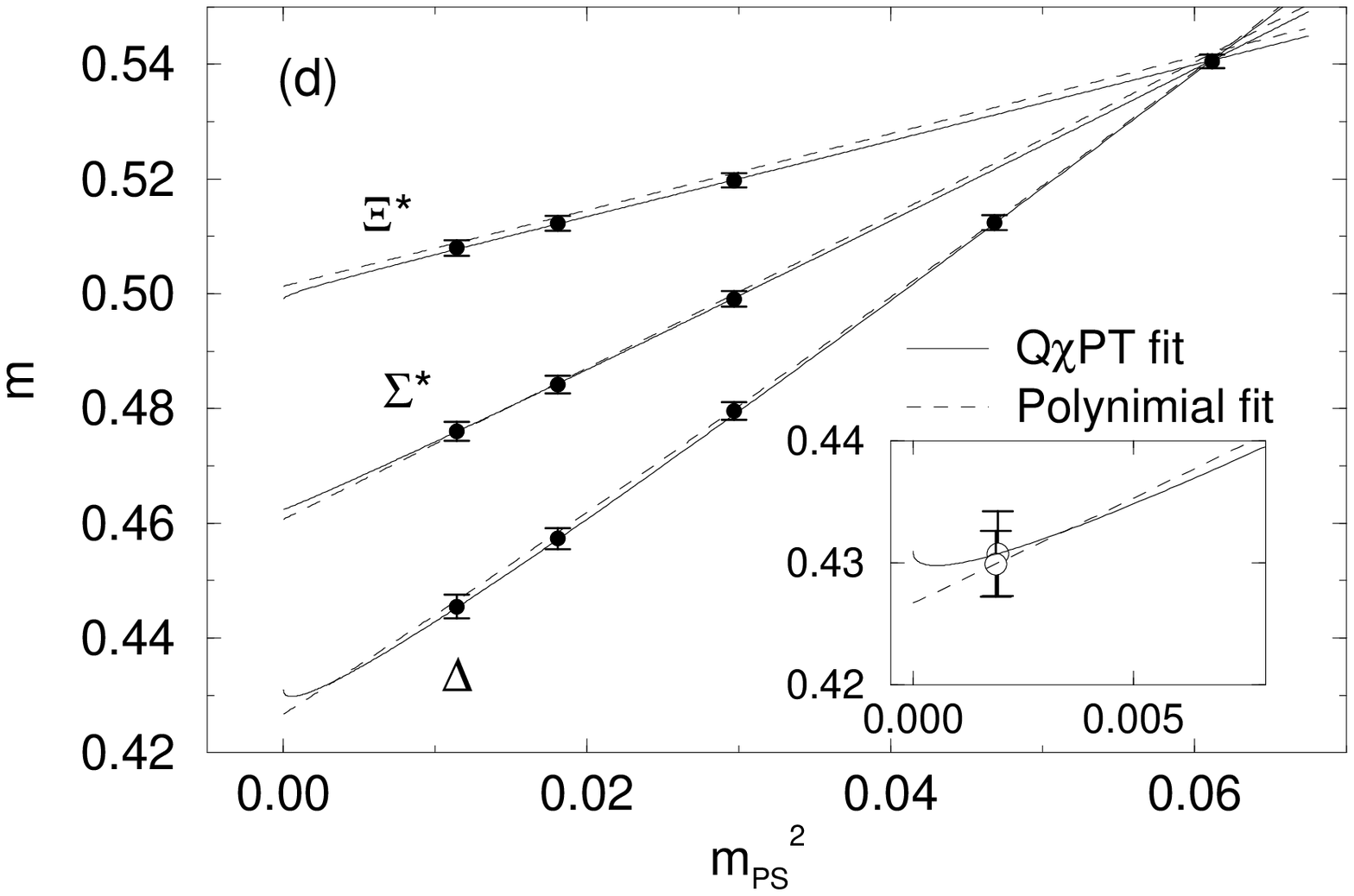}}
\caption{The same for Fig.~\protect\ref{fig:chiralFit590}
at $\beta=$ 6.25.}
\label{fig:chiralFit625}
\end{figure}

\begin{figure}[htb]
\centerline{\epsfxsize=8.5cm \epsfbox{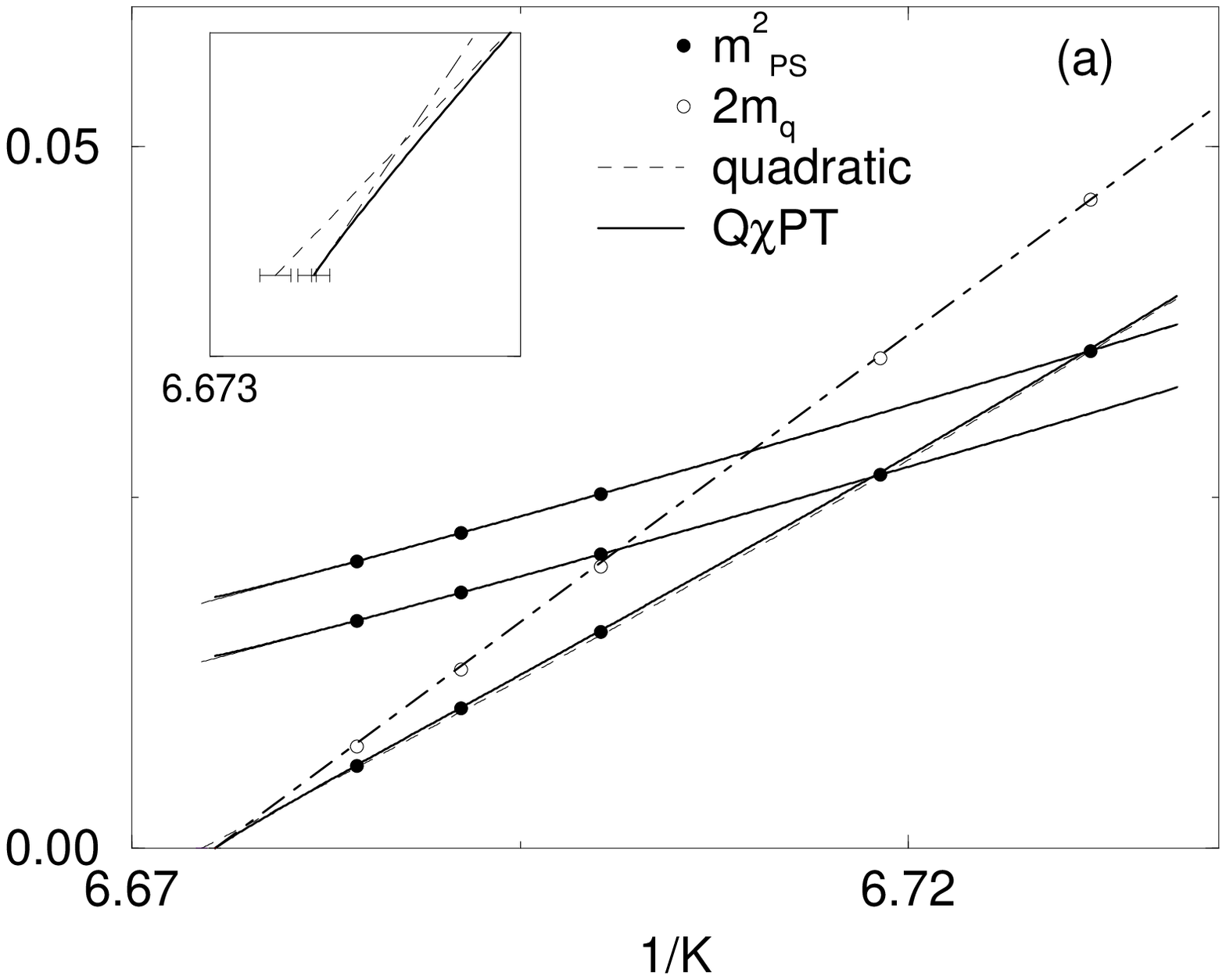}}
\begin{minipage}{8.5cm}
\centerline{\epsfxsize=8.5cm \epsfbox{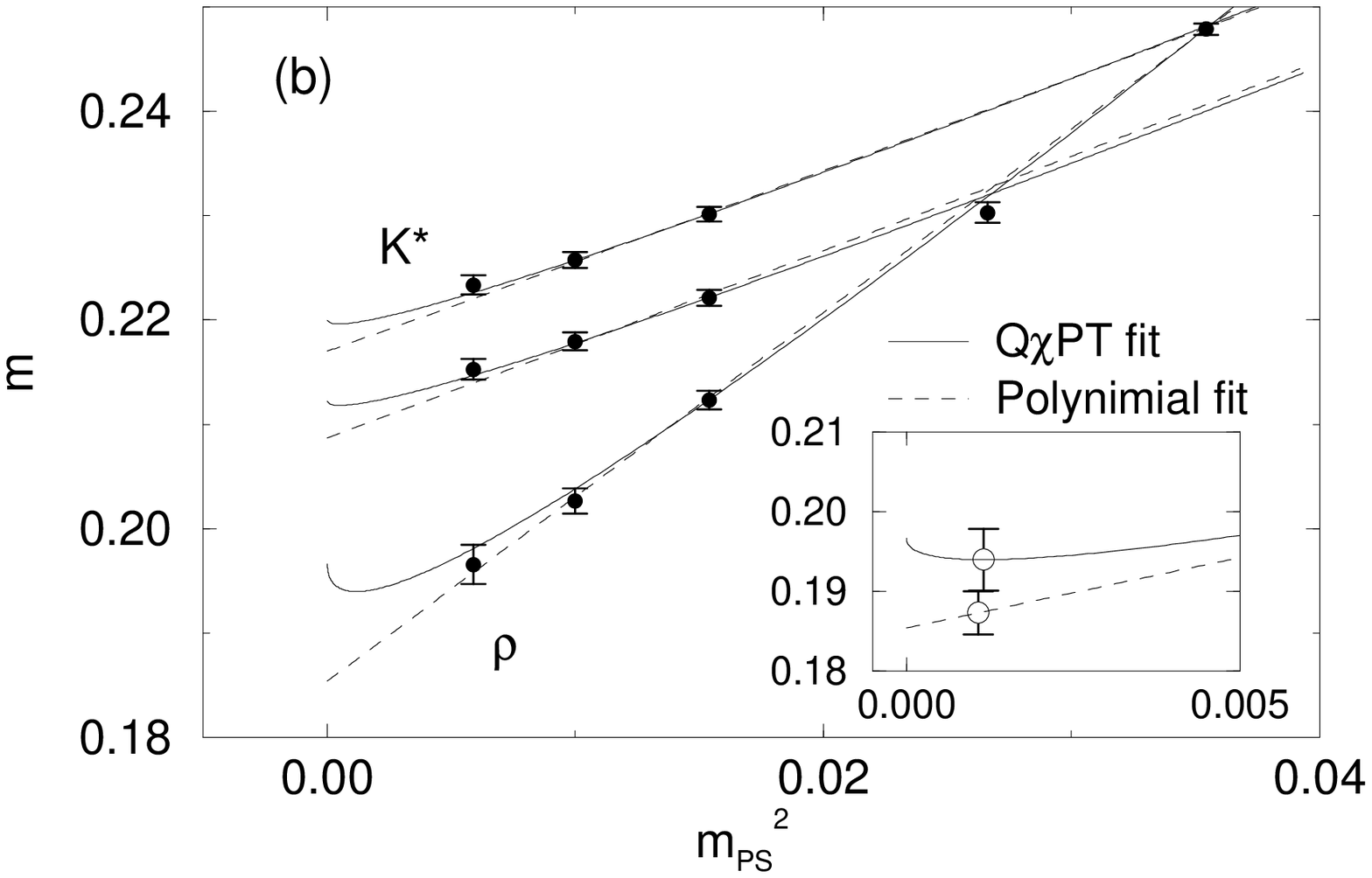}}
\end{minipage}
\begin{minipage}{8.5cm}
\centerline{\epsfxsize=8.5cm \epsfbox{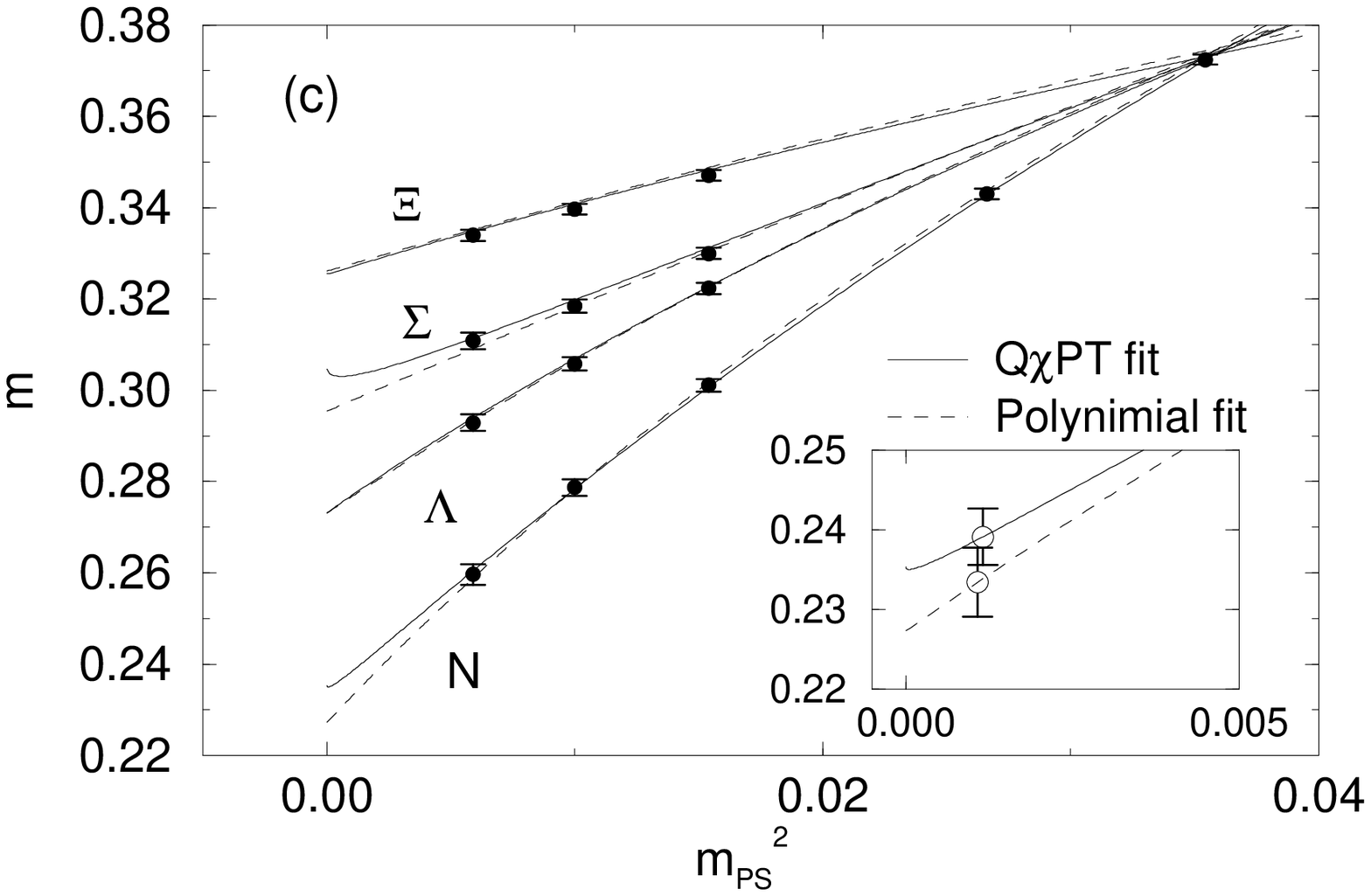}}
\end{minipage}
\centerline{\epsfxsize=8.5cm \epsfbox{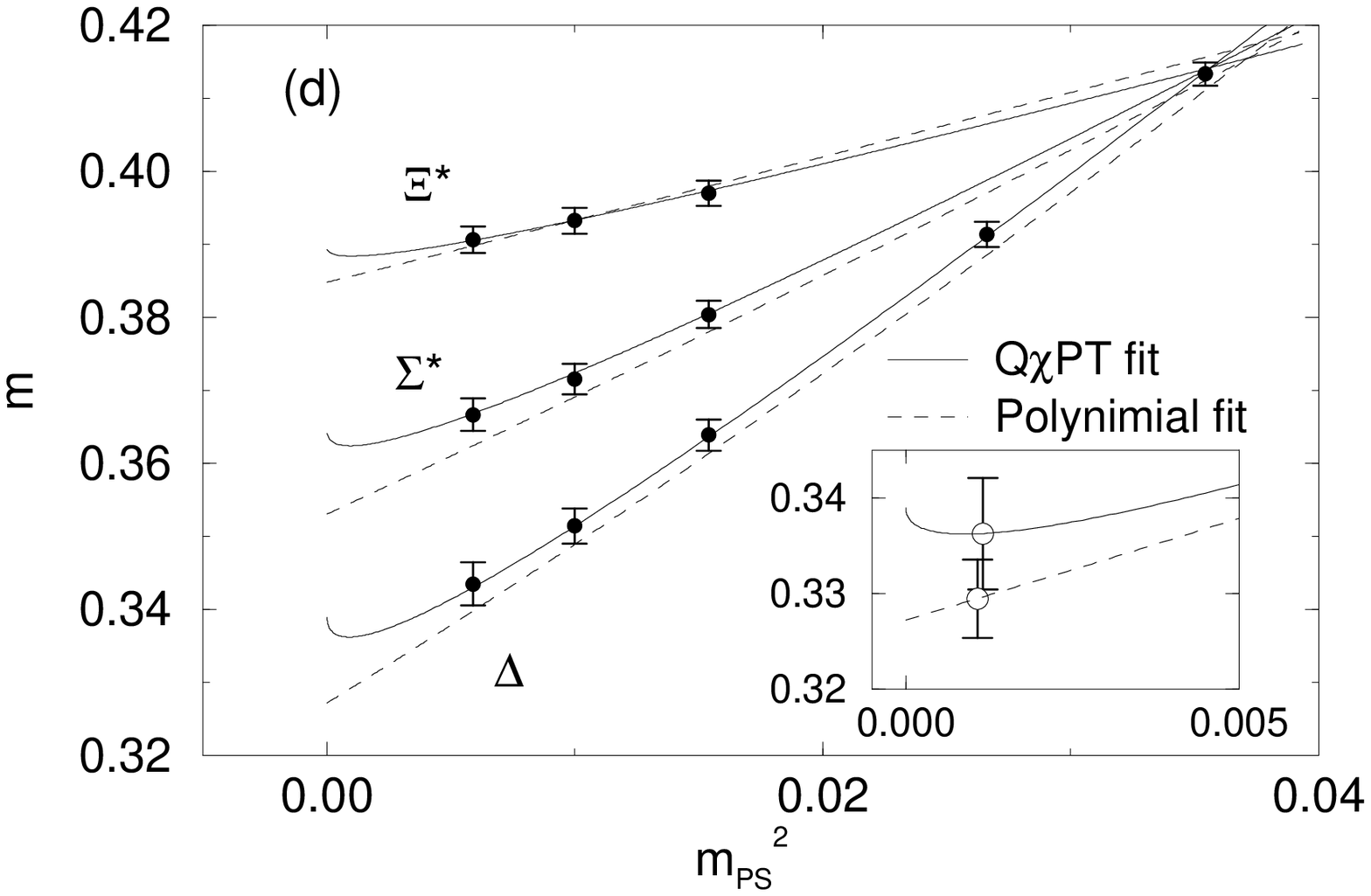}}
\caption{The same for Fig.~\protect\ref{fig:chiralFit590}
at $\beta=$ 6.47.}
\label{fig:chiralFit647}
\end{figure}

\begin{figure}[htb]
\centerline{\epsfxsize=8.5cm \epsfbox{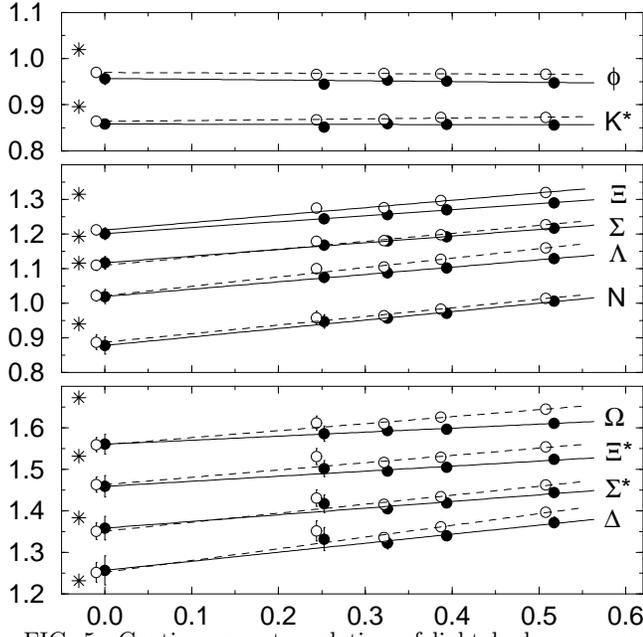}}
\caption{Continuum extrapolation of light hadron masses
from $m_K$-input. 
Filled symbols and solid lines are the results 
from the Q$\chi$PT chiral fits, 
while open symbols and dashed lines are from the polynomial fits.
Experimental values are shown by the stars.}
\label{fig:ContinuumFP}
\end{figure}

\begin{figure}[htb]
\centerline{\epsfxsize=8.5cm \epsfbox{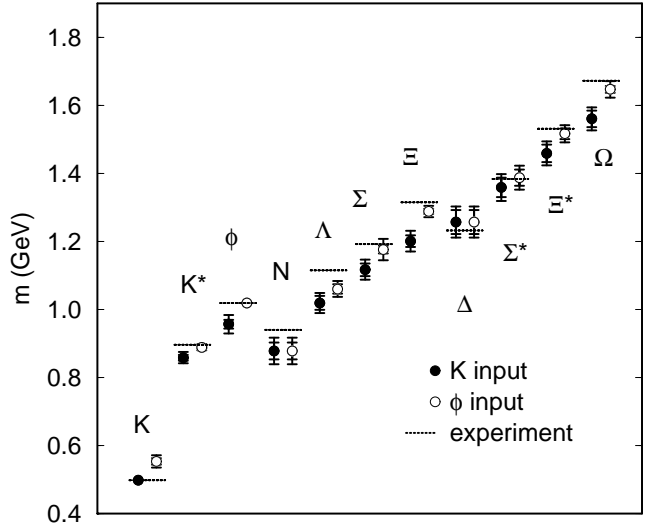}}
\caption{Quenched light hadron spectrum compared with experiment.
The statistical error and the sum of the statistical and systematic
errors are indicated.}
\label{fig:MassFinal}
\end{figure}

\begin{figure}[htb]
\centerline{\epsfxsize=8.5cm \epsfbox{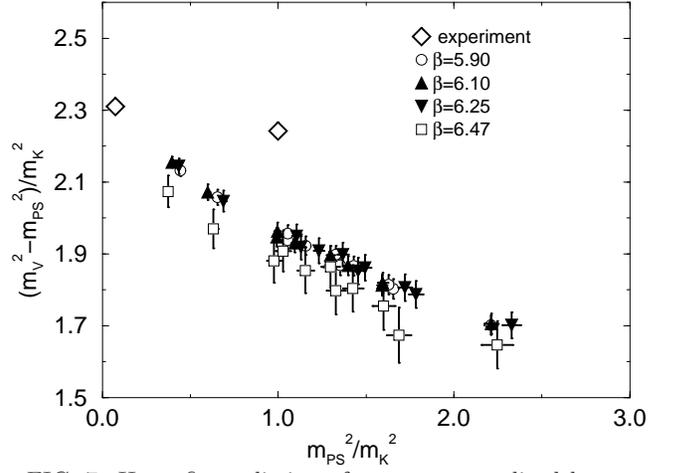}}
\caption{Hyperfine splitting of mesons normalized by $m_K$.
Diamonds represent the experimental points corresponding to 
$(m_{PS},m_V) = (m_\pi,m_\rho)$ and $(m_K,m_K^*)$, 
where the former is the input.}
\label{fig:hyperfine}
\end{figure}

\begin{figure}[htb]
\centerline{\epsfxsize=8.5cm \epsfbox{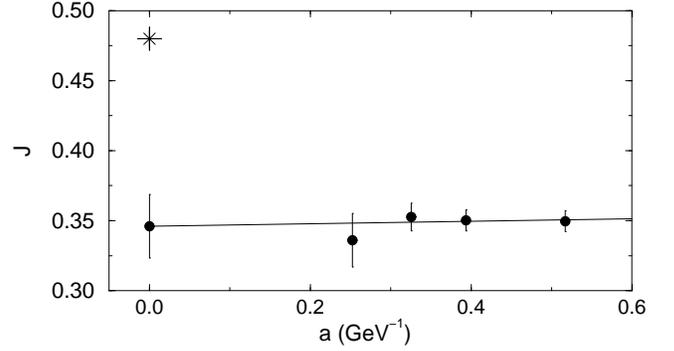}}
\caption{$J$ parameter.
The star represents the experimental value.}
\label{fig:valJ}
\end{figure}

\begin{figure}[htb]
\centerline{\epsfxsize=8.5cm \epsfbox{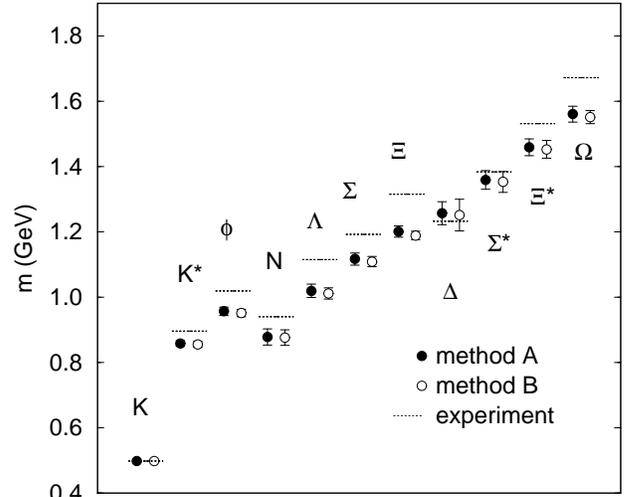}}
\caption{Comparison of the spectra from method A
(black circles) and method B (open circles).
$m_K$ is taken as input.}
\label{fig:ReverseK}
\end{figure}

\begin{figure}[htb]
\centerline{\epsfxsize=8.5cm \epsfbox{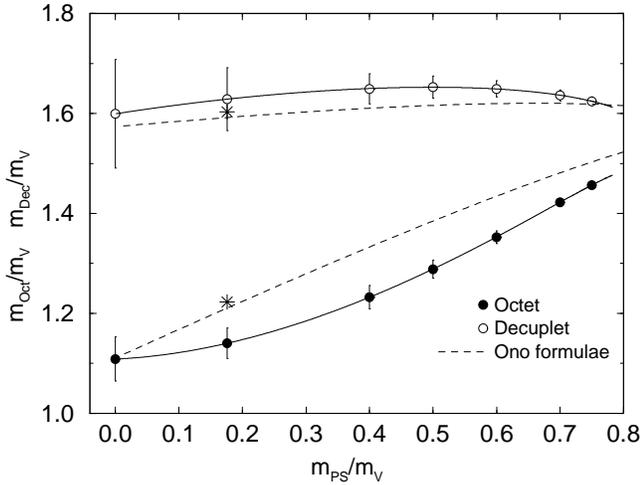}}
\caption{Edinburgh plot in the continuum limit. 
The stars represent experimental values.
Dashed curves illustrate the phenomenological 
mass formulae by Ono~\protect\cite{ref:Ono}.}
\label{fig:EDcont}
\end{figure}

\begin{figure}[htb]
\centerline{\epsfxsize=8.5cm \epsfbox{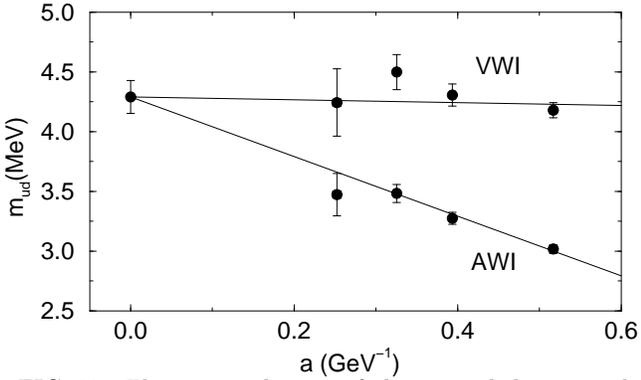}}
\caption{The averaged mass of the up and down quarks and its continuum 
extrapolation. The leftmost point is the value extrapolated
to the continuum limit.}
\label{fig:Mnormal-MeV}
\end{figure}

\begin{figure}[htb]
\centerline{\epsfxsize=8.5cm \epsfbox{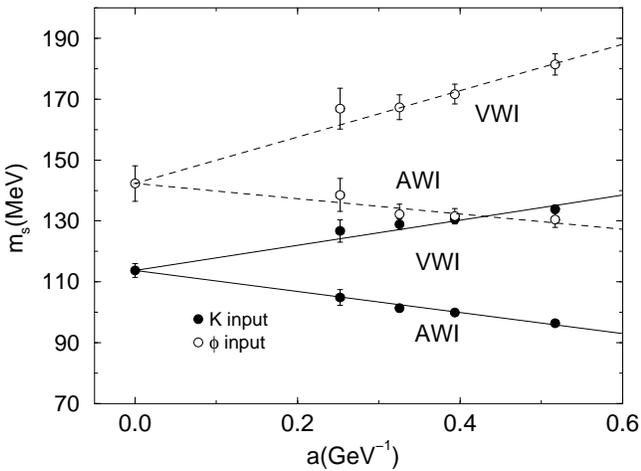}}
\caption{The strange quark masses and their continuum 
extrapolation. The leftmost points are the values extrapolated
to the continuum limit.}
\label{fig:Mstrange-MeV}
\end{figure}

\begin{figure}[htb]
\centerline{\epsfxsize=8.5cm \epsfbox{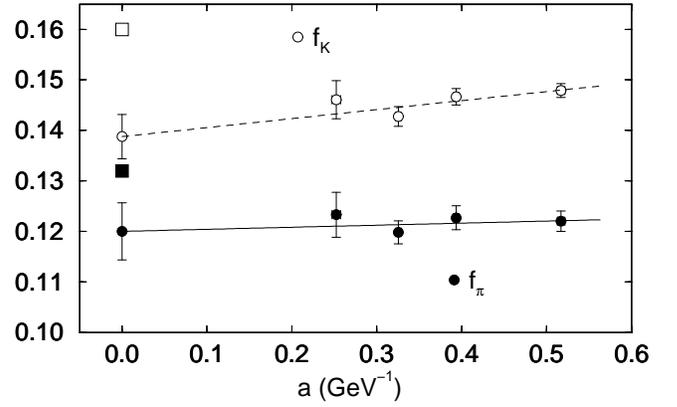}}
\caption{Continuum extrapolations of pseudoscalar meson decay constants 
$f_\pi$ and $f_K$ ($m_K$-input). 
Large squares at $a=0$ represent experimental values.}
\label{fig:fPSCont}
\end{figure}

\begin{figure}[htb]
\centerline{\epsfxsize=8.5cm \epsfbox{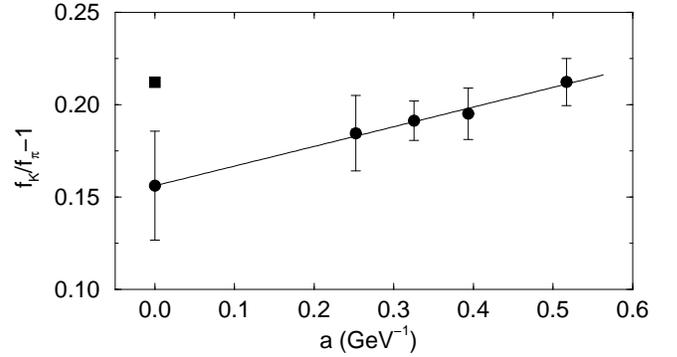}}
\caption{Continuum extrapolation of $f_K/f_\pi-1$ from the $m_K$-input. 
The square at $a=0$ represents the experimental value.}
\label{fig:fPSRatCont}
\end{figure}

Chiral fits of the pseudoscalar meson mass are already described 
in Sec.~\ref{sec:qChPT:PSfit}, 
and are shown in Figs.~\ref{fig:chiralFit590}-(a) -- 
~\ref{fig:chiralFit647}-(a)
with parameters summarized in Table~\ref{tab:PSQchPTPrm}. 
Comparisons of various fit functions for pseudoscalar meson masses
are given in Appendix~\ref{appendix:PSChiralFits}.

For vector mesons and baryons, we choose the pseudoscalar meson mass 
as the variable to represent the quark mass dependence. 
For vector mesons we adopt\cite{ref:Booth}
\bea
 m_{V,12} &=&   m_V^0 \nonumber \\
      &&+ \frac{C_{1/2}}{6} \{\frac{3}{2}(m_{11}+m_{22})
        + 2\frac{m_{22}^3 - m_{11}^3}{m_{22}^2 - m_{11}^2} \} \nonumber \\
      &&+ \frac{C_1}{2} (m_{11}^2 + m_{22}^2), 
\label{eq:qchptVfinal}
\eea
where $m_{fg}$ is the pseudoscalar meson mass
with the quark flavor combination $(f,g)$.  
The coefficient $C_{1/2}$ is proportional to $\delta$ while the $C_1$
term is present in ordinary $\chi$PT.  
For octet baryons, we employ\cite{ref:LabrenzSharpe}
\bea
m_\Sigma &=& m_O^0 \nonumber \\
  &&+ \frac{1}{2}\{4F^2 w_{uu}-4(D-F)F w_{ud}+(D-F)^2 w_{ss}\} \nonumber \\
  &&- 4b_F m_{uu}^2 + 2(b_D-b_F)m_{ss}^2  \nonumber \\
  &&+ (2D^2/3-2F^2)v_{uu} \nonumber \\
  &&+ (2D^2/3-4DF+2F^2)v_{ud}, 
\label{eq:qchptSfinal} \\
m_\Lambda &=& m_O^0 \nonumber \\
   &&+ \frac{1}{2}\{ (4D/3-2F)^2 w_{uu} + (D/3+F)^2w_{ss} \nonumber \\
   &&- 2(4D/3-2F)(D/3+F)w_{ud}\} \nonumber \\
   &&+ 4(2b_D/3-b_F)m_{uu}^2 - 2(b_D/3+b_F)m_{ss}^2 \nonumber \\
   &&+ (2D^2/9-8DF/3+2F^2)v_{uu} \nonumber \\
   &&+ (10D^2/9-4DF/3-2F^2)v_{ud}, 
\label{eq:qchptLfinal}
\eea
for $\Sigma$-type and $\Lambda$-type cases. For decuplet, the formula 
reads,  
\bea
m_D &=& m_D^0 + \frac{5H^2}{162}(4w_{uu}+4w_{ud}+w_{ss}) \nonumber \\
   &&+ \frac{C^2}{18}(w_{uu}-2w_{ud}+w_{ss}) \nonumber \\
   &&+ c(2m_{uu}^2+m_{ss}^2).
\label{eq:qchptDfinal}
\eea
Here 
\bea
w_{fg} &=& -2\pi\delta \frac{m_{ff}^3-m_{gg}^3}{m_{ff}^2-m_{gg}^2} \label{eq:Wus} \\ 
v_{fg} &=& \frac{m_{fg}^3}{8\pi f^2}. \label{eq:Vus}
\eea

The $O(m_{PS}^3)$ terms are not included in Eq.~(\ref{eq:qchptVfinal}) for vector 
mesons and in Eq.~(\ref{eq:qchptDfinal}) for decuplet baryons.   
The octet-decuplet coupling terms are also ignored in 
Eqs.(\ref{eq:qchptSfinal}) and (\ref{eq:qchptLfinal}) for octet baryons. 
These choices are made because fitting parameters are not well
determined if these terms are introduced (see Appendix
\ref{appendix:QChPT}),   
and the dropped terms have small effects for the spectrum.
Fittings with and without them are compared in
Fig.\ref{fig:chiralB590Deg} for degenerate masses at $\beta=5.90$.  
The two types of fittings reproduce data equally well.
The difference remains small at the physical point, 
at most 5\% ($5\sigma$) at finite lattice spacings 
and at most 1.2\% ($1.3\sigma$) after the continuum extrapolation.

We set $\delta=0.1$ and $\alpha_\Phi=0$ as suggested from 
the pseudoscalar case.  
These choices do not affect the fits for vector mesons and
decuplet baryons: a non-vanishing $\alpha_\Phi$ leads to $O(m_{fg}^3)$
effect which is not included in the fit function, 
and a change of $\delta$ is absorbed by a redefinitions of 
the parameters. 

For the nucleon mass, 
dropping $O(m_{PS}^3)$ terms in Eq.~(\ref{eq:qchptSfinal}) would lead to 
a positive curvature (concave function), which contradicts the 
data that show a negative curvature (convex function) in
Figs.~\ref{fig:chiralFit590}-(c) -- \ref{fig:chiralFit647}-(c). 
Therefore we include $O(m_{PS}^3)$ terms 
for octet baryons. The coefficient of $O(m_{PS}^3)$ terms is affected
by the choice of $\delta$ and $\alpha_\Phi$. We study the effect of
the uncertainty of $\delta$ and $\alpha_\Phi$ on the resulting octet
masses by varying $\delta$ from 0.08 to 0.12 and $\alpha_\Phi$ from 
$-0.7$ to $+0.7$. The change of $\delta$ in this range results in 
$0.4$\% (1.3$\sigma$) difference at finite lattice spacings and 0.3\%
(0.3$\sigma$) in the continuum limit. The change of $\alpha_\Phi$ 
leads to differences of 2.9\%(4.7$\sigma$) and 2.2\%(1.4$\sigma$),
respectively.
We also fix $f=f_\pi=$ 132 MeV. Changing $f$ to $f_K=226$~MeV affects 
octet baryon masses by at most 2.5$\sigma$ at finite lattice spacings, 
and 0.5$\sigma$ in the continuum limit.
Artifacts of fixing these parameters are sufficiently small at least in
the continuum limit.

Fits are made to degenerate and non-degenerate data together.
Because the size of covariance matrix becomes too large and 
the matrix elements cannot be determined reliably, we do not include
correlations among different quark masses. 

Fits for vector mesons and baryons are shown 
in Figs.~\ref{fig:chiralFit590} to \ref{fig:chiralFit647}.
Parameters are summarized in Table~\ref{tab:VQchPTPrm} for vector
mesons, in Table~\ref{tab:OQchPTPrm} for octet baryons and in
Table~\ref{tab:DQchPTPrm} for decuplet baryons. 
As we see in the figures and tables, Q$\chi$PT fits reproduce data 
at all $\beta$.

\subsection{Hadron masses at the physical point}

The extrapolation and interpolation to the physical point is made as follows. 
For the case of the $m_K$-input, we determine
$m_{PS}(\kappa_{ud},s_1)$ and $m_{PS}(\kappa_{ud},s_2)$ from 
non-degenerate fits to pseudoscalar mesons,  and interpolate them
linearly in terms of the $s$ quark mass so that
$m_{PS}(\kappa_{ud},\kappa_s)$ takes the experimental value of $m_K$.
For the $m_\phi$-input, we first determine the mass $m_{ss}$ of the
degenerate pseudoscalar meson consisting of two strange quarks from 
the vector meson mass fit and evaluate $\kappa_s$. We then make a
linear interpolation of  
$m_{PS}(\kappa_{ud},s_1)$ and $m_{PS}(\kappa_{ud},s_2)$
to find $m_K=m_{PS}(\kappa_{ud},\kappa_s) \equiv m_{us}$.
Values of $m_{uu}$, $m_{ss}$ and $m_{us}$ then lead to the predictions 
for other hadron masses.

Hadron masses in lattice units are listed in Table~\ref{tab:MassQchPTLat}.
We include results for fictitious
hadrons such as ``$\eta_s=\bar s s$'' (pseudoscalar meson consisting of two
strange quarks), $\Lambda_{ss}$ ($\Lambda$-like baryon consisting of
two quarks with $m_s$ and a light quark with $m_{ud}$), 
and $N_{sss}$ (octet baryon consisting of three quarks with $m_s$). 
Hadron masses translated to physical units are compiled in 
Table~\ref{tab:MassQchPTPhys}.

\subsection{Continuum extrapolation}
\label{sec:spectrum:context}
For our lattice actions scaling violation is given by
\be
m(a) = m(0) [ 1 + S a + (S'a)^2 + O(a^3)]. 
\label{eq:extrapform}
\ee
Hadron mass data in Fig.~\ref{fig:ContinuumFP} are fitted 
well without the $O(a^2)$ and higher order terms ($\chi^2/N_{df} < 1.6$).  
Hadron masses in the continuum limit are given 
in Table~\ref{tab:MassFinal}. 
Statistical errors are about 1--3\%.
Table~\ref{tab:MassSlope} lists $S$ of the term linear in $a$.
The $S$ for baryons are larger than those for mesons.
Lighter baryons have larger values of $S$ for both octet and decuplet.
The nucleon has the largest $S$ of about 280 MeV, 
with which the scaling violation $Sa$ is about 10\% in the middle of
our range of lattice spacing, $a = 0.075$ fm.

A fit retaining a quadratic term leads to $S$ and $S'$ ill determined
with the magnitude of errors comparable to the central values.  
The masses in the continuum limit have large errors of 13\% which are 5 times 
those from the linear extrapolation. Within these errors, the continuum results
from the quadratic fit are consistent with those of the linear one.
With only four points of lattice spacings, we are not able to 
test effects of higher order terms further. 

We therefore estimate the systematic errors from the $O(a^2)$ terms 
by an order estimate, assuming $S' \sim S$ and substituting $a = 0.075$~fm 
as above.\footnote{ 
For an estimate of $S'$, we fit deviations of $m(a)$ from the linear
fit using a pure quadratic function of $a$. This method gives
$S'$ of $O(50\mbox{\ MeV})$ with error of $O(100\mbox{\ MeV})$.}
The errors estimated in this way normalized by the central values 
are summarized in Table ~\ref{tab:deviations} under 
the column ``$O(a^2)$ error''.
We find their magnitude to be quite small.
Even for the nucleon with the largest scaling violation, the $O(a^2)$
error is about 1\%. 
Thus, unless $S'$ is unduly large, 
$O(a^2)$ systematic errors would not 
exceed a percent level.  This is much smaller than 
the deviation between the calculated quenched spectrum and 
experiment.

\subsection{Results from polynomial chiral fits}
\label{sec:spectrum:poly}

Polynomial chiral fits are carried out to degenerate and
non-degenerate data separately, fully incorporating the correlation
among different quark masses. We employ quadratic polynomials in terms 
of the VWI quark mass, except a cubic polynomial for degenerate octet
baryons. 
The fitting procedure is described in Appendix~\ref{appendix:CorrFits}. 
The fits are plotted in Figs.~\ref{fig:chiralFit590} --
\ref{fig:chiralFit647} by dashed lines.    
Hadron masses at the physical points are listed in 
Table~\ref{tab:MassPolyPhys}.
 
Extrapolating the results to the continuum limit,  
hadron masses from polynomial chiral fits are also 
fitted well by linear functions in $a$ with $\chi^2/N_{df}\le$ 1.7.
The continuum extrapolation is shown in Fig.~\ref{fig:ContinuumFP} by dashed 
lines.  
Masses in the continuum limit are given in Table ~\ref{tab:MassPoly}. 

At the four $\beta$ values, the difference in hadron masses at the 
physical point between the Q$\chi$PT fit and the polynomial fit is 
at most 3\%.   In the continuum limit the differences are 
within 1.5\%  (1.6$\sigma$),  
as listed in the column ``chiral fit error'' in Table~\ref{tab:deviations}. 
This difference is sufficiently small so that it does not
alter the pattern of deviation between the quenched spectrum 
calculated with Q$\chi$PT chiral extrapolation and the experimental 
spectrum as shown in Fig.~\ref{fig:MassPoly}. 

\subsection{Systematic error and final results}
\label{sec:spectrum:spctrum}
Total systematic error for the mass spectrum is estimated by adding 
in quadrature the error from continuum extrapolation (``$O(a^2)$ error''
in Table ~\ref{tab:deviations}), that from chiral extrapolations
(``chiral fit error'' in the table), and a 0.6\% error for finite size effects.

Our final results for the quenched light hadron spectrum including the
systematic error are summarized in 
Fig.~\ref{fig:MassFinal} and Table~\ref{tab:MassFinal}. 

\subsection{Comparison with previous results}

\subsubsection{Meson hyperfine splitting}
The GF11 Collaboration~\cite{ref:GF11mass}
calculated hadron masses with the $m_K$-input 
using lattices with $L_sa \approx 2.3$ fm. 
The chiral and continuum extrapolations are made with a linear form.
Based on a finite size study at $\beta=5.7$ with 
$L_sa \approx 2.3$ fm ($L_s=16$) and $\approx 3.4$ fm ($L_s=24$), 
they corrected the continuum results for finite size effects.
They claimed that the hyperfine splitting between $K$ and $K^*$ 
is consistent with experiment.

This differs from our small hyperfine splitting.
We compare 
our data (filled symbols) and the GF11 data (open symbols), 
both with $m_K$-input, in Fig.~\ref{fig:ContinuumGF11}.
For the GF11 data,  the results from the larger lattice with $L_sa \approx 3.4$~fm 
are also shown (open squares) at $a \approx 0.7$ GeV$^{-1}$,  and 
the continuum estimates before and after the the finite size correction 
are shown at $a=0$.

We observe that all data for $K^*$ and $\phi$ at finite $a$ are nearly 
consistent with each other. 
The difference in the continuum limit is due to a steeper slope 
of the GF11 data for the continuum extrapolation, 
arising from small values of $m_{K^*}$ and $m_\phi$ 
at $\beta=5.7$ on the lattice of $L_sa \approx 2.3$ fm 
(the rightmost triangles). If we adopted the data from the $L_sa\approx 3.4$ 
lattice (open squares), we would obtain a continuum value in agreement 
with our result.  

In Ref.~\cite{ref:GF11mass}, 
the discrepancy between $L_sa \approx 3.4$ and 2.3 fm is considered 
as finite size effects. 
However, since data at smaller $a$ are consistent between 
$L_sa \approx 3.0$ fm (our data) and 2.3 fm (Ref.~\cite{ref:GF11mass}), 
it is not clear whether we can attribute the difference simply 
to finite size effects. 
The conclusion of the GF11 critically depends on their data at $\beta=5.7$,
for which we suspect an underestimation of errors.

\subsubsection{Nucleon mass}

In previous calculations at $\beta\approx5.7$--6.2 with 
$m_{PS}/m_V \simgt 0.5$, 
nucleon masses are significantly higher than experiment 
at finite $a$ \cite{ref:GF11mass,ref:LANLmass}.
The GF11 claimed 
agreement with experiment after the continuum extrapolation 
and the finite size correction. 
In the present study, however, we find the nucleon mass to be 
{\it smaller} than the previous estimates even at finite $a$. 
Extrapolating to the continuum limit, we find the nucleon mass 
to be smaller than experiment by 7\% ($2.5\sigma$).
See Fig.~\ref{fig:ContinuumGF11} where our data and those of 
the GF11 are compared.

The origin of our small nucleon mass at a finite $a$ is 
the negative curvature in $1/\kappa$ toward small quark masses, 
as observed in Figs.~\ref{fig:chiralFit590} -- \ref{fig:chiralFit647}.
This trend becomes manifest only when the quark mass 
is reduced to $m_{PS}/m_V\approx 0.4$ while sustaining statistical precisions.
In fact, a linear fit of our data at $m_{PS}/m_V \simgt 0.5$ 
gives a larger nucleon mass consistent with the previous results.

\subsubsection{Masses of $\Xi^*$ and $\Omega$}
The GF11 reported 
the masses of $\Xi^*$ and $\Omega$ from the $m_K$-input 
higher than experiment by 3--5\%. 
In contrast, Fig.~\ref{fig:MassFinal} shows that 
our masses are smaller than experiment by a similar magnitude. 

The origin of these differences can be seen in the top panel
in Fig.~\ref{fig:ContinuumGF11}.
While the results from the two groups are consistent at
$a\approx$ 0.5 GeV$^{-1}$, 
the continuum extrapolation is different, especially for $\Omega$. 
As in the case of the meson hyperfine splitting, the continuum 
extrapolation of the GF11 is critically affected by 
the data at $\beta=5.7$ and $L_sa\approx2.3$ fm.

\subsubsection{Comparison with staggered quark results}

The negative curvature of the nucleon mass has also been
reported in Ref.\cite{ref:MILC-KS},
in which the nucleon mass for the staggered quark action 
is calculated down to $m_{PS}/m_V \approx0.3$--0.4.
However, 
our result $m_N = 878(25)$ MeV in the continuum limit obtained from 
the Wilson quark action is smaller than 
$m_N = 964(35)$ MeV \cite{ref:MILC-KS} from 
the staggered quark action by about 2.5 $\sigma$. 

It has been pointed out in Ref.\cite{ref:Aoki2000} that
the difference in the Wilson and Kogut-Susskind results for the nucleon mass 
exists not only at the physical point but even at heavier quark masses 
for which the discrepancy is statistically more significant. 
In Ref.\cite{ref:Aoki2000}, the nucleon to $\rho$ mass ratio
off the physical quark mass is calculated in the continuum limit,
using the same method as that explained in Sec.\ref{sec:methodB} below.
The ratios for the staggered action are larger than ours
by about 8\% for the whole range of the quark mass 
(see Fig.~4 in Ref.\cite{ref:Aoki2000}). 

The origin of the difference is not explained by 
finite size effects since both calculations employ sufficiently large 
lattices, nor by the chiral extrapolation since the difference exists even 
for heavy quarks.
The continuum extrapolation is also improbable as the cause, because masses 
calculated at finite $\beta$ are well reproduced by lowest order scaling 
violation of $O(a)$ for our
Wilson results, and $O(a^2)$ for the staggered quarks.
For the moment the origin of the difference is an open issue.

\section{Light quark masses}
\label{sec:Mq}
\subsection{Renormalization factors} 

We calculate renormalized quark masses in the 
${\overline {\rm MS}}$ scheme at the scale $\mu=2$~GeV. 
They are given by
\bea
m_q^{VWI} & = & Z_m m_q^{VWI(0)} a^{-1},
\label{eq:VWImq} \\
m_q^{AWI} & = &
\frac{Z_A}{Z_P} m_q^{AWI(0)} a^{-1}. 
\label{eq:AWImq}
\eea
for VWI and AWI masses.
The renormalization factors 
$Z_m$~\cite{ref:Zm}, $Z_A$~\cite{ref:ZaE}, and $Z_P$~\cite{ref:ZpaL} are 
estimated with tadpole-improved one-loop perturbation 
theory\cite{ref:LepageMackenzie}, by matching the lattice scheme
to the $\overline{\rm MS}$ scheme at $\mu=1/a$. 
They read 
\bea
Z_m & = &  8\kappa_c [ 1 + 0.01\alpha_P(1/a) ], \\
Z_A & = & (1 + (0.448 - (4\pi/12))\alpha_P(1/a))/u_0, \\
Z_P & = & 1 - 1.0335\alpha_P(1/a),
\eea
with $u_0 = \langle U_P \rangle^{1/4}$, where $\langle U_P \rangle$ 
is the plaquette average.
For $\alpha_P(1/a)$, we first compute $\alpha_P(q^*)$ according to  
\be
 -\ln(\langle U_P \rangle) = 
 \frac{4\pi}{3} \alpha_P(q^*)
 \left[ 1 - 1.18969\,\alpha_P(q^*)\right],
\label{eq:alphaV}
\ee
where $q^*=3.4018/a$, and use the renormalization group
equation to two-loop order. 
The running of the quark mass from $\mu=1/a$ to 2 GeV is made
employing the three-loop renormalization group equation \cite{ref:mqrun}. 

\subsection{Chiral and continuum extrapolations} 

For AWI quark masses, we need to carry out 
chiral extrapolation and/or interpolation to the physical point. 
Polynomials in $1/\kappa$ are used for this since 
quenched chiral singularities are absent 
\cite{ref:SharpePC,ref:GoltermanPC}.
In fact, as shown in Fig.~\ref{fig:PertWIFP}, 
the ratio of renormalized quark masses 
\be
y=\frac{m_1^{VWI}+m_2^{VWI}}{m_1^{AWI}+m_2^{AWI}}
\ee
is flat as a function of $x=m_1^{AWI}+m_2^{AWI}$, 
suggesting linear behavior of the AWI quark mass in $1/\kappa$. 

A comparison of $\kappa_c$ and $\kappa_c^{AWI}$ where $m_q^{AWI}$
vanishes suggests the presence of Q$\chi$PT singularity for 
the pseudoscalar meson mass $m_{PS}$ and its absence for $m_q^{AWI}$;
see inset plots of
Figs.~\ref{fig:chiralFit590}(a) -- ~\ref{fig:chiralFit647}(a). 
The value of $\kappa_c$ from the Q$\chi$PT fit to $m_{PS}$ 
and $\kappa_c^{AWI}$ from 
a linear fit to $m_q^{AWI}$ agree well, whereas a quadratic fit of
$m_{PS}$ clearly fails to do so. 
Values for $\kappa_c^{AWI}$ and $\kappa_c$ from various 
fits are compiled in Table~\ref{tab:Kc}.
Numerically, $\kappa_c$ from the Q$\chi$PT fit of $m_{PS}^2$ agrees with
$\kappa_c^{AWI}$ obtained from a linear or quadratic fit 
with at most 2.8$\sigma$.
On the other hand, if we adopt fits with the quadratic (cubic) extrapolation
of $m_{PS}^2$, the difference between $\kappa_c$ and $\kappa_c^{AWI}$
increases to as much as 17$\sigma$ (12$\sigma$).

For actual chiral extrapolation of $m_q^{AWI}$, 
we employ a quadratic fit and enforce $\kappa_c^{AWI}$ to agree with $\kappa_c$ 
obtained from the Q$\chi$PT fit of $m_{PS}$, having confirmed their agreement
as described above. This constraint is imposed because even a small difference 
between $\kappa_c^{AWI}$ and $\kappa_c$ affects estimates of $m_q^{AWI}$ at
the physical point.
We employ 
\be
2m_q^{AWI(0)} = B_1^{dg}(1/\kappa - 1/\kappa_c) + B_2^{dg}(1/\kappa - 1/\kappa_c)^2.
\label{eq:mud}
\ee
Fitting parameters $B_1^{dg}$ and $B_2^{dg}$ are given 
in Table~\ref{tab:mqextrap}.
Fits with various functional forms 
are hardly distinguishable, as 
shown in Fig.~\ref{fig:mqchiral}.

For calculating the $s$ quark mass, we first make a linear fit to 
non-degenerate combination of quark masses, 
\be
m_{q}^{AWI(0)}+m_{s_i}^{AWI(0)} = A_0^{s_i} + A_1^{s_i}/\kappa
\ee
keeping $\kappa_{s_i}$ fixed. We then set $\kappa=\kappa_{ud}$, and calculate  
$m_{ud}^{AWI(0)}+m_{s}^{AWI(0)}$ by a linear interpolation
in terms of $1/\kappa_{s_i}$.
We do not employ a quadratic extrapolation
since the effect of the quadratic term is negligibly small in
$s$-quark mass but increases errors of fitting parameters 
significantly. 

Values for $m_q^{AWI(0)}$ and $m_q^{VWI(0)}$ at the physical 
quark mass point are presented in Table~\ref{tab:MQLat}.
Quark masses translated to ${\overline {\rm MS}}$ scheme at
$\mu=2$ GeV are given in Table~\ref{tab:MQPhys}, and are 
plotted in Figs.~\ref{fig:Mnormal-MeV} and \ref{fig:Mstrange-MeV}.
They are well reproduced by linear functions in $a$.
The AWI and VWI quark masses,
which are different at finite $a$, 
extrapolate to a universal value in the continuum limit, 
as they should. 
We determine the quark masses in the continuum limit by a combined linear 
extrapolation of $m_q^{VWI}$ and $m_q^{AWI}$ (Table \ref{tab:MQCont}).

\subsection{Systematic errors and final results}
To estimate systematic errors from chiral extrapolations, 
we consider a quadratic fit to $m_q^{AWI}$ taking $\kappa_c^{AWI}$ as
a fit parameter. 
We then carry out a Q$\chi$PT fit to $m_{PS}$ with $\kappa_c$ set
to the $\kappa_c^{AWI}$, and evaluate VWI and AWI quark masses. 
Chiral fits up to here employ $\kappa$ as an independent variable.
To evaluate errors in quark masses from fits in terms of $\kappa$,
we consider an independent Q$\chi$PT fit to 
$m_{PS}^2$ as a function of $m_q^{AWI}$ without referring to $\kappa$. 

Figure~\ref{fig:Mud-chiral} shows that $m_{ud}$ is sensitively affected 
by the treatment of the chiral limit. 
At finite $a$, both $m_q^{VWI}$ and $m_q^{AWI}$ from the alternative
fit above shown by triangles differ from the original ones (circles)
far beyond statistical errors.  Linearly extrapolated to the continuum, 
the alternative methods lead to $m_{ud}\approx$ 4.1 -- 4.8 MeV, 
depending on the choice of fits.
The fit to $m_{PS}$ as a function of $m_q^{AWI}$ (shown by 
diamonds) gives the lowest value of $\approx$ 3.5 MeV. 
Taking the maximum of the differences between five results and 
the value obtained in the preceding subsection, 
we estimate the systematic error to be $+0.51$ MeV and
$-0.79$ MeV. 

Systematic errors from chiral fits are not large for $m_s$.
As Figs.~\ref{fig:Ms-chiral.K} and \ref{fig:Ms-chiral.P} show, 
results from various chiral fits at finite $a$ agree with each other within
at most 3$\sigma$. 
We estimate the systematic error in the continuum limit by the same method
as for $m_{ud}$. 
We obtain $+5.8$ MeV and $-2.9$ MeV for the $m_K$ input and 
$+22.0$ MeV and $-0$ MeV for the $m_\phi$ input.

We also investigate uncertainties from various definitions 
of the axial-vector current 
and higher order effects in the renormalization factors.
For the former, we test for the local axial current defined by
\be
A_\mu^{local}(n) = {\bar f}_n i\gamma_5\gamma_\mu g_n
\ee 
with the tadpole improved renormalization factor
\be
Z_A^{local}  =  1 - 0.316 \alpha_P(1/a). \label{eq:ZAlocal}
\ee
The procedure to calculate the AWI quark mass $m_q^{AWI,local(0)}$ is
described in Appendix~\ref{appendix:mqlocal}. 
For the latter, we repeat analyses using the ${\overline {\rm MS}}$
coupling 
\be
      \frac{1}{g_{\overline {\rm MS}}^2(1/a)} =
      \frac{\langle U_p \rangle}{g^2} - 0.134868,
\ee
instead of $\alpha_P(1/a)$.

Figures~\ref{fig:ELAM-Mud}, \ref{fig:ELAM} and ~\ref{fig:ELAM-Pinp} show
the data and continuum extrapolations.
Values of $m_q^{AWI}$ (triangles) and $m_q^{AWI,local}$ (squares)
are in good agreement. 
The difference is about 5\% on the coarsest lattice and 
is smaller on finer lattices for all cases of $m_{ud}$, 
$m_{s}$ ($m_K$-input) and $m_{s}$ ($m_\phi$ input).
The two values agree in the continuum limit within 1.5$\sigma$
of the statistical error.
The results with $\alpha_P$ and $\alpha_{\overline {\rm MS}}$ are
compared using filled and open symbols.
The small difference in $m_q^{VWI}$ reflects a small value of one-loop
coefficients. 
On the other hand, the difference in $m_q^{AWI}$ is 
about 5\% on the coarsest lattice and about 3\% on the finest lattice, 
which leads to a difference of 2\% in the continuum limit, 
to be compared with the statistical error of 2--4\%.

As shown in Figs.~\ref{fig:ELAM-Mud}, \ref{fig:ELAM} and ~\ref{fig:ELAM-Pinp},
the central values in the continuum limit are contained within the 
error band given by the sum of statistical error and systematic
one from chiral extrapolations.
Therefore we do not add the errors from the definition of current
and higher order effects in the renormalization factors
to the estimate of the systematic error above.

Final results are given in Eqs.~(\ref{eq:Fmud}),(\ref{eq:FmsK}) and  
(\ref{eq:FmsP}).
We note that these numbers are different from those given
in our earlier publication~\cite{ref:CPPACSletter}, in which we
employed a linear chiral extrapolation of $m_q^{AWI}$
and corrected for a small difference between
$\kappa_c$ and $\kappa_c^{AWI}$.
Values here obtained by constrained quadratic chiral fits
are our final results.

\section{QCD scale parameter}
\label{sec:AlphaLambda}

\subsection{Methods and results}
We calculate the QCD scale parameter $\Lambda_{\overline{\rm MS}}$ 
in the $\overline{\rm MS}$ scheme. 
In this scheme, the renormalization group coefficients 
are known to four-loop order. 
Since the relation between the lattice coupling and the $\overline{\rm MS}$
coupling is known only up to two loops~\cite{ref:LuscherWeisz}, 
we employ the expression to three-loop order given by 
\begin{equation}
\Lambda_{\overline{\rm MS}}/\mu = \big(\beta_0 y\big)^{-\frac{\beta_1}{2\beta_0^2}}
     \exp(-\frac{1}{2\beta_0y}) 
     \big( 1+ \frac{\beta_1^2 - \beta_0\beta_2}{2\beta_0^3}y\big), 
\label{eq:defLambda}
\end{equation}
where $y=\alpha_{\overline{\rm MS}}(\mu)/4\pi$ and 
$\beta_2^{\overline {\rm MS}} = 2857/2$ for the $\overline {\rm MS}$
scheme of quenched QCD \cite{ref:betaMS3}.  

We estimate the $\overline{\rm MS}$ coupling in three ways, using the mean-field
coupling $\alpha_{MF}$~\cite{ref:ElKhadra}, 
the plaquette coupling $\alpha_P$~\cite{ref:LepageMackenzie} and the potential 
coupling $\alpha_V$. 

The mean field (or tadpole improved) coupling $\alpha_{MF}$ is
defined by $\alpha_{MF} = \alpha_0/\langle U_P\rangle$, where $\alpha_0$ is the
bare lattice coupling.
Using the relation between the ${\overline {\rm MS}}$ coupling and the 
bare coupling up to two loop~\cite{ref:LuscherWeisz} and the
perturbative expansion of the plaquette, one obtains 
\begin{equation}
\frac{1}{\alpha_{\overline {\rm MS}}(\pi/a)} = \frac{1}{\alpha_{MF}} 
   + 0.30928 - 1.95683 \alpha_{MF}.
\end{equation}
Substituting $\alpha_{\overline{\rm MS}}(\pi/a)$ into (\ref{eq:defLambda}) 
with $\mu=\pi/a$ yields $\Lambda_{\overline{\rm MS}}$. 
Our measurements of plaquette are listed in Table \ref{tab:plaquette}. 
The very small statistical error of plaquette 
is ignored in our analyses. 

The potential coupling $\alpha_V$ is defined through the static $q\bar q$ 
potential \cite{ref:LepageMackenzie}. 
Using the plaquette coupling $\alpha_P$ defined by 
\begin{equation}
-\ln \langle U_P\rangle = \frac{4\pi}{3}\alpha_P(q^*) \big(1 - 1.18969\alpha_P(q^*)\big),
\label{eq:alphaPP}
\end{equation}
with the optimal value of 
$q^* = 3.4018/a$\cite{ref:LepageMackenzie,ref:Klassen},
the potential coupling is evaluated by \cite{ref:qq3,ref:pp3} 
\begin{equation}
\alpha_V(q^*) = \alpha_P(q^*) + 2.81404 \alpha_P^2(q^*).
\label{eq:alphaVV}
\end{equation}
We evolve $\alpha_V(q^*)$ from $\mu=q^*$ to $\mu=\pi/a$ 
using the three-loop renormalization group
with $\beta_2^{V} = 6178.36$ \cite{ref:qq3,ref:pp3}, and calculate
\begin{eqnarray}
&&\alpha_{\overline {\rm MS}}(\pi/a) = \alpha_V(\pi/a) \nonumber \\ 
&& - 0.82230 \alpha_V(\pi/a)^2 - 2.66504 \alpha_V(\pi/a)^3 .
\end{eqnarray}

We also evaluate $\alpha_{\overline {\rm MS}}$ directly from the $\alpha_P$
coupling using the BLM~\cite{ref:BLMimp} improved relation
\begin{equation}
\alpha_{\overline {\rm MS}}(\pi/a) = 
\alpha_P(\mu) + \frac{2}{\pi}\alpha_P(\mu)^2 + 0.95465\alpha_P^2(\mu)^3,
\end{equation}
where $\mu=e^{5/6}q^*$. 

Values of $\Lambda_{\overline {\rm MS}}$ are 
given in Table~\ref{tab:Lambda} and Fig.~\ref{fig:Lambda}.
They are fitted well with 
a function linear in $a$ with small $\chi^2/N_{df} = 0.3$ -- 0.5.
The difference among continuum values of $\Lambda_{\overline {\rm MS}}$ 
is smaller than statistical errors.
This suggest that possible higher order terms in $a$ are not important 
within our accuracy.
Since $\Lambda_{\overline {\rm MS}}$ from $\alpha_V$ coupling
exhibits the smallest scaling violation, we quote Eq.(\ref{eq:Flmd}) as our best
estimate of $\Lambda_{\overline {\rm MS}}$ in quenched QCD.
We note that the scale is determined from $m_\rho$.

\subsection{Comparison with other results}
There are a number of ways to determine the QCD scale parameter.
The SCRI~\cite{ref:SCRILambda} group obtained 
$\Lambda_{\overline {\rm MS}}=$ 247(16) MeV
from the measurement of $\sqrt\sigma/\Lambda$ using the  
string tension $\sqrt\sigma=$ 465 MeV estimated from the
charmonium level splitting.
With a recursive finite-size technique using the Schr\"odinger
functional the ALPHA Collaboration~\cite{ref:ALPHALambda}
obtained $\Lambda_{\overline {\rm MS}}=$ 238(19) MeV,
where the physical scale is fixed by the Sommer scale~\cite{ref:Sommer-r0}
$r_0=$ 0.5 fm. 
In what follows we consider $\Lambda$ values that
would come out if we determine the string tension or the Sommer scale
with the $\rho$ meson mass obtained in our simulation.  

We borrow a parameterization~\cite{ref:SCRILambda} 
(proposed in Ref.~\cite{ref:Allton}) of $\sqrt\sigma$ and $1/r_0$ obtained
with high statistics data, and
evaluate them at $\beta=$ 5.9, 6.1, 6.25 and 6.47, the points of
our simulation. Using our $m_\rho$ we evaluate
$\sqrt\sigma/m_\rho$ and $1/r_0m_\rho$ as depicted in 
Fig.~\ref{fig:pot-rho}, where the errors arising from $\sqrt\sigma$ 
and $1/r_0$ are ignored. These values 
show a linear dependence in $a$, leading to 
\begin{eqnarray}
\sqrt\sigma/m_\rho &=& 0.494(16), \\
1/(r_0m_\rho) &=& 0.409(13).
\end{eqnarray}
With $m_\rho=$ 768.4 MeV, we obtain in the continuum limit
$\sqrt\sigma=$ 380(12) MeV and $r_0=$ 0.628(20) fm,
which exhibit 10$-$20\% deviations from the 
usually accepted values. 

With these scales
the SCRI result, $\Lambda_{\overline {\rm MS}}$ = 247(16) MeV,
is converted to $\Lambda_{\overline {\rm MS}}$ = 202(13)(7) MeV, while
the ALPHA result, $\Lambda_{\overline {\rm MS}}$ = 238(19) MeV, to
$\Lambda_{\overline {\rm MS}}$ = 189(15)(6) MeV.
Here, the first errors come from those quoted in the original
literature, and the second are from the error of our $\rho$ mass
measurement. Figure~\ref{fig:Lambda} compares our estimate of $\Lambda$
with those obtained by SCRI and ALPHA, and with
those we have re-evaluated consistently using the $\rho$ mass 
(labelled as ``translated''). It seems that there are discrepancies
somewhat beyond errors.
Three values of $\Lambda$ obtained with the same scale ($m_\rho$ in this case)
are supposed to be those in the continuum limit 
and hence discrepancy, if it exists, should be resolved within quenched QCD.

\section{Meson decay constants}
\label{sec:decay}

\subsection{Pseudo-scalar meson decay constants}

We extract pseudoscalar meson decay constants $f_{PS}$ from the
correlation function $\langle A_4^{local}(t) P(0)\rangle$.
The value of $f_{PS}$ is related to $m_q^{AWI,local(0)}$ defined 
in Eq.~(\ref{eq:AWImqloc0}): 
\bea
f_{PS} a &=& \tilde Z_A f_{PS}^{(0)}, \\
f_{PS}^{(0)} &=&
\frac{\sqrt{2m_{PS} a C_{PS}^P}}{(m_{PS}a)^2} \nonumber \\
&& \times (m_f^{AWI,local(0)} + m_g^{AWI,local(0)})
\label{eq:fpsbare}
\eea
for the flavor combination $(f,g)$, 
where $C_{PS}^P$ is the amplitude of propagator
with the point quark source defined by
\be
\langle P^P(t) P^P(0) \rangle \approx  C_{PS}^P \exp(-m_{PS}t).
\ee

To avoid the direct use of point source propagators, which are noisier 
than the corresponding smeared source propagators, we apply the 
following procedure:
the amplitude $C_{PS}^S$ for the smeared source propagator 
is already obtained from the meson mass fit 
\be
\langle P^P(t) P^S(0) \rangle \approx C_{PS}^S \exp(-m_{PS}at).
\ee
The ratio $\eta = C_{PS}^P / C_{PS}^S$ can be obtained 
from an additional fit 
\be
\frac{\langle P^P(t) P^P(0) \rangle}{\langle P^P(t) P^S(0) \rangle} 
 \approx \eta.
\label{eq:PPPS}
\ee
Note that Eq.~(\ref{eq:PPPS}) is the only new fit which is necessary 
to calculate $f_{PS}$.
We illustrate a typical fit in Fig.~\ref{fig:PPPS}.

For the renormalization constant $\tilde Z_A$, we employ
the tadpole-improved\cite{ref:LepageMackenzie,ref:ZKtp}
one-loop formula~\cite{ref:ZpaL}, which is factorized into two parts, 
\be
\tilde Z_A = Z_A^{local} Z_\kappa, 
\ee
where
\be
Z_\kappa = \sqrt{1-\frac{3\kappa_f}{4\kappa_c}} 
           \sqrt{1-\frac{3\kappa_g}{4\kappa_c}}.
\label{eq:ZK}
\ee
Table~\ref{tab:fp} summarizes the results for $Z_\kappa f_{PS}^{(0)}$ 
at the simulation points.

For $f \equiv Z_\kappa f_{PS}^{(0)}$, 
non-degenerate data are well reproduced by a linear
function of $1/\kappa_{u_i}$, 
where $\kappa_{u_i}$ is the hopping parameter for the $u,d$ quark, 
while degenerate data show 
a slight negative curvature (Fig.~\ref{fig:fPSchiral}).
Therefore, we employ a quadratic polynomial function of $1/\kappa$
\be
f = A_0^{dg} +  A_1^{dg}/\kappa + A_2^{dg}/\kappa^2
\ee
for the degenerate case, and a linear function of $1/\kappa_{u_i}$
\be
f = A_0^{s_j} +  A_1^{s_j}/\kappa_{u_i}
\ee
for the non-degenerate case $(u_i,s_j)$. 
Table~\ref{tab:fPSLat} presents $Z_\kappa f_{PS}^{(0)}$ at the physical point.

Multiplying $Z_A^{local}$ and $a^{-1}$, 
we obtain $f_{PS}$ in physical units (Table~\ref{tab:fPSPhys}).
Data are well reproduced by a linear function in $a$, as shown in 
Fig.~\ref{fig:fPSCont}. 
Accordingly, we obtain small $\chi^2/N_{df} = 0.45$, 0.86 and 1.38
for $f_\pi$, $f_K$ ($m_K$-input) and $f_K$ ($m_\phi$-input), respectively.
Our final results for $f_{PS}$ are summarized 
in Eqs.~(\ref{eq:fpifnl}), (\ref{eq:fKfnl}) and Table~\ref{tab:fFinal}.
For the ratio $f_K/f_\pi-1$, we obtain 0.156(29) in the continuum limit 
from a linear fit of Fig.~\ref{fig:fPSRatCont}.

\subsection{Vector meson decay constants}

Vector meson decay constants $F_V$ are extracted from the correlation
function $\langle V_i(n) V_i(0) \rangle$ of the local vector current
\be
V_i(n) = \bar f_n \gamma_i g_n. 
\ee
Since $V_i$ is the vector meson operator, 
we obtain 
\bea
F_V a &=& Z_V \sqrt{2\kappa_f} \sqrt{2\kappa_g} F_V^{(0)}, \\
F_V^{(0)} &=& \sqrt{{2C_V^P}/{m_Va}},
\eea
where $C_V^P$ is the amplitude of vector meson propagator with 
the point source. 
Employing a method similar to that used in the previous subsection, 
we first obtain the amplitude of smeared propagator
$C_V^S$ and then fit the ratio $\eta$ of the point and smeared propagators
to calculate $C_V^P=\eta C_V^S$.

We adopt a non-perturbative definition\cite{ref:MaianiMartinelli} given by
\be
 Z_V = \lim_{t\to\infty} 
       \frac{\langle V^C(t)V(0)\rangle}{\langle V(t)V(0)\rangle},
\label{eq:Zv}
\ee
where 
\be
V_i^C(n) = \frac{1}{2}\Big\{
   \bar f_{n+\hat\mu}U^\dagger_{n,\mu}(\gamma_i+1) g_n
 + \bar f_{n}U_{n,\mu}(\gamma_i-1) g_{n+\hat\mu} \Big\}
\label{eq:consV}
\ee
is the conserved vector current. 
Examples of the fit for Eq.~(\ref{eq:Zv}) are shown in Fig.~\ref{fig:ZV}.
Values of $Z_V$ depend little on the quark mass, as shown 
in Table~\ref{tab:ZV}.  
We provide $F_V a$ at the simulation points in Table~\ref{tab:FVR}. 

\begin{figure}[htb]
\centerline{\epsfxsize=8.5cm \epsfbox{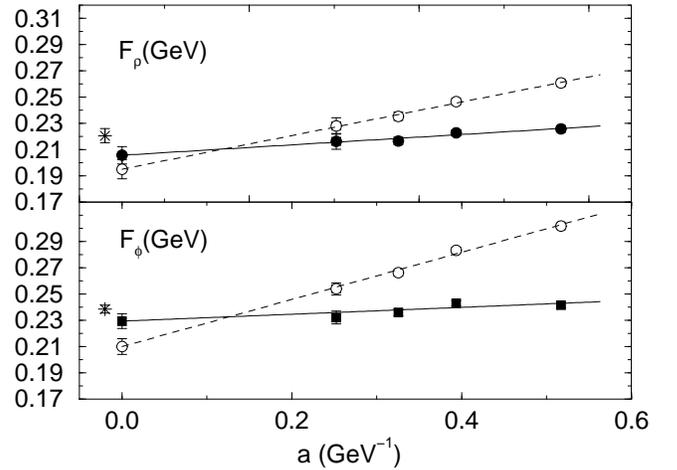}}
\caption{$F_\rho$ (top panel) and $F_\phi$ (bottom panel) 
as a function of $a$. 
Fits for the continuum extrapolation are also shown. 
Solid symbols are for non-perturbatively renormalized 
decay constants, 
and open symbols are for decay constants renormalized 
by tadpole-improved one-loop perturbation theory. 
Stars represent experimental values.}
\label{fig:FVcont}
\end{figure}

\begin{figure}[htb]
\centerline{\epsfxsize=7cm \epsfbox{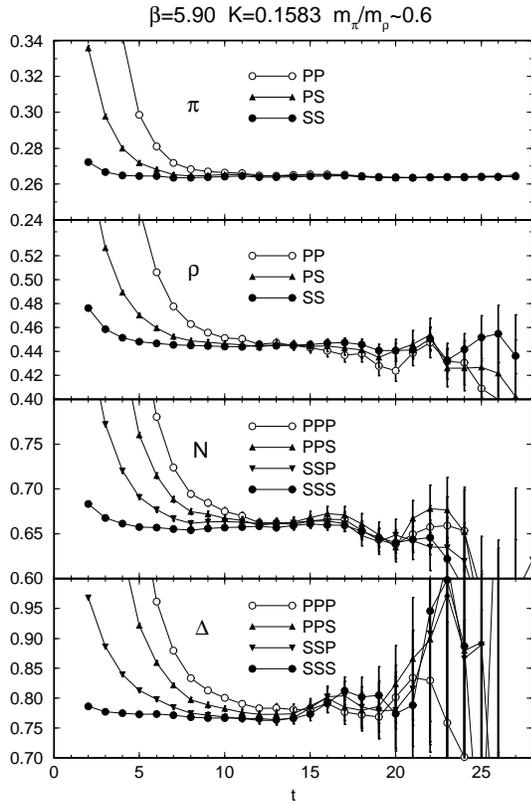}}
\caption{Typical effective masses
obtained with various combinations of quark sources.}
\label{fig:smear}
\end{figure}

\begin{figure}[htb]
\centerline{\epsfxsize=7cm \epsfbox{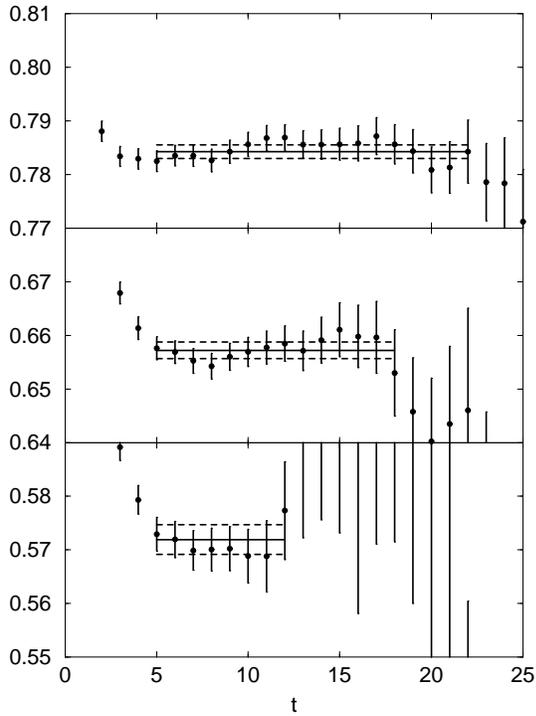}}
\caption{Effective mass plots of degenerate octet baryons at 
$m_{PS}/m_V\approx$ 0.75 (top), $m_{PS}/m_V\approx$ 0.6 (middle), 
and $m_{PS}/m_V\approx$ 0.4 (bottom) at $\beta=5.90$.}
\label{fig:pro590}
\end{figure}

\begin{figure}[htb]
\centerline{\epsfxsize=7cm \epsfbox{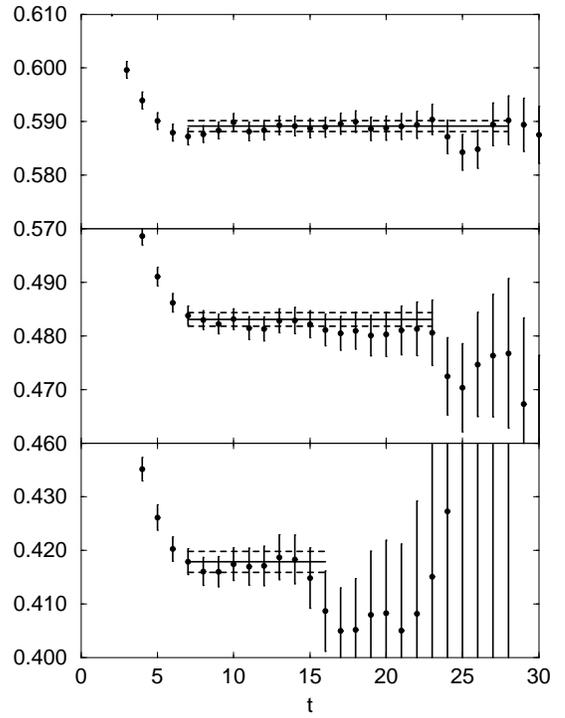}}
\caption{The same as Fig.\protect\ref{fig:pro590} at $\beta=$ 6.10.}
\label{fig:pro610}
\end{figure}

\begin{figure}[htb]
\centerline{\epsfxsize=7cm \epsfbox{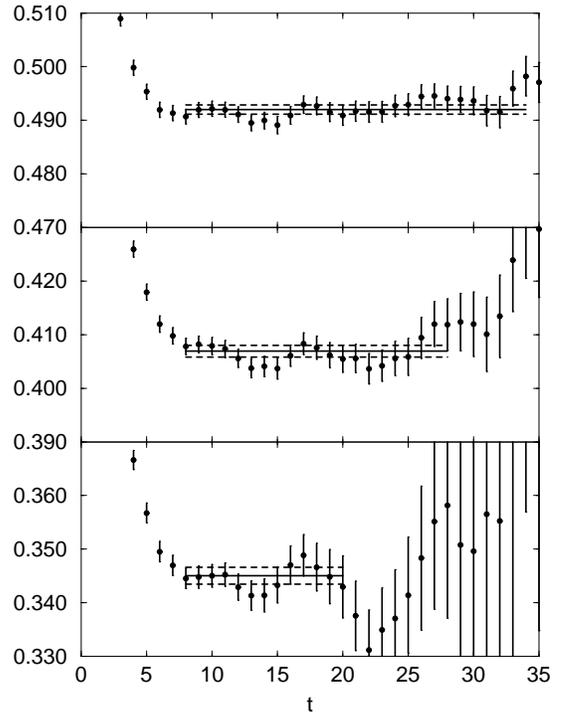}}
\caption{The same as Fig.\protect\ref{fig:pro590} at $\beta=$ 6.25.}
\label{fig:pro625}
\end{figure}

\begin{figure}[htb]
\centerline{\epsfxsize=7cm \epsfbox{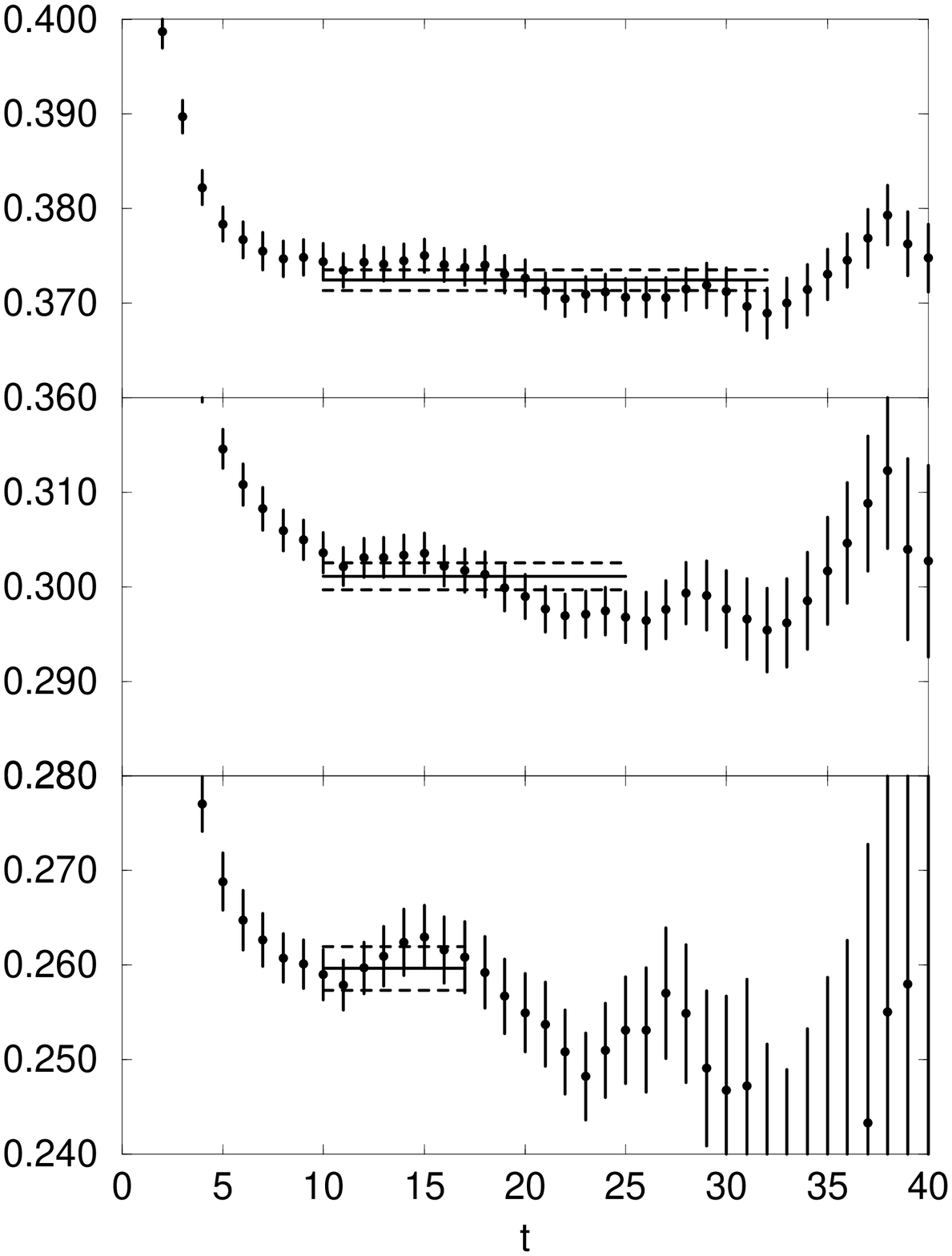}}
\caption{The same as Fig.\protect\ref{fig:pro590} at $\beta=$ 6.47.}
\label{fig:pro647}
\end{figure}

\begin{figure}[htb]
\centerline{\epsfxsize=7cm \epsfbox{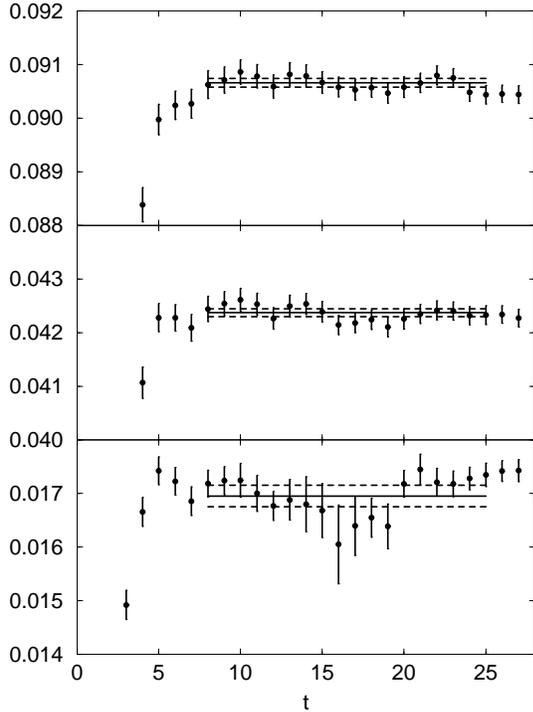}}
\caption{Effective mass plots for twice the AWI quark mass, 
$2 m_q^{AWI(0)}$, at $\beta=5.90$ 
for the degenerate cases, corresponding to 
$m_{PS}/m_V\approx$ 0.75 (top), 0.6 (middle) and 0.4 (bottom).}
\label{fig:e-mq590}
\end{figure}

\begin{figure}[htb]
\centerline{\epsfxsize=7cm \epsfbox{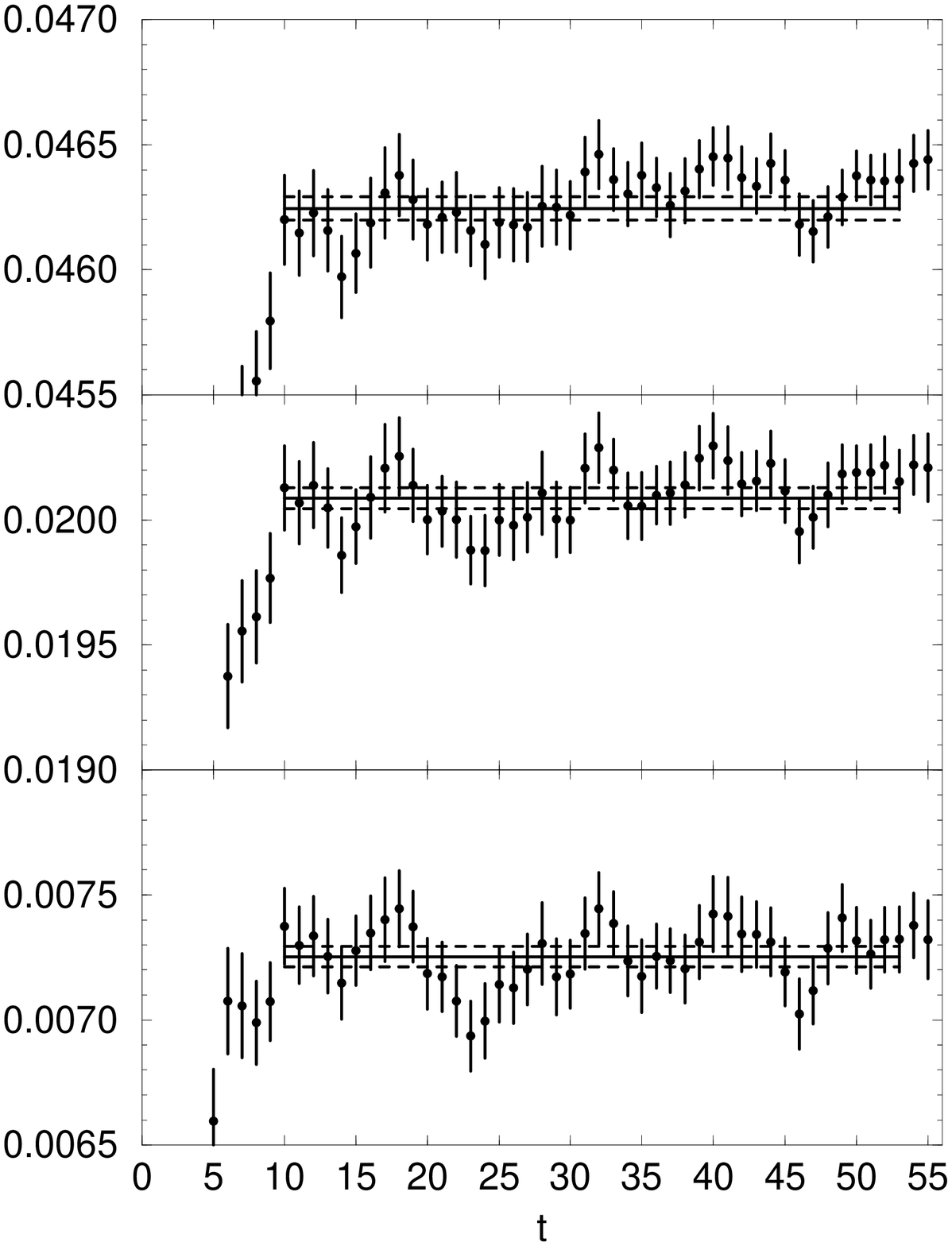}}
\caption{The same as Fig.~\protect\ref{fig:e-mq590} at $\beta=$ 6.47.}
\label{fig:e-mq647}
\end{figure}

\begin{figure}[htb]
\centerline{\epsfxsize=8.5cm \epsfbox{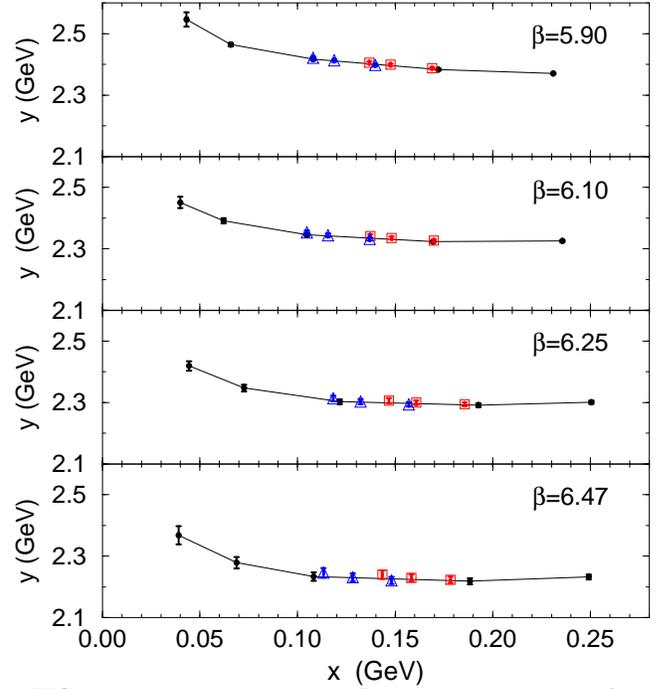}}
\caption{Deviations from the chiral relation 
$y \equiv m_{PS,12}^2/(m_1^{AWI}+m_2^{AWI}) = $ constant. 
The horizontal axis is $x \equiv m_1^{AWI}+m_2^{AWI}$.
Filled and open symbols represent degenerate and
non-degenerate quark mass cases, respectively. 
The error from the lattice spacing is not included.}
\label{fig:PS2mq}
\end{figure}

\begin{figure}[htb]
\centerline{\epsfxsize=8.5cm \epsfbox{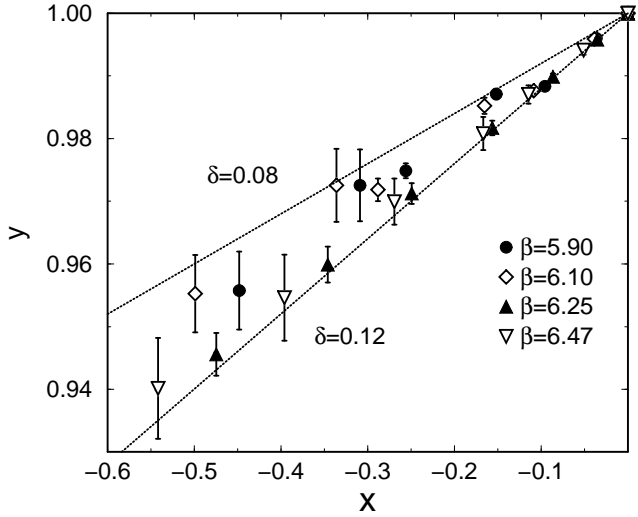}}
\caption{Test of quenched chiral logarithms for 
pseudoscalar meson masses.}
\label{fig:PSratio}
\end{figure}

\begin{figure}[htb]
\centerline{\epsfxsize=8.5cm \epsfbox{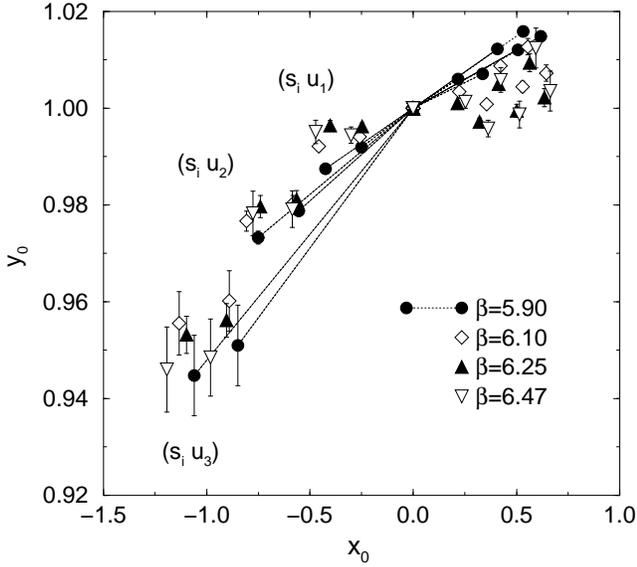}}
\caption{Ratio test proposed in Ref.~\protect\cite{ref:QChPTBGmass}.
Data at $\beta =5.90$ are connected to 
the point $(0.0,1.0)$ to guide the eyes.}
\label{fig:PSratioOrg}
\end{figure}

\begin{figure}[htb]
\centerline{\epsfxsize=8.5cm \epsfbox{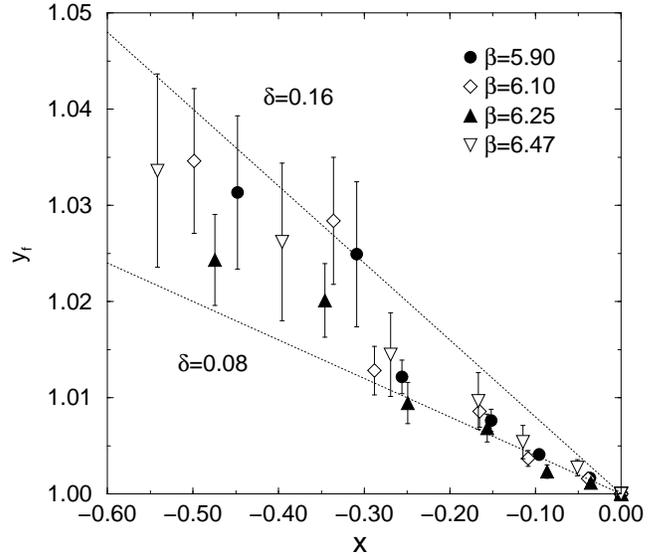}}
\caption{Ratio test of quenched chiral logarithms for 
pseudoscalar meson decay constants.}
\label{fig:PSdecayRatio}
\end{figure}

\begin{figure}[htb]
\centerline{\epsfxsize=8.5cm \epsfbox{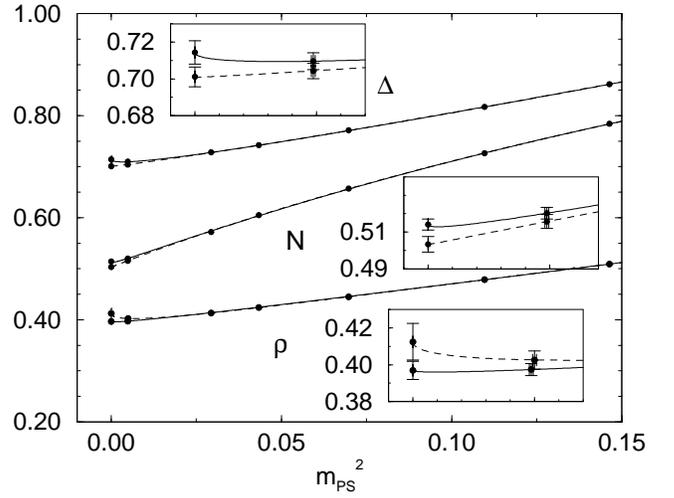}}
\caption{Degenerate hadron masses vs. $m_{PS}^2$ at $\beta$=5.9.
The leftmost points are values extrapolated to the chiral limit, 
and the second ones from the left are those at the physical point. 
Fits from two types of chiral extrapolations based on 
Q$\chi$PT are shown. See text for details.}
\label{fig:chiralB590Deg}
\end{figure}

\begin{figure}[htb]
\centerline{\epsfxsize=8.5cm \epsfbox{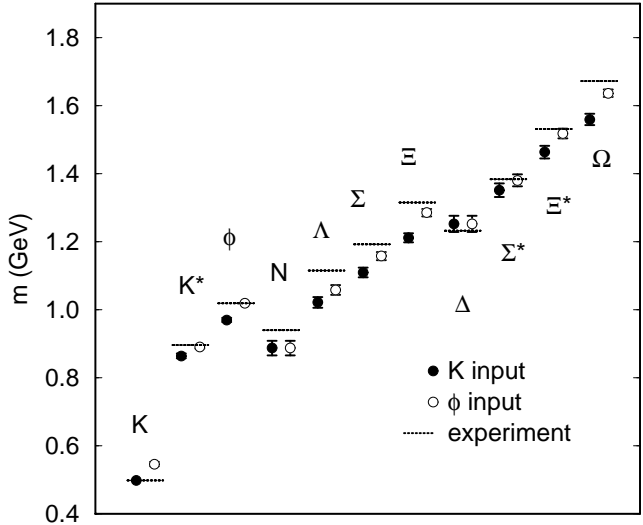}}
\caption{Light hadron spectrum from polynomial chiral fits.}
\label{fig:MassPoly}
\end{figure}

\begin{figure}[htb]
\centerline{\epsfxsize=8.5cm \epsfbox{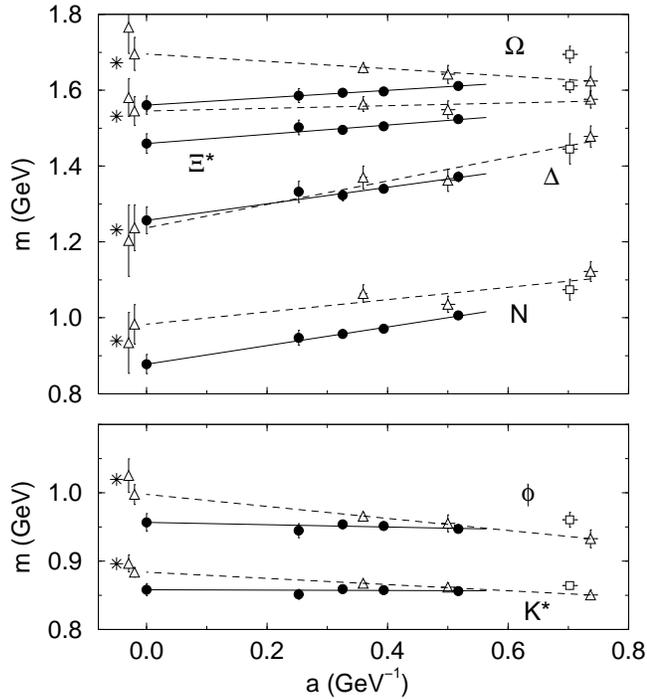}}
\caption{Comparison of our hadron masses (filled symbols) with those of
Ref.~\protect\cite{ref:GF11mass} (open symbols).
Open triangles (squares) are data obtained on lattices with 
$L_sa \approx 2.3$ fm ($L_sa\approx 3.4$ fm). 
Open symbols at $a\approx 0$ represent data on finite lattice (right)
and those after finite size corrections (left). 
Experimental values are shown by stars.
}
\label{fig:ContinuumGF11}
\end{figure}

\begin{figure}[htb]
\centerline{\epsfxsize=8.5cm \epsfbox{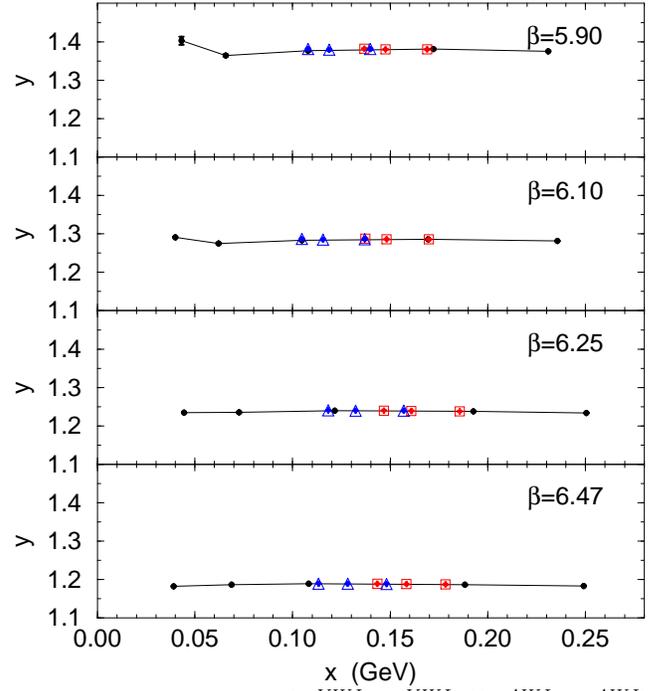}}
\caption{
The ratio $y=(m_1^{VWI}+m_2^{VWI})/(m_1^{AWI}+m_2^{AWI})$
as a function of $x=m_1^{AWI}+m_2^{AWI}$.
Filled and open symbols are for degenerate and non-degenerate cases, 
respectively.
$\kappa_c^{AWI}$ is used to calculate $m_{q_i}^{VWI}$.
Errors from the lattice spacing are not included.}
\label{fig:PertWIFP}
\end{figure}

\begin{figure}[htb]
\centerline{\epsfxsize=8.5cm \epsfbox{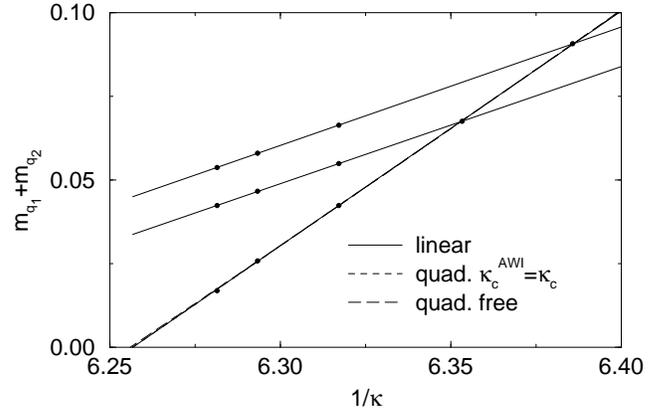}}
\caption{Chiral extrapolations of the quark mass based on the axial-vector
Ward identity at $\beta=5.90$.}
\label{fig:mqchiral}
\end{figure}

\begin{figure}[htb]
\centerline{\epsfxsize=8.5cm \epsfbox{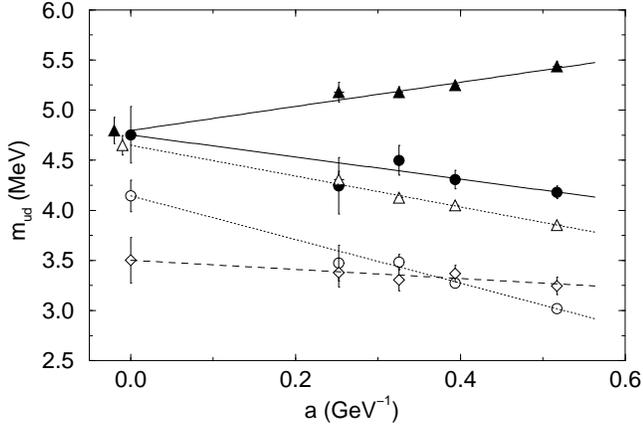}}
\caption{Comparison of the light quark masses determined from 
various chiral fits. 
The VWI (AWI) quark masses are shown by filled (open) symbols. 
Circles are results obtained from our main analysis, 
while triangles are from the alternative chiral fits discussed in the text.
Open diamonds are obtained from fits to $m_{PS}^2$ as a 
function of $m_q^{AWI}$.} 
\label{fig:Mud-chiral}
\end{figure}

\begin{figure}[htb]
\centerline{\epsfxsize=8.5cm \epsfbox{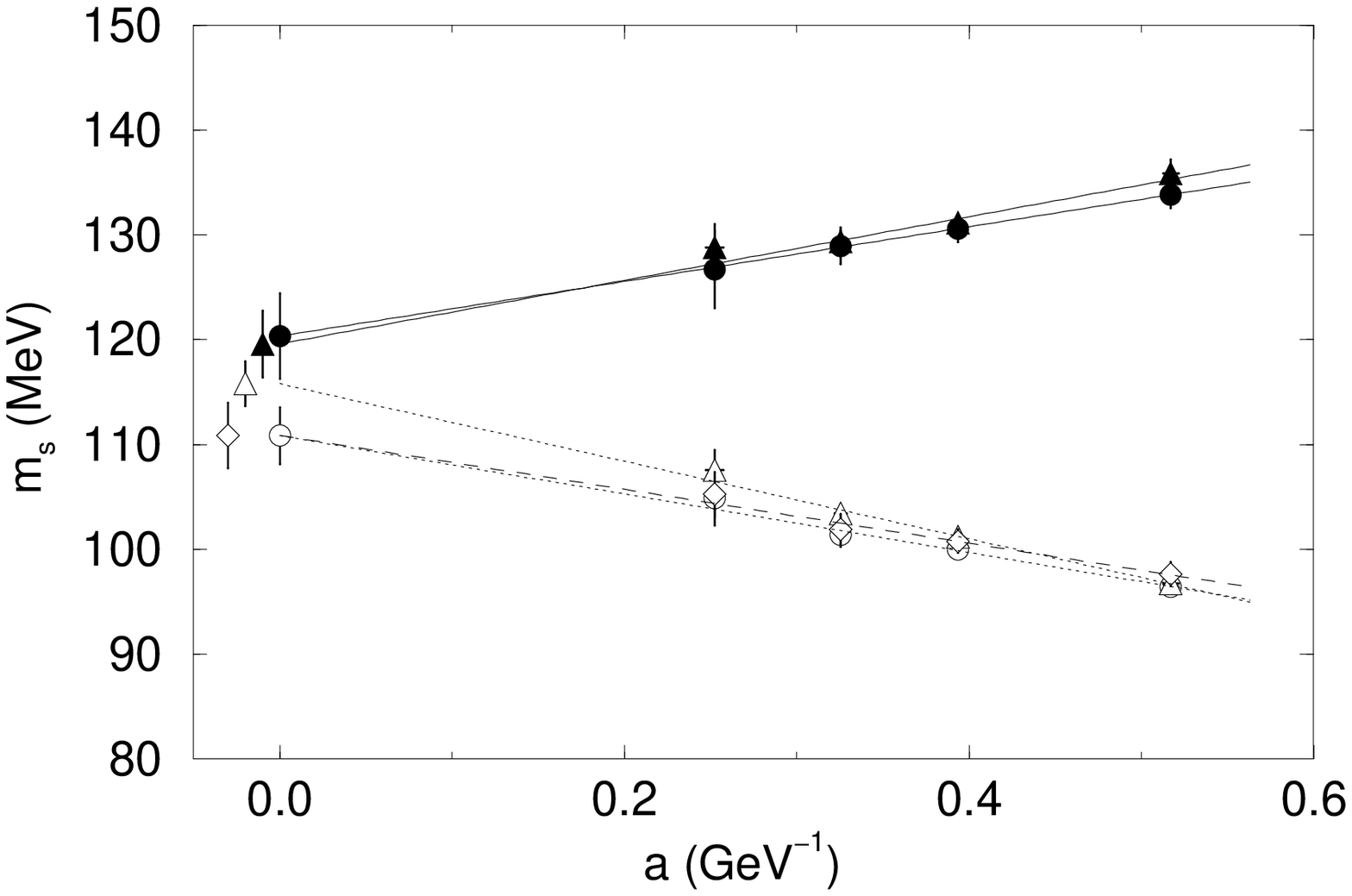}}
\caption{The same as Fig.~\protect\ref{fig:Mud-chiral} for 
the strange quark mass with $m_K$ input.}
\label{fig:Ms-chiral.K}
\end{figure}

\begin{figure}[htb]
\centerline{\epsfxsize=8.5cm \epsfbox{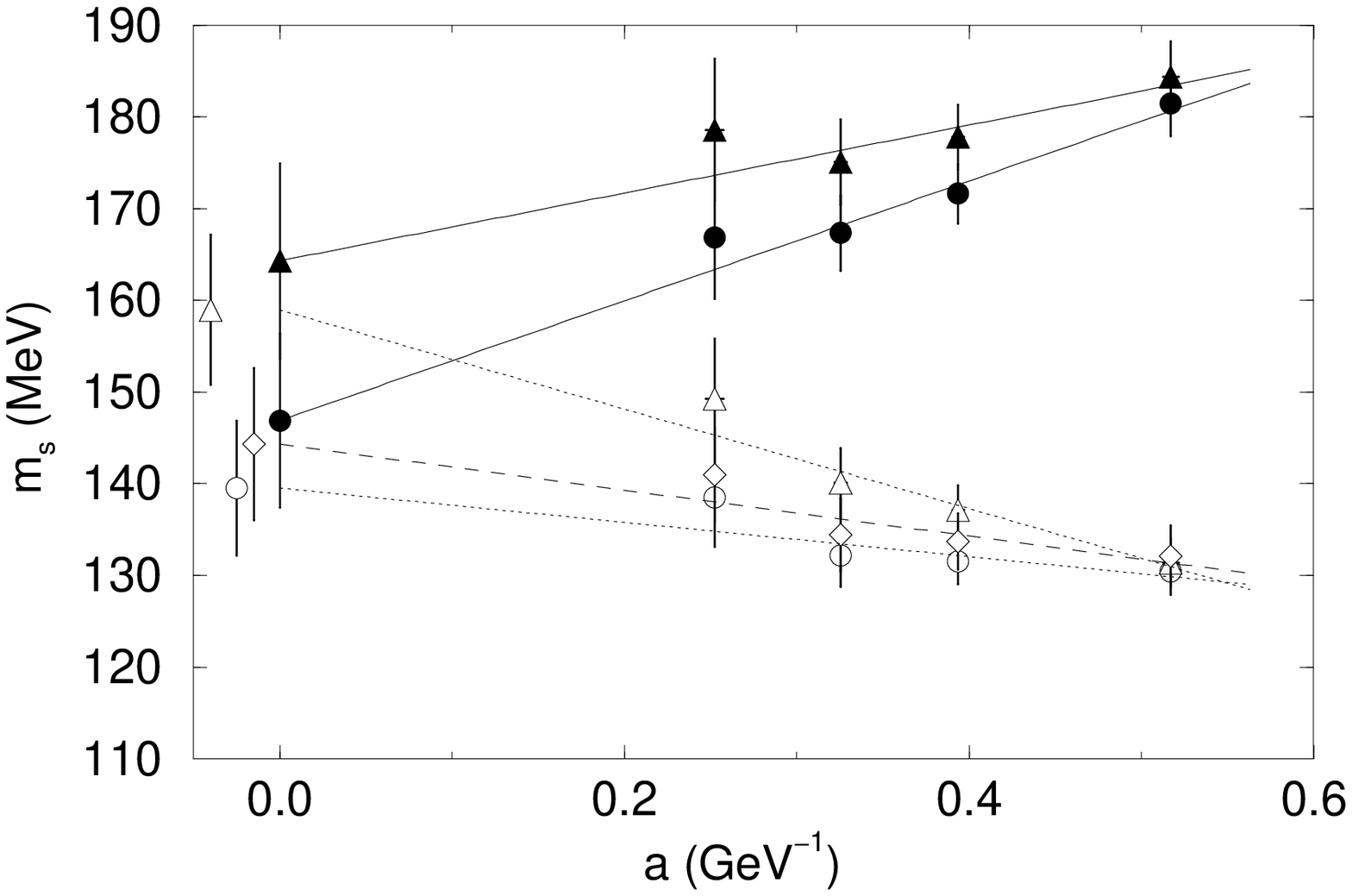}}
\caption{The same as Fig.~\protect\ref{fig:Mud-chiral} for 
the strange quark mass with $m_\phi$ input.}
\label{fig:Ms-chiral.P}
\end{figure}

\begin{figure}[htb]
\centerline{\epsfxsize=8.5cm \epsfbox{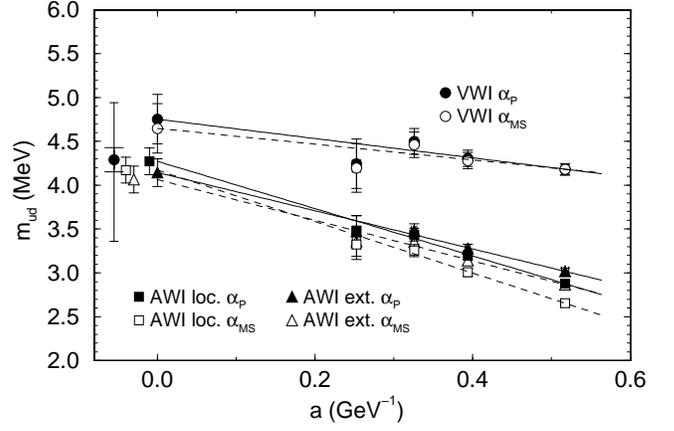}}
\caption{Comparison of the $u,d$ quark mass.
Circles are the VWI quark masses. 
Triangles (squares) are the AWI quark masses derived from the extended 
(local) axial vector current. 
Filled (open) symbols indicate that masses are calculated using 
$\alpha_P$ ($\alpha_{\overline {\rm MS}}$).
The leftmost data shows the result from a combined fit with 
the statistical error and the sum of statistical and systematic
errors. See Sec.~\protect\ref{sec:Mq} for detail.}
\label{fig:ELAM-Mud}
\end{figure}

\begin{figure}[htb]
\centerline{\epsfxsize=8.5cm \epsfbox{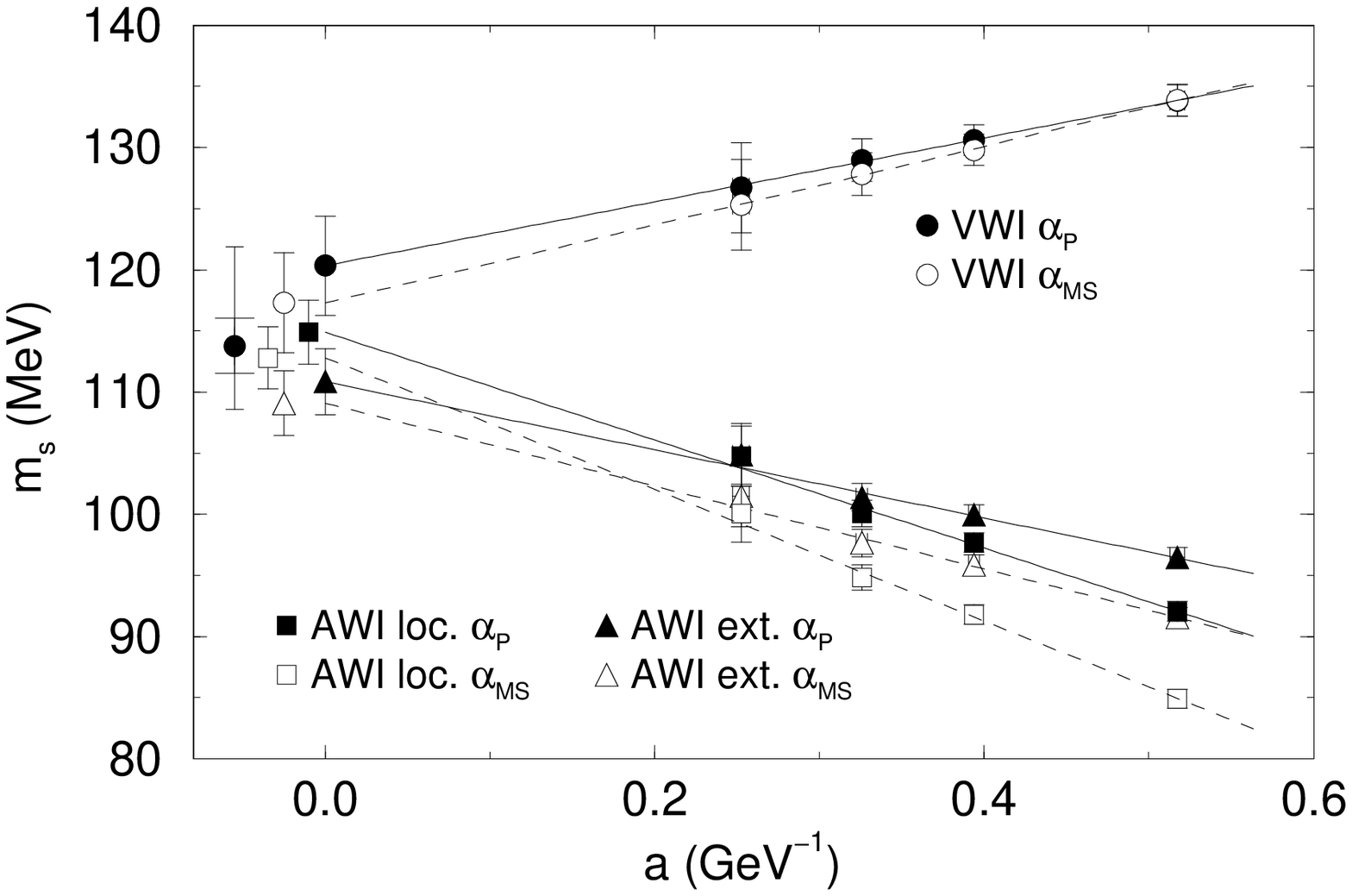}}
\caption{The same as Fig.~\protect\ref{fig:ELAM-Mud} for $m_s$
with $m_K$ input.}
\label{fig:ELAM}
\end{figure}

\begin{figure}[htb]
\centerline{\epsfxsize=8.5cm \epsfbox{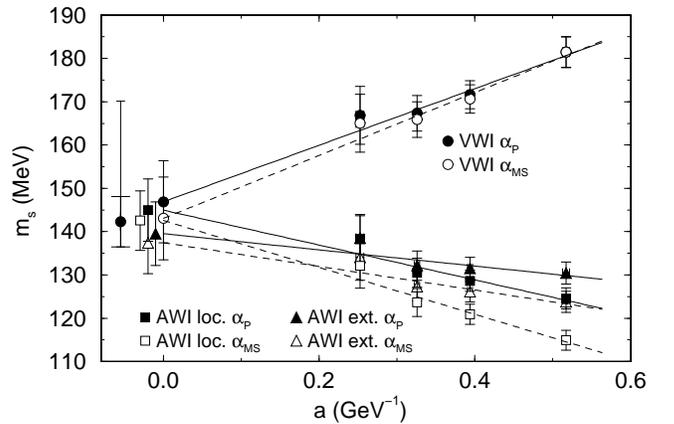}}
\caption{The same as Fig.~\protect\ref{fig:ELAM-Mud} for $m_s$
with $m_\phi$ input.}
\label{fig:ELAM-Pinp}
\end{figure}

\begin{figure}[htb]
\centerline{\epsfxsize=8.5cm \epsfbox{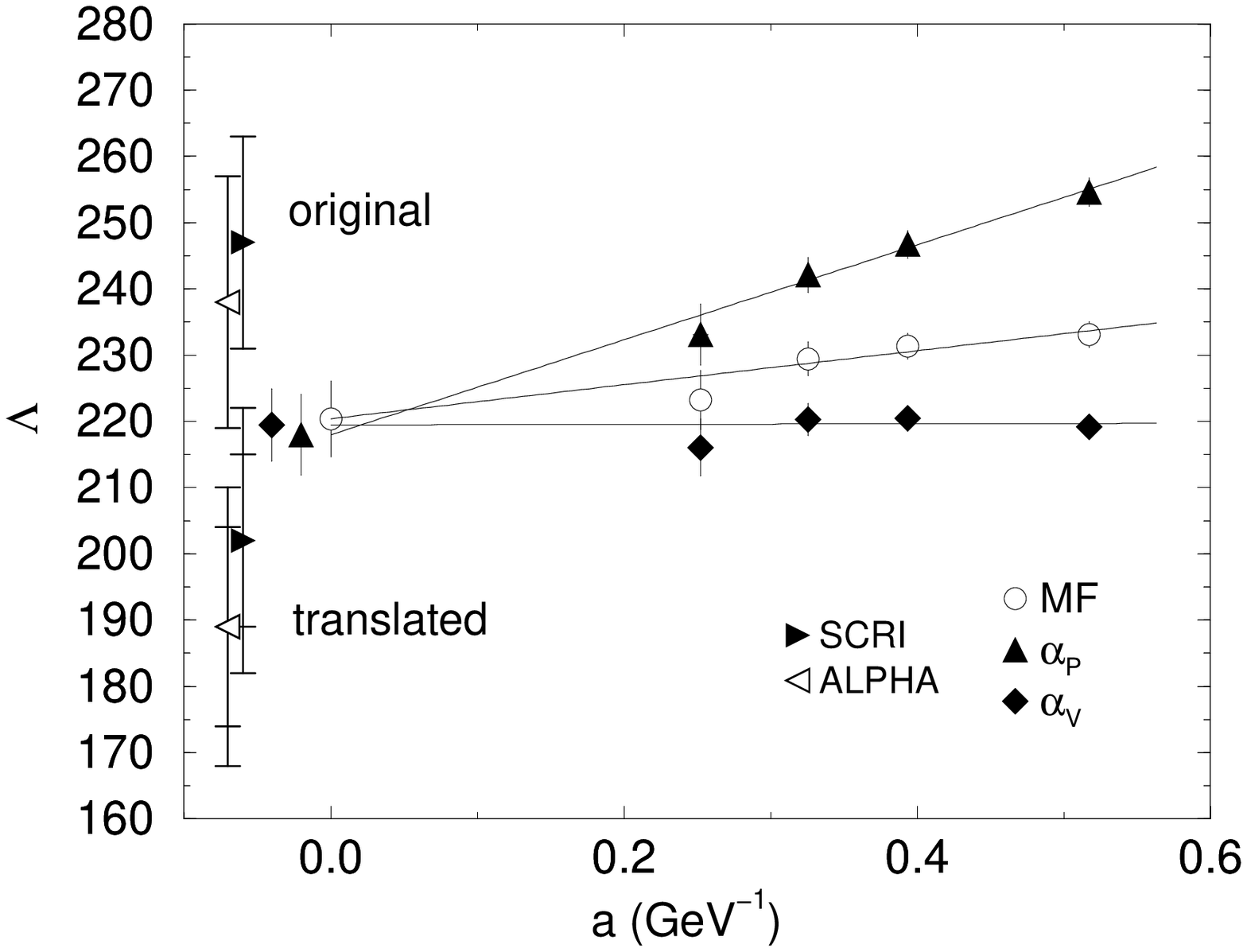}}
\caption{$\Lambda_{\overline {\rm MS}}$ vs $a$.}
\label{fig:Lambda}
\end{figure}

\begin{figure}[htb]
\centerline{\epsfxsize=8.5cm \epsfbox{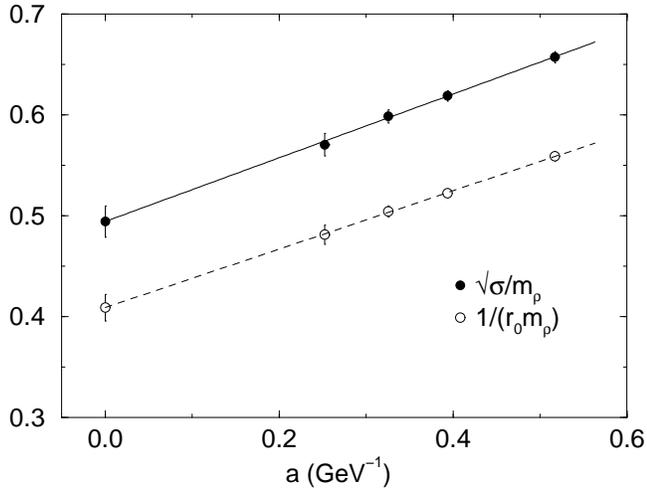}}
\caption{$\sqrt\sigma/m_\rho$ and $1/r_0m_\rho$ vs. $a$.}
\label{fig:pot-rho}
\end{figure}

\begin{figure}[htb]
\centerline{\epsfxsize=7cm \epsfbox{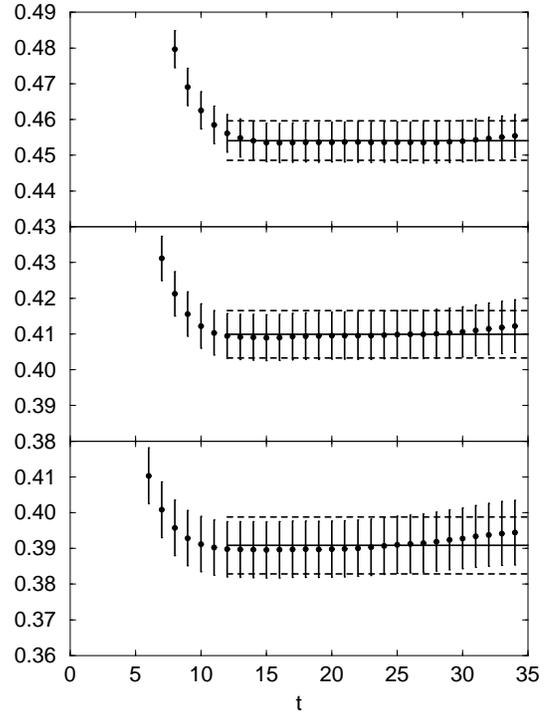}}
\caption{Examples of fits for Eq.~(\protect\ref{eq:PPPS})
at $\beta$ = 6.10 in units of $10^5$.
Three panels show the data for $m_{PS}/m_V\approx$ 0.75 (top),
0.6 (middle) and 0.4 (bottom).}
\label{fig:PPPS}
\end{figure}

\begin{figure}[htb]
\centerline{\epsfxsize=8.5cm \epsfbox{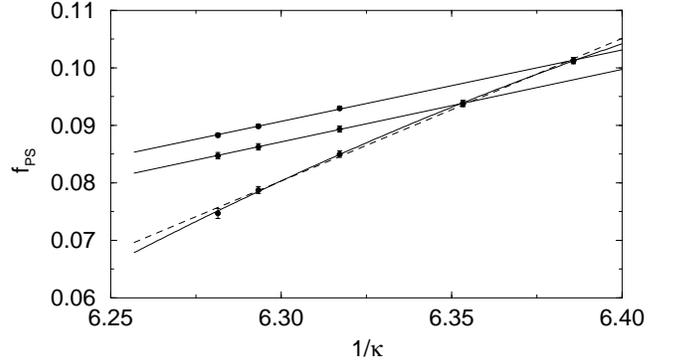}}
\caption{Pseudoscalar meson decay constants versus $1/\kappa$
at $\beta=5.90$. Solid curves show chiral fits with quadratic polynomial
(linear) for the degenerate (non-degenerate) case.
Dashed line is for a linear fit to the degenerate case.} 
\label{fig:fPSchiral}
\end{figure}

\begin{figure}[htb]
\centerline{\epsfxsize=7cm \epsfbox{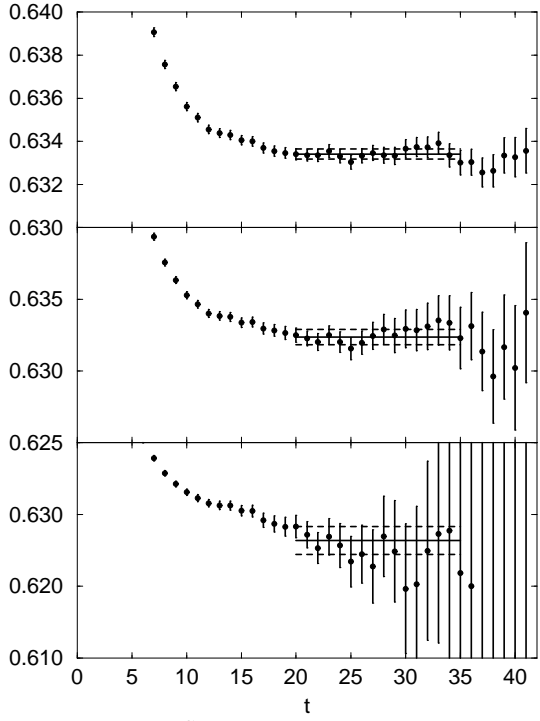}}
\caption{Ratios $\langle V^CV^L\rangle /\langle V^LV^L\rangle$ and
fits to obtain $Z_V$ at $\beta=6.25$.
Parameters are $m_{PS}/m_V\approx$ 0.75 (top), $m_{PS}/m_V\approx$ 0.6 (middle), 
and $m_{PS}/m_V\approx$ 0.4 (bottom) at $\beta=5.90$.}
\label{fig:ZV}
\end{figure}

\begin{figure}[htb]
\centerline{\epsfxsize=8.5cm \epsfbox{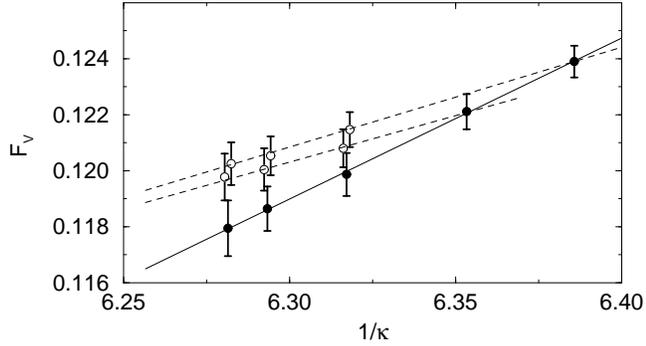}}
\caption{$F_V a$ vs. $1/\kappa$ at $\beta=5.90$ with chiral extrapolations.
Non-degenerate data are slightly shifted in $x$ for clarity.}
\label{fig:FVchiral}
\end{figure}

Chiral extrapolations are carried out in a similar manner to $f_{PS}$. 
Data are fit by a linear function of $1/\kappa$ (see Fig.~\ref{fig:FVchiral}).
Table~\ref{tab:FVPhys} presents $F_\rho$ and $F_\phi$ ($m_\phi$ input)
in units of GeV. 
We extrapolate them to the continuum limit linearly in $a$,
as shown in Fig.~\ref{fig:FVcont}.
Final results for $F_V$ using the non-perturbative renormalization 
constant are summarized 
in Eqs.~(\ref{eq:frhofnl}), (\ref{eq:fphifnl}) and Table~\ref{tab:fFinal}. 

For comparison, we study the renormalization factor estimated by 
tadpole-improved one-loop perturbation theory;
\bea
F_V^{TP} a = &&
\sqrt{1-{3\kappa_f}/{4\kappa_c}} \sqrt{1-{3\kappa_g}/{4\kappa_c}} \nonumber \\
&& \times (1 - 0.82\alpha_P(1/a)) F_V^{(0)}. \label{eq:FvTP}
\eea
Continuum extrapolations, plotted in Fig.~\ref{fig:FVcont}, show that
$F_V^{TP}$ exhibits scaling violation much larger than $F_V$.
Although the difference of $F_V^{TP}$ and $F_V$ at finite $a$
becomes smaller towards the continuum limit, 
$F_V^{TP}$ extrapolates to a value slightly smaller than $F_V$.
The coefficient of $\alpha_P$ in Eq.~(\ref{eq:FvTP}) is rather large,
when compared to that for $Z_A^{local}$ in Eq.~(\ref{eq:ZAlocal}) used 
for $f_{PS}$.
Higher order terms may be important to extract $F_V^{TP}$ in the
continuum limit from the region of our lattice spacing.
Therefore we take $F_V$ determined with the non-perturbative
renormalization factor as our best estimate.

\section{Light hadron spectrum as functions of quark masses}
\label{sec:methodB}

We carry out an additional analysis in which the continuum
extrapolations are made before the chiral fits, and   
obtain the hadron spectrum in the continuum limit as functions of
quark masses. 
A motivation of this analysis concerns the question of whether Q$\chi$PT mass
formulae at finite $a$ may suffer from lattice artifacts.  
Because Q$\chi$PT parameters obtained in Sec.~\ref{sec:spectrum} are
smooth in $a$, we expect that $O(a)$ terms vanish smoothly toward the
continuum limit.
The alternative procedure provides us with a more direct test of
the Q$\chi$PT formulae in the continuum limit.
Furthermore, quark mass or $m_{PS}/m_V$ dependence of
hadron masses in the continuum limit can be used to estimate the size of the
scaling violation in future calculations with improved actions at fixed $a$,
without difficult chiral extrapolations.

\subsection{Continuum extrapolation}
We take the continuum extrapolation of hadron mass data 
at finite quark mass.
To do this, we first interpolate or slightly extrapolate data to 
$m_{PS}/m_V= 0.75$, 0.7, 0.6, 0.5, and 0.4 at each $\beta$. 
These values of $m_{PS}/m_V$ are close to our raw data points 
such that the errors and uncertainties from fits are small.
In practice, we use the Q$\chi$PT formulae adopted
in Sec.~\ref{sec:spectrum}.
We also repeat the whole 
procedure using polynomial chiral fits to estimate systematics of choosing
different fit functions.
Errors for interpolated/extrapolated values are a factor 1--3 larger
than those for the raw data.

We then extrapolate the hadron masses to the continuum limit
at each $m_{PS}/m_V$. 
In order to calculate a relative value of $a$ as a function of $\beta$, 
we use the vector meson mass $m_V$ at $m_{PS}/m_V = 0.75$, which we 
denote as $m_V^{(0.75)}$,  
{\it i.e.}, masses at each $m_{PS}/m_V$ are normalized by $m_V^{(0.75)}$ 
and extrapolated linearly in $m_V^{(0.75)}a$ to the continuum limit.
A value of $m_V^{(0.75)}$ in physical units is not necessary 
at this step. 
We provide in 
Tables \ref{tab:PS075} -- \ref{tab:D075} normalized hadron masses at each 
$\beta$ obtained by the Q$\chi$PT formulae and those in the continuum limit.
Values of $\chi^2/N_{df}$ for the continuum extrapolations are
$\chi^2/N_{df} \approx 0.5$ for mesons, $\approx$ 1.0 for octet baryons and
$\approx$ 3.0 for decuplet baryons.

\subsection{Q$\chi$PT fits in the continuum limit}
We fit the continuum hadron spectrum, normalized by $m_V^{(0.75)}$, 
using the Q$\chi$PT formulae and following the procedure 
given in Sec.~\ref{sec:spectrum},
and obtain the hadron spectrum 
as well as the value of $m_V^{(0.75)}= 0.981(18)$ GeV.
The quark mass in the continuum limit at each $m_{PS}/m_V$ is necessary
for a chiral fit of pseudoscalar meson masses.  
We first interpolate the AWI quark mass 
$m_q^{AWI(0)}$ linearly in $1/\kappa$ to each value of $m_{PS}/m_V$ at each $\beta$. 
We then convert $m_q^{AWI(0)}$ to the values in the ${\overline {\rm MS}}$ scheme at
$\mu=2$ GeV, using $a$ determined from $m_V^{(0.75)}$.
The AWI quark masses are linearly extrapolated to the continuum 
limit with a reasonably small $\chi^2/N_{df}<0.5$. 
For completeness, we provide in Table \ref{tab:mq075} 
lattice spacings at each $\beta$ and quark masses for 
each $m_{PS}/m_V$ normalized by $m_V^{(0.75)}$.

\begin{figure}[htb]
\centerline{\epsfxsize=8.5cm \epsfbox{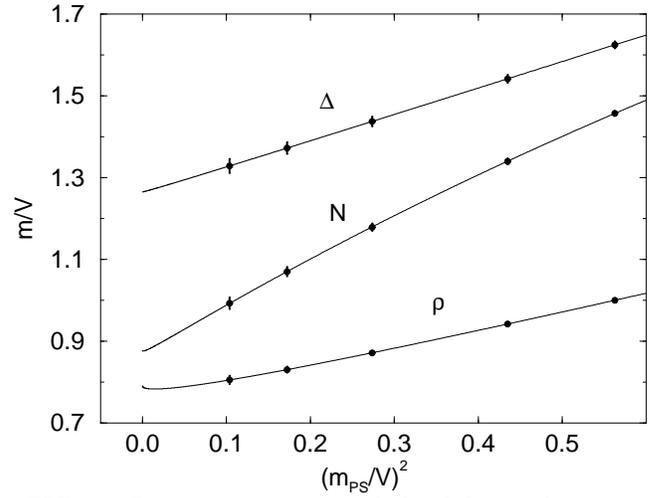}}
\caption{Degenerate masses and chiral fits in the continuum limit.
$V=m_V^{(0.75)}$. See text for details.}
\label{fig:FitInCont}
\end{figure}

\begin{figure}[htb]
\centerline{\epsfxsize=8.5cm \epsfbox{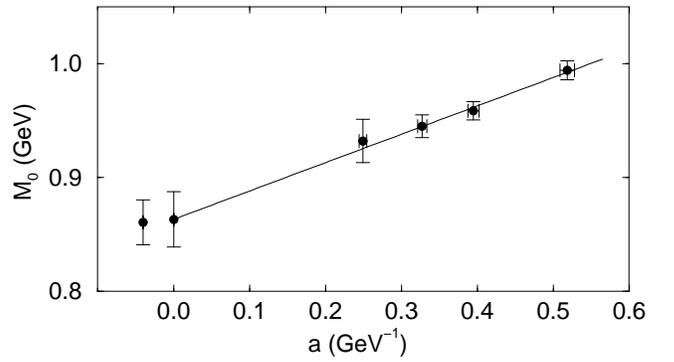}}
\caption{Octet baryon mass in the chiral limit versus 
the lattice spacing. The leftmost point represents 
the value in the continuum limit but determined from
the chiral fit in the continuum limit.}
\label{fig:OctM0}
\end{figure}

The Q$\chi$PT fits in the continuum limit look quite
similar to those at finite $a$ as shown in Fig.~\ref{fig:FitInCont}.
The values of the parameters are consistent 
with those obtained from an extrapolation of the parameters at finite $\beta$,
as given in Table \ref{tab:ContChiralPrm}.
See also, for example, Fig.~\ref{fig:OctM0} in which we plot
the octet baryon mass in the chiral limit.
A comparison of the coefficients of the singular terms of the Q$\chi$PT formulae
is made in Appendix \ref{appendix:QChPT}.

\subsection{Universality of the light hadron spectrum}
Table \ref{tab:ContChiralMass} summarizes the light hadron masses
thus obtained together with deviations from those from the original procedure.
The two spectra are consistent with each other with differences
smaller than 1$\sigma$ of the statistical error.  
See also Fig.~\ref{fig:ReverseK} in which we compare masses from 
the two methods for the case of the $m_K$ input. 
For interpolation/extrapolation of hadron masses, 
we test polynomial chiral fits instead of the Q$\chi$PT formulae.
The deviations remain within 1$\sigma$ for pseudoscalar mesons and
baryons and 1.5$\sigma$ for vector mesons. 

We therefore conclude that the two spectra determined by the two methods
are consistent with each other with differences smaller than 1.5$\sigma$. 
This confirms that both chiral and continuum extrapolations are under control.
We take the differences as systematic errors due to the chiral and continuum
extrapolations.

\section{Conclusions}
\label{sec:conclusions}

In this article we presented details of our calculation of the light 
hadron spectrum and quark masses in quenched QCD.
The computational power provided by the CP-PACS computer
enabled an exploration of hadron masses at lighter quark masses 
than hitherto attempted.

The high precision data for pseudoscalar meson masses revealed evidence 
supporting the presence of chiral singularities as predicted by
quenched chiral perturbation theory.
In the vector meson and baryon sectors the precision 
of our data is not sufficient to draw conclusive statements on quenched 
chiral singularities.  However, simulations covered the range of quark
masses sufficiently small to obtain a stable result of the spectrum
also for vector mesons and baryons.
Predictions do not depend on the choice of conventional polynomial
chiral fits or fits based on quenched chiral perturbation theory.
Since scaling violation also turned out to be mild for the
plaquette gluon and Wilson quark actions, the hadron  
masses keep the statistical precision of 
1--2\% for mesons and 2--3\% for baryons
after the continuum extrapolation.
The systematic error is estimated to be at most 1.7\%. 

The chief finding 
is the pattern and magnitude of the breakdown of the quenched
approximation for the light hadron spectrum.
In the meson sector the quenching error manifests itself 
in a small hyperfine splitting when compared with experiment. A small mass 
splitting is also seen in the decuplet baryons, and masses themselves
are small for octet baryons.
The magnitude of deviation, typically 5--10\%, is much larger
than the statistical and systematic errors.

The quenched approximation poses a limitation of our ability to 
predict fundamental parameters of QCD. 
The strange quark mass $m_s$ depends on hadron mass input, with 
a difference as large as 25\%.
The QCD scale parameter has uncertainty of the order of 15\% depending 
on inputs, {\it i.e.}, $m_\rho$ or phenomenological values of 
$\sqrt{\sigma}$ or $r_0$.

It appears to us that it is not worthwhile to pursue precision further,
and the effort should rather be directed
toward QCD simulations incorporating the effects of dynamical sea
quarks.  We in fact started an attempt~\cite{ref:cppacs-full} in this
direction using improved gluon and quark actions. 
In the course of this work we recalculated the quenched hadron spectrum 
using the improved actions.  The light hadron spectrum obtained in the 
continuum limit is in good agreement with the results reported in this 
article, providing further confirmation of the success and limitation 
of quenched QCD. 

\acknowledgements

We thank all the members of the CP-PACS Project with whom the
CP-PACS computer has been developed in the years 1992 -- 1996. 
Valuable discussions with M.~Golterman and S.~Sharpe on quenched chiral 
perturbation theory are gratefully acknowledged. 
We thank V.~Lesk for valuable suggestions on the manuscript.
This work is supported in part by the Grants-in-Aid of Ministry of
Education (Nos. 08NP0101, 
09304029, 
11640294, 
12304011, 
12640253, 
13640260, 
).
GB, SE, and KN are JSPS Research Fellows for 1996 -- 1997, 
1998 -- 2000 and 1998 -- 2000, respectively. 
HPS is supported by JSPS Research for Future Program in 1998.

\appendix

\section{Gauge fixing}
\label{appendix:gfix}

For the measurement of hadronic observable, we fix gauge 
configurations to the Coulomb gauge by maximizing the quantity
\be
 H = \sum_n\ \sum_{\mu=1}^3 \mbox{Re} [\mbox{Tr}\ U_{n,\mu}].
\ee
To achieve this, we combine two methods: (a) an SU(2) subgroup
method, which is similar to the pseudo heat-bath algorithm 
for SU(3) gauge theories\cite{ref:HeatBath}, 
and (b) an over-relaxed steepest descent method\cite{ref:Mino}.
Both methods can be vectorized and parallelized by splitting 
the lattice sites into even and odd sites.

\subsection{SU(2) subgroup method}

Under a gauge transformation 
\be
U_{n,\mu} \to U'_{n,\mu} = G_n U_{n,\mu} G^\dagger_{n+\hat\mu}
\ee
with SU(3) matrices $G_n$, $H$ transforms as 
\bea
 H \to 
 H' &=& \sum_{\mu=1}^3 
    \mbox{Tr}\ [G_n U_{n,\mu} + U_{n-\hat\mu,\mu}G^\dagger_n] 
\nonumber \\
&&    + \mbox{terms indep. of}\ G_n.
\eea
If the gauge group is SU(2), it is easy to find the solution $G_n$ 
which maximizes $H'$ for a given site $n$. 
The global maximum can be achieved iteratively by repeating the 
maximization at all sites.
For the SU(3) case, we can gradually increase $H$ by applying 
the maximization for different SU(2) subgroups of SU(3). 
In our simulations, we maximize three SU(2) subgroups per iteration.

\subsection{Over-relaxed steepest descent method}

Alternatively, $H$ can be maximized by iterative gauge 
transformations with
\bea
&& G_n = \exp (\alpha_n \Delta_n) \\
&& \Delta_n = \sum_{\mu=1}^3 \Delta_{n,\mu}\\
&& \Delta_{n,\mu} = [U_{n-\hat\mu,\mu} - U_{n,\mu} - {\rm h.c.} - {\rm trace}]
\eea
with a suitable real number $\alpha_n$, 
because the gradient of $H$
with respect to $\theta_n^a$ defined by $G_n=\exp(i\lambda^a \theta_n^a)$
is given by
\be
\frac{\delta H}{\delta \theta^a_n} = \mbox{Re\,Tr}
[ - i\lambda^a \Delta_n ].
\ee
Note that ${\delta H}/{\delta \theta^a_n}$ vanishes at the supremum of $H$
and is proportional to $\sum_\mu \partial_\mu A_\mu^a(x)$ in the continuum limit.

A candidate of $\alpha_n$ is obtained by solving the supremum condition of $H$
\be
\mbox{Re} \sum_{\mu=1}^3 \mbox{Tr}
[e^{\alpha_n\Delta_n}\Delta_nU_{n,\mu} - U_{n-\hat\mu,\mu}\Delta_n e^{-\alpha_n\Delta_n}]=0
\ee
to the leading order of $\alpha_n$, which gives 
\be
\alpha_n = \frac{\mbox{Re}\sum_\mu \mbox{Tr}\ [\Delta_n\{U_{n-\hat\mu,\mu}-U_{n,\mu}\}]}
                {\mbox{Re}\sum_\mu \mbox{Tr}\ [\Delta_n^2\{U_{n-\hat\mu,\mu}+U_{n,\mu}\}]}.
\ee
An over-relaxation is introduced through a parameter
$1<\omega<2$ in the gauge transformation: 
\be
G_n = \exp(\omega \alpha_n \Delta_n).
\ee

\subsection{Implementation on the CP-PACS}

We find that the steepest descent algorithm with the over-relaxing 
parameter $\omega \approx 1.98$--1.99 converges much faster than 
the SU(2) subgroup method, 
when the configuration is already close to the maximum. 
When the configuration is far from the maximum, however, 
this method sometimes fails to converge. 
Therefore, we first apply the SU(2) subgroup method for several hundred 
iterations to drive the configuration close to the maximum.
In our simulations, we adopt the SU(2) subgroup method for the first 
200, 500, 1000 and 6000 iterations at $\beta=$5.9, 6.1, 6.25 and 6.47,
respectively, before applying the steepest descent method. 

For a convergence check, we monitor 
\be 
  h = H / (9 L_s^3 L_t) 
\ee
and 
\be
 \Delta = \frac{1}{L_s^3L_t} \sum_n\  \frac{1}{3} \sum_{\mu=1}^3 
          \frac{1}{3} \mbox{Tr} [\Delta_{n,\mu}\Delta_{n,\mu}^\dagger].
\ee
Note that
\be
 \Delta \propto \int dx \sum_{\mu=1}^3 \sum_{a=1}^8 (\partial_\mu A_\mu^a)^2
\ee
in the continuum limit.
We truncate iterations at $i$-th iteration when the conditions
\bea
  \vert h_i - h_{i-1} \vert & < & 10^{-10}, \\
                 \Delta_i   & < & 10^{-14}, 
\eea
are both satisfied.
We have checked that stronger convergence criterion does not lead to
a significant difference in hadron propagators. 

\section{Quark propagators}
\label{appendix:qprop}

In order to solve Eq.(\ref{eq:qprop}), we use a red/black-preconditioned 
minimal residual (MR) algorithm. 
We accelerate the convergence by applying successive over-relaxations. 
For the over-relaxation factor, we adopt $f=1.1$ from a test study 
made at $f=0.9$ 1.0, 1.1 and 1.2.
The number of iterations for $f=1.1$ is smaller than 
those for $f=0.9$ ($f=1.2$) by 20\% (5\%). 

At each MR step, we monitor the residual sum of squares
$R^2$, where $R=S-D(\kappa)G$, and
truncate iterations when the condition $R^2 < 10^{-12}$
($R^2 < 10^{-14}$ at $\beta=6.47$) is satisfied for the point 
source, and $R^2 < 10^{-7}$ for the smeared source.
Hadron propagators obtained with this stopping condition
are compared with those with much stronger one on several configurations.
From this test we estimate that the truncation error in hadron
propagators on each configuration is smaller than 5\% of
our final statistical error for any particle at any time slice.

The number of iterations needed to calculate quark propagators is
listed in Table~\ref{tab:param3}.
The number is approximately proportional to the inverse
of the quark mass defined by $(1/\kappa - 1/\kappa_c)/2$, 
where $\kappa_c$ is the critical hopping parameter.

Our exponentially smeared source Eq.~(\ref{eq:expsmear}) is motivated 
from a result of the JLQCD Collaboration for the pion wave function,
\be
\Psi(r) = \frac{\langle 0|\sum_n \bar\psi(n) \gamma_5 \psi(n+r)|\pi\rangle}
               {\langle 0|\sum_n \bar\psi(n) \gamma_5 \psi(n)|\pi\rangle},
\ee
which was well reproduced by a single exponential function
$\Psi(r)=A \exp( -Br)$ except at the origin $\Psi(0) \approx 1.0$\cite{ref:JLQCD-smear}. 
The coefficient $A$ and 
the slope $B$ of the JLQCD Collaboration can be parametrized as 
\bea
A(m_\rho a,m_q a) &=& a_0 + a_1 m_\rho a + (a_2 + a_3 (m_\rho a)^2) m_q a, \\  
B(m_\rho a,m_q a) &=& b_0 + b_1 m_\rho a + (b_2 + b_3 (m_\rho a)^2) m_q a,
\eea
where $m_\rho a$ is the dimensionless $\rho$ meson mass in the chiral limit, 
$m_q a=(1/\kappa-1/\kappa_c)/2$ is the bare quark mass, and 
\bea
&& a_0 =  0.915,\ \   a_1 =  0.576,  \nonumber \\
&& a_2 =  0.2127,\ \  a_3 = -0.644,  \nonumber \\
&& b_0 = -0.0537,\ \  b_1 =  0.978,  \nonumber \\
&& b_2 =  0.2146,\ \  b_3 = -0.5123. \nonumber 
\eea
Applying the results of test runs for $m_\rho$ and $\kappa_c$, we 
adopt $A$ and $B$ listed in Table \ref{tab:param3}. 
The smearing radius is approximately constant,
$a/B \approx 0.33$ fm at our four $\beta$ values.

\section{Hadron propagators}
\label{appendix:hadron}

For mesons we employ the operators defined by
\be
M_A^{fg}(n)=\overline{f}_n\Gamma_A g_n  \label{eq:mesonop}
\ee
where $f$ and $g$ are quark fields with flavors $f$ and $g$, and
$\Gamma_A$ is one of the 16 spin matrices
\bea
&& \Gamma_S=I,\ \Gamma_P=\gamma_5,\ \Gamma_{\tilde P}=i\gamma_0\gamma_5,\
\Gamma_V=\gamma_i,\ \Gamma_{\tilde V}=i\gamma_0\gamma_i   \nonumber \\
&& \Gamma_A=i\gamma_5\gamma_i \ \mbox{and}\ \Gamma_T=i[\gamma_i,\gamma_j]/2
\ \ (i,j=1,2,3).
\eea
With these operators, we calculate 16 meson propagators, 
$
\langle M_A(n)M_A(0)\rangle
$.

For the spin $1/2$ octet baryons we take the operators defined by
\be
O^{fgh}_\alpha(n)=\epsilon^{abc}(f_n^{Ta}C\gamma_5g_n^b)h^c_{n\alpha},
\label{eq:octetop}
\ee
where $a,b,c$ are color indices, 
$C=\gamma_4\gamma_2$ is the charge conjugation matrix, 
and $\alpha=1,2$ represents the spin state, up or down, of the octet baryon. 
To distinguish $\Sigma$- and $\Lambda$-like octet baryons, 
we antisymmetrize flavor indices, written symbolically as
\bea
\Sigma&=&-\frac{[fh]g+[gh]f}{\sqrt{2}}, \\
\Lambda&=&\frac{ [fh]g-[gh]f-2[fg]h}{\sqrt{6}},
\eea
with $[fg]=fg-gf$.

The spin $3/2$ decuplet baryon operators are given by
\be
D^{fgh}_{\mu,\alpha}(n) = 
\epsilon^{abc}(f_n^{Ta}C\gamma_\mu g_n^b)h^c_{n\alpha}.
\label{eq:decupletop}
\ee
Writing out the spin structure $(\mu,\alpha)$ explicitly, we obtain
\bea
 D_{3/2}  = && \epsilon^{abc}(f^{Ta}C\Gamma_+g^b)h^c_1, \\
 D_{1/2}  = && \epsilon^{abc}[(f^{Ta}C\Gamma_0g^b)h^c_1 - 
               (f^{Ta}C\Gamma_+g^b)h^c_2]/3, \\
 D_{-1/2} = && \epsilon^{abc}[(f^{Ta}C\Gamma_0g^b)h^c_2 - 
               (f^{Ta}C\Gamma_-g^b)h^c_1]/3, \\
 D_{-3/2} = && \epsilon^{abc}(f^{Ta}C\Gamma_-g^b)h^c_2,
\eea
where $\Gamma_{\pm} =(\gamma_1 \mp i\gamma_2)/2$, $\Gamma_0 = \gamma_3$, and
the subscript of $D$ denotes the $z$-component of the spin.

With these operators, we calculate 8 baryon propagators given by
\bea
&&\langle \Sigma_\alpha(n)\Sigma_\alpha(0)\rangle, \quad \alpha=1,2 \\
&&\langle \Lambda_\alpha(n)\Lambda_\alpha(0)\rangle, \quad \alpha=1,2 \\
&&\langle D_S(n)D_S(0)\rangle, \quad S=3/2,1/2,-1/2,-3/2
\eea
together with 8 antibaryon propagators defined by the same expressions 
with the baryon operators replaced by antibaryon operators.
To enhance the signal, we average zero momentum propagators on a 
configuration over all states with the same quantum numbers; 
three polarization states for the vector meson 
and two (four) spin states for the octet (decuplet) baryon.
We also average the propagators for the particle and the antiparticle,
{\it i.e.}, meson propagators at $t$ and $L_t-t$ are averaged, 
and baryon propagators for particle at $t$ and those for antiparticle at
$L_t-t$ are averaged.
Errors of propagators are estimated treating the data thus obtained
being statistically independent.

\section{Correlated fits for chiral extrapolation}
\label{appendix:CorrFits}

A difficulty in a correlated chiral extrapolation is that the size of 
the full covariance matrix (error matrix) 
$C(t,\kappa_1\kappa_2;t',\kappa_1'\kappa_2')$ is very large 
and the matrix becomes close to a singular matrix 
so that $C^{-1}$ necessary for
$\chi^2$ fits cannot be estimated reliably.
When we make a fit for both degenerate and non-degenerate data
simultaneously, the size of $C$ becomes of order 200, {\it e.g.}, 
$(28-10+1)\times 11$ = 209
for fitting range $[10,28]$ used for 11 combinations of quark masses.
We find that the condition number of $C$ is far beyond $10^{15}$ so that
one cannot handle the matrix within our numerical accuracy.
Instead of the simultaneous fits, we make independent fits
for degenerate and non-degenerate cases. 
Namely we make three fits for mesons: 1) degenerate fit with 5 data at
$\kappa_1\kappa_2= s_1s_1,s_2s_2,u_1u_1,u_2u_2,u_3u_3$,
2) non-degenerate fit with one of the hopping parameter $\kappa_1=s_1$ 
using 4 data at $\kappa_1\kappa_2= s_1s_1,s_1u_1,s_1u_2,s_1u_3$, 
and 3) the same with $\kappa_1=s_2$ using 
$\kappa_1\kappa_2= s_2s_2,s_2u_1,s_2u_2,s_2u_3$.
Fits for baryons are carried out similarly, because two quarks in baryons
are taken to be degenerate.

For each correlated fit, we employ the procedure 
adopted in Ref.\cite{ref:QCDPAX96}.
We first minimize $\chi^2_{\it full}$ defined by
\bea
\chi^2_{\it full} = \sum_{t,t',\kappa_2,\kappa_2'}
\{ G(t,\kappa_2) - && G_0(t,\kappa_2)\}C^{-1}(t,\kappa_2; t'\kappa_2')  
\nonumber \\
  && \{ G(t',\kappa_2') - G_0(t',\kappa_2')\}, 
\label{eq:chi2full}
\eea
where $G(t,\kappa_2)$ are data of hadron propagators and 
$G_0(t,\kappa_2)$ is a fitting function {\it e.g.} 
$G_0(t,\kappa_2)=A(\kappa_2)\exp(-m(\kappa_2)t)$ for baryons.
The masses $m(\kappa_2)$ thus determined are in general different from
those obtained by individual $\chi^2$ fits for each $\kappa_2$.
The difference is small for most cases, though it occasionally amounts to
1.2$\sigma$.
We use masses from the full correlated fits for later analyses.
Results remain essentially the same 
if masses from individual fits are used.
We then calculate an error matrix $\Sigma$ for the fit parameters by
\be
\Sigma = (D^TC^{-1}D)^{-1}, \label{eq:fullSigma}
\ee
where $D$ is the Jacobian defined by
\bea
&&D_{t,\kappa_2;A(\kappa_2'),m(\kappa_2')} 
\nonumber \\
&& = [ \partial G_0(t,\kappa_2)/\partial A(\kappa_2'),
    \partial G_0(t,\kappa_2)/\partial m(\kappa_2')]. 
\eea
Note that $D$ is diagonal with respect to $\kappa_2$.
For the chiral extrapolation, we minimize $\chi^2_{\it ext}$ given by
\be
\chi^2_{\it ext} = \sum_{\kappa_2}
\{m(\kappa_2) - f(\kappa_2)\} \Sigma^{-1}(\kappa_2,\kappa_2')
\{m(\kappa_2') - f(\kappa_2')\}, 
\label{eq:chiext}
\ee
where $f(\kappa_2)$ is the fitting function we try and the matrix
$\Sigma(\kappa_2,\kappa_2')$ is the sub-matrix among the masses of the 
full error matrix $\Sigma$ in Eq.(\ref{eq:fullSigma}).

This procedure works well only when the full covariance matrix $C$
is reliably determined.
Although the condition number of $C$ is as large
as $10^{14}$-- $10^{15}$ for degenerate fits and $10^{12}$ --$10^{13}$ 
for non-degenerate fits, small eigenvalues and the corresponding
eigenvectors responsible for $\chi^2_{\it full}$ are determined well.
The jackknife error for these quantities is of the order of 10\%  (20\%
at $\beta=6.47$). 
The error sometimes increases to 50\% for large eigenvalues.
We think that it causes no problem because corresponding eigenmodes 
have no significant contribution to $\chi^2_{\it full}$.

In fully correlated chiral fits, $\chi^2_{\it full}$ should
be close to $N_{df}$. 
For pseudoscalar mesons, we choose the fitting range carefully 
to satisfy this condition (see Table \ref{tab:chi2fullPS}). 
We use the common ranges for 
both Q$\chi$PT fits in Sec.~\ref{sec:spectrum:chiral}
and quadratic polynomial fits in Sec.~\ref{sec:spectrum:poly}.

For vector mesons and baryons, we employ uncorrelated Q$\chi$PT 
chiral extrapolations made simultaneously to hadron masses with degenerate and
non-degenerate quark masses. 
In addition, we perform fully correlated polynomial chiral fits. 
With our choice of fitting ranges for the former, we observe 
that $\chi^2_{\it full}/N_{df}$ for the latter are much larger 
than unity although errors of $\chi^2_{\it full}/N_{df}$ are also large. 
After trial and error, discarding data around $t\approx t_{min}$
in degenerate propagators at $\kappa = s_1$ and $s_2$ and/or $u_1$ 
leads to $\chi^2_{\it full}/N_{df} \approx 1$.
We therefore use different ranges for polynomial chiral fits
from those for Q$\chi$PT fits,  
noting that masses for these cases do not change significantly.

For the mass-mass covariance matrix $\Sigma$, the condition
number is $O(10^4)$ and all errors for eigenvalues and eigenvectors are 
contained within 25\% of central values. 
Hence we are able to perform numerically reliable full
correlated chiral extrapolations.

\section{Chiral fits for pseudoscalar mesons}
\label{appendix:PSChiralFits}

We compare eight
chiral fit functions for pseudoscalar meson masses 
listed in Table~\ref{tab:chi2extPS}, using  
fully-correlated fits described in Appendix~\ref{appendix:CorrFits}.
The first five are for the degenerate cases, $s_is_i$ and $u_iu_i$.
The fits 1--3 are polynomials, while the fits 4 and 5 are based on 
Q$\chi$PT using Eq.~(\ref{eq:mps0}) with $\alpha_\Phi=0$, 
with or without the quadratic term $B(m_1+m_2)^2$. 
The remaining three functions are for the non-degenerate cases,
$s_iu_j$.
Two of them are polynomials, and the last one is the 
Q$\chi$PT formula (\ref{eq:mps0}) with $\alpha_\Phi=0$. 
For non-degenerate fits, we fix $\kappa_c$ to a value determined 
from a degenerate fit.

Values of $\chi^2_{ext}/N_{df}$
for chiral extrapolations are very large irrespective of the choice of
fitting functions,  
as shown in Table~\ref{tab:chi2extPS}.
A similar phenomenon was observed also in previous studies.
See {\it e.g.}\ Ref.~\cite{ref:MILC-KS}. 
This may be due to the fact that higher order terms are required
to reproduce our data. 
Because the number of our data points is limited, inclusion of 
such terms is not possible, and hence we choose a functional form 
from overall consistency.

Concerning the relative magnitude of $\chi^2_{ext}/N_{df}$, we find,
for the degenerate fits, that $\chi^2_{ext}/N_{df}$ is the smallest with the
Q$\chi$PT formulae keeping the $O(m_q^2)$ term (Fit 5). 
When we remove the $O(m_q^2)$ term, $\chi^2_{ext}/N_{df}$ becomes much larger. 
This observation is consistent with the presence of the $O(m_q^2)$ term
expected from the mass ratio test given in Sec.~\ref{sec:qChPT}.
For non-degenerate fits, similar values of $\chi^2_{ext}/N_{df}$ are obtained 
from both quadratic (Fit 2n) and Q$\chi$PT (Fit 5n) fits.

To keep consistency with the presence of 
Q$\chi$PT singularity shown by the ratio tests in Sec.~\ref{sec:qChPT:mPS}, 
we decide to employ Q$\chi$PT fits (Fit 5 and Fit 5n) for the main
course of our analyses and use quadratic fits (Fit 2 and Fit 2n) for
estimations of systematic errors from the chiral extrapolation.

\section{Test of Q$\chi$PT mass formula for vector mesons and baryons}
\label{appendix:QChPT}

Lowest order Q$\chi$PT mass formulae for vector mesons\cite{ref:Booth}
and baryons\cite{ref:LabrenzSharpe} can be written as
\begin{equation}
m_H(m_{PS}) = m^0 + C_{1/2} m_{PS} + C_1 m_{PS}^2 + C_{3/2}m_{PS}^3,
\label{eq:mvb}
\end{equation}
where $C_i$ are polynomials of the couplings in the quenched chiral 
Lagrangian.  
We find that it is difficult to constrain all coupling parameters 
(6 for vector mesons and 11 for baryons in addition to $\delta$ and 
$\alpha_\Phi$) under the limitation of the accuracy of our mass data and
the number of data points. 
We, therefore, set $\delta=0.1$ and $\alpha_\Phi=0$, and drop the couplings 
of the flavor-singlet pseudoscalar meson to vector mesons and baryons.
We also set $f=132$ MeV unless otherwise stated.

\subsection{Vector Mesons}
\label{appendix:QChPTv}
The lowest order Q$\chi$PT formula for vector mesons\cite{ref:Booth} is given by
\bea
 m_V  &=&   m_V^0
        + \frac{C_{1/2}}{6} \{\frac{3}{2}(m_{uu}+m_{ss})
        + 2\frac{m_{ss}^3 - m_{uu}^3}{m_{ss}^2 - m_{uu}^2} \} 
\nonumber \\
      &&+  \frac{C_1}{2} (m_{uu}^2 + m_{ss}^2)
\nonumber \\
      &&+ C_D (m_{uu}^3 + m_{ss}^3) + C_N m_{us}^3,
\label{eq:qchptV}
\eea
where $m_{fg}$ is the pseudoscalar meson mass. 
The coefficients are written in terms of the couplings 
$g_i$ of the quenched chiral Lagrangian; 
\bea
    C_{1/2} &=&  -4 \pi g_2^2 \delta, \\
    C_D  &=& -2 g_2 g_4 / (12 \pi f^2), \\
    C_N  &=&  (-4 g_1 g_2 + 4 g_2^2 A_N) / (12 \pi f^2).
\eea
The coefficient $C_{1/2}$ of the term linear in $m_{PS}$ is 
proportional to $\delta$, and represents the quenched singularity.
Q$\chi$PT predicts a negative value for $C_{1/2}$.
A phenomenological estimate is $C_{1/2}\approx -0.71$ 
using $\delta=0.1$ and $g_2=0.75$ \cite{ref:Booth}.

\subsubsection{Ratio test}

We perform ratio tests for vector meson masses independently
for degenerate and non-degenerate cases.
For the degenerate case, the mass formula Eq.~(\ref{eq:qchptV}) 
reduces to Eq.~(\ref{eq:mvb}) with 
\be
C_{3/2}=C_D+C_N.
\label{eq:c32v}
\ee
Hence, we obtain a relation 
\be
y= C_{1/2} + C_1 x + O(m_{PS}^2)
\label{eq:ratioVB}
\ee
for
\bea
y&=&\frac{m_{V,22}-m_{V,11}}{m_{PS,22}-m_{PS,11}},
\\
x&=& m_{PS,22} + m_{PS,11}.
\eea
We calculate $y$ and $x$ for all 10 combinations of
$\kappa_1$ and $\kappa_2$.
Eq.~(\ref{eq:ratioVB}) is obtained 
also for non-degenerate cases, with $y$ and $x$ replaced by 
more complicated expressions.
We obtain 15 data points for $y$ and $x$ from all combinations 
satisfying $(u_is_j) \ne (u'_is'_j)$.

\begin{figure}[htb]
\centerline{\epsfxsize=8.5cm \epsfbox{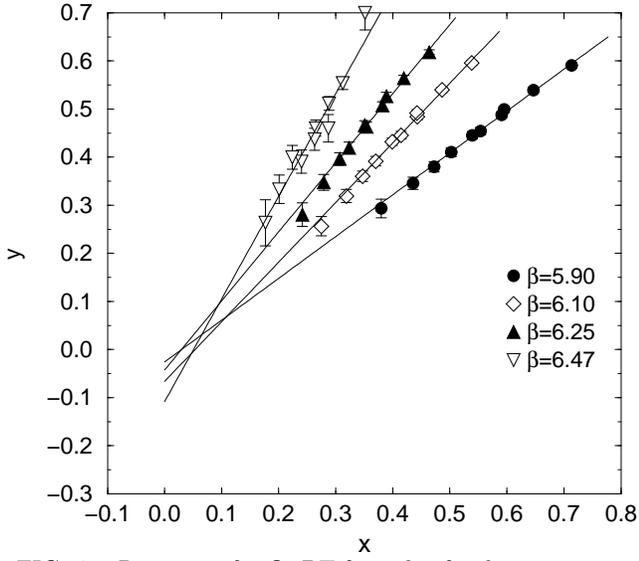}}
\caption{Ratio test for Q$\chi$PT formulae for 
degenerate vector mesons.}
\label{fig:RhoRatioDeg}
\end{figure}

\begin{figure}[htb]
\centerline{\epsfxsize=8.5cm \epsfbox{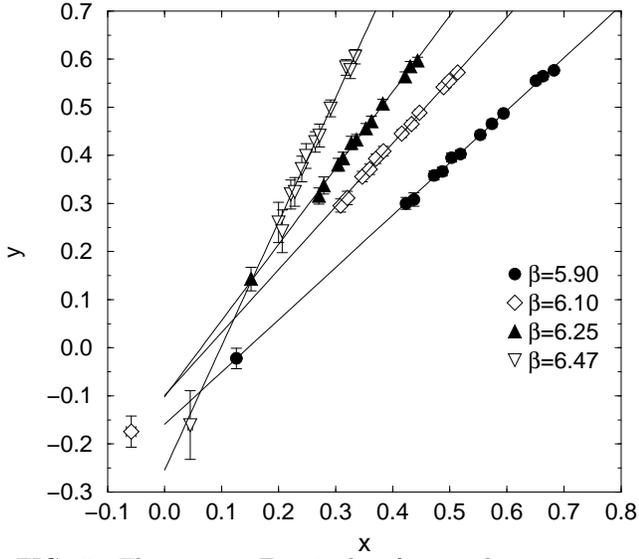}}
\caption{The same as Fig.~\protect\ref{fig:RhoRatioDeg}, 
but for non-degenerate vector mesons.}
\label{fig:RhoRatioND}
\end{figure}

In Figs.~\ref{fig:RhoRatioDeg} and \ref{fig:RhoRatioND} we show
plots of $y$ versus $x$ for degenerate and non-degenerate cases,
respectively.
Data for $y$ are fitted well by a linear function of $x$ and intercepts
are negative taking a value in the range $-0.3$ -- 0.0.
These results suggest that the $O(m_{PS}^2)$ term in Eq.(\ref{eq:ratioVB})
and hence $C_{3/2}$ in Eq.(\ref{eq:c32v}) or
$C_D$ and $C_N$ in Eq.(\ref{eq:qchptV}) are small. 
We find that 
$C_{1/2}$ is negative but much smaller in magnitude than
the phenomenological estimate $\approx -0.71$.

\subsubsection{Chiral fit}

We make a fit (\ref{eq:qchptV}) directly to the vector meson mass data, 
treating the degenerate and non-degenerate cases simultaneously. 
We ignore correlations among masses for different quark masses,
or else the size of full covariance matrix becomes too large to obtain
reliable matrix elements.

We find that the Q$\chi$PT fit keeping all five fitting parameters 
is unstable; the covariance matrix for the fit parameters 
becomes close to singular
with the condition number of $O(10^8$--$10^9)$. 

Dropping the $O(m_{PS}^3)$ terms, the fit become more stable 
with the condition numbers of $O(10^4$--$10^6)$. 
The fit reproduces data equally well as that including 
the $O(m_{PS}^3)$ terms 
as illustrated in Fig.~\ref{fig:chiralB590Deg} for the degenerate data at 
$\beta=5.90$ (the baryon fits in this figure are discussed below). 
Equivalently, $\chi^2/N_{df}$ of at most 0.8 obtained without
the $O(m_{PS}^3)$ terms are comparable to 0.9 including the 
$O(m_{PS}^3)$ terms. 
Taking the stability of fits as a guide, we adopt the fit without 
the $O(m_{PS}^3)$ terms for vector mesons.
This choice also agrees with a small value of $C_{3/2}$
observed in the ratio test.

For the coefficient of the leading chiral singularity, we obtain
$C_{1/2}= -0.058(27)$, $-0.075(28)$, $-0.065(35)$ and $-0.155(72)$
at $\beta= 5.90$, 6.10, 6.25 and 6.47, respectively, which are stable 
under variation of $\beta$. 
Taking a weighted average, we find $C_{1/2} = -0.071(8)$ 
which is much smaller than a phenomenological estimate 
$C_{1/2}= -4\pi g_2^2\delta = -0.71$ ($\delta=0.1$). 
The value $C_{1/2}=-0.13(10)$ obtained from chiral fits in the 
continuum limit (Sec.\ref{sec:methodB}) also supports this conclusion.
A much larger value of $C_{1/2}\approx$ $-$0.6 --
$-$0.7 is reported in Ref.~\cite{ref:QCDSF}. 
One possible origin of the difference is finite size effects in 
Ref.~\cite{ref:QCDSF}.

\subsection{Decuplet baryons}
\label{appendix:QChPTd}

Keeping terms up to $O(m_{PS}^3)$, 
the Q$\chi$PT formula for decuplet baryon masses is given 
by\cite{ref:LabrenzSharpe}
\bea
m_D &=& m_D^0 + \frac{5H^2}{162}(4w_{uu}+4w_{us}+w_{ss}) \nonumber \\
   &&+ \frac{C^2}{18}(w_{uu}-2w_{us}+w_{ss}) \nonumber \\
   &&+ c(2m_{uu}^2+m_{ss}^2) \nonumber \\
   &&+ (-10H^2/81+C^2/9)(v_{uu}+2v_{us}),
\label{eq:qchptD}
\eea
where $w_{ij}$ and $v_{ij}$ are defined in Eqs.~(\ref{eq:Wus}) and (\ref{eq:Vus}).
The terms proportional to $v_{ij}$ are $O(m_{PS}^3)$. 

\subsubsection{Ratio test}

For the degenerate case,
the formula (\ref{eq:qchptD}) reduces 
to the cubic polynomial as in Eq.(\ref{eq:mvb}),
where
\be
C_{1/2} = -(5\pi\delta/6) H^2.
\label{eq:ChalfBd}
\ee
Therefore, we can perform a ratio test by constructing $y$ and $x$ 
which are similar to those for vector mesons. 

\begin{figure}[htb]
\centerline{\epsfxsize=8.5cm \epsfbox{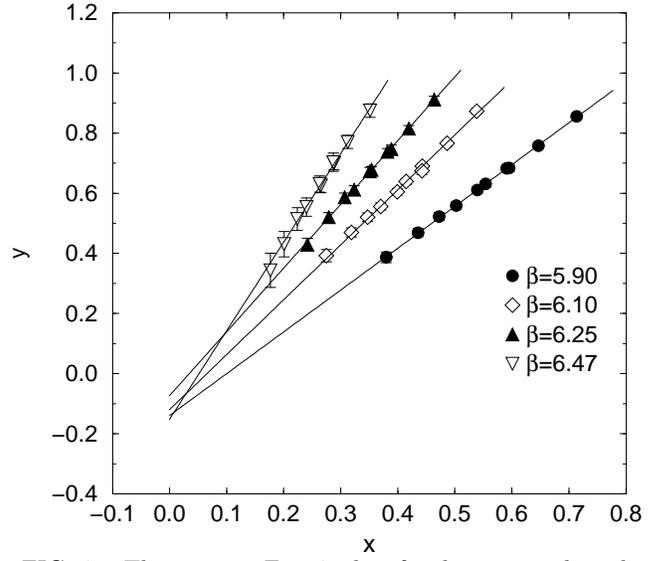}}
\caption{The same as Fig.~\protect\ref{fig:RhoRatioDeg},
but for degenerate decuplet baryons.}
\label{fig:DecRatioDeg}
\end{figure}

As Fig.~\ref{fig:DecRatioDeg} shows,  
we observe $C_{1/2}\approx -0.1$ from the intercept and a small value of
$C_{3/2}$ from the linearity of data.
The negative value for $C_{1/2}$ is consistent with the negative sign
in Eq.(\ref{eq:ChalfBd}).

\subsubsection{Chiral fit}

We study direct fits of decuplet baryon mass data to the mass formula 
(\ref{eq:qchptD}) using degenerate and non-degenerate masses simultaneously.
Since $C_{3/2}$ is small in the ratio test, 
we consider fits with and without the $O(m_{PS}^3)$ terms in 
Eq.(\ref{eq:qchptD}). We note that the number of parameters is the same
(four).  

The two types of fits are indistinguishable. 
Both yield $\chi^2/N_{df} \approx 0.5$ with 
the condition number of the covariance matrix of $O(10^6\mbox{\rm --}10^7)$, 
and the fitting curves are nearly identical in the range of 
our data points. See Fig.\ref{fig:chiralB590Deg} for the results at 
$\beta=5.90$. 

The fits without the $O(m_{PS}^3)$ terms give  
$H^2=0.65(13)$, 0.55(15), 0.39(15) and 0.70(40),
and $C_{1/2}= -(5\pi\delta/6) H^2=-0.169(33)$, $-0.145(39)$, 
$-0.101(40)$ and $-0.18(10)$ for $\beta=5.90$, 6.10, 6.25, and 6.47, 
respectively.
Results at different $\beta$ are consistent with each other, 
and the weighted average $C_{1/2}=-0.14(1)$ 
is consistent with the estimate $\approx -0.1$ from the ratio test.

Including the $O(m_{PS}^3)$ terms reduces the $C_{1/2}$ coefficient to
$C_{1/2}=-0.020(29)$, $-0.019(32)$  $-0.031(28)$ and $-0.204(73)$
for $\beta=5.90$, 6.10, 6.25, and 6.47, which are much smaller than 
the values from the ratio test.

Considering the consistency with the ratio test, we employ
the mass formula without the $O(m_{PS}^3)$ terms in the main analyses, 
and use the fits with the $O(m_{PS}^3)$ terms for error estimations.

For the adopted fit, 
$C^2$ is consistent with zero within $\sim 2 \sigma$.  
Setting $C^2=0$ does not change the decuplet baryon masses at 
the physical point by more than 0.5$\sigma$ for all $\beta$ values. 
We keep, however, the term proportional to $C^2$ since it is the 
same order as the term proportional to $H^2$.

\subsection{Octet baryon}
\label{appendix:QChPTo}

The Q$\chi$PT mass formulae we consider for octet baryons are written as
\bea
m_\Sigma &=& m_O^0 \nonumber \\
  &&+ \frac{1}{2}\{4F^2 w_{uu}-4(D-F)F w_{us}+(D-F)^2 w_{ss}\} \nonumber \\
  &&+ \frac{C^2}{9} (w_{uu} - 2 w_{us} + w_{ss})  \nonumber \\
  &&- 4b_F m_{uu}^2 + 2(b_D-b_F)m_{ss}^2       \nonumber \\
  &&+ (2D^2/3-2F^2-C^2/9)v_{uu} \nonumber \\
  &&+ (2D^2/3-4DF+2F^2-5C^2/9)v_{us},
\label{eq:qchptS} \\
m_\Lambda &=& m_O^0 \nonumber \\
   &&+ \frac{1}{2}
    \{ (4D/3-2F)^2 w_{uu} + (D/3+F)^2w_{ss} \nonumber \\
   &&- 2(4D/3-2F)(D/3+F)w_{us}\} \nonumber \\
   &&+ 4(2b_D/3-b_F)m_{uu}^2 - 2(b_D/3+b_F)m_{ss}^2 \nonumber \\
   &&+ (2D^2/9-8DF/3+2F^2-C^2/3)v_{uu} \nonumber \\
   &&+ (10D^2/9-4DF/3-2F^2-C^2/3)v_{us}.
\label{eq:qchptL}
\eea
The notation is the same as that for decuplet baryons.
  
\subsubsection{Ratio test}
The Q$\chi$PT mass formula (\ref{eq:qchptS}) reduces to 
the cubic polynomial in Eq.~(\ref{eq:mvb}) with 
\be
C_{1/2} = -(3\pi\delta/2) (D-3F)^2.
\label{eq:ChalfBo}
\ee
The negative sign of $C_{1/2}$ suggests that the degenerate octet 
mass is a concave function of $m_{PS}^2$ for sufficiently light quarks.
On the other hand, our data exhibits a convex ({\it i.e.}, negative) 
curvature as shown in Fig.~\ref{fig:chiralB590Deg}.
A negative curvature is also observed in Ref.~\cite{ref:MILC-KS} for
the Kogut-Susskind action.
We consider that the negative curvature is due to $O(m_{PS}^3)$ terms, 
and that they have a large effect for the range of quark masses 
covered by our data.

\begin{figure}[htb]
\centerline{\epsfxsize=8.5cm \epsfbox{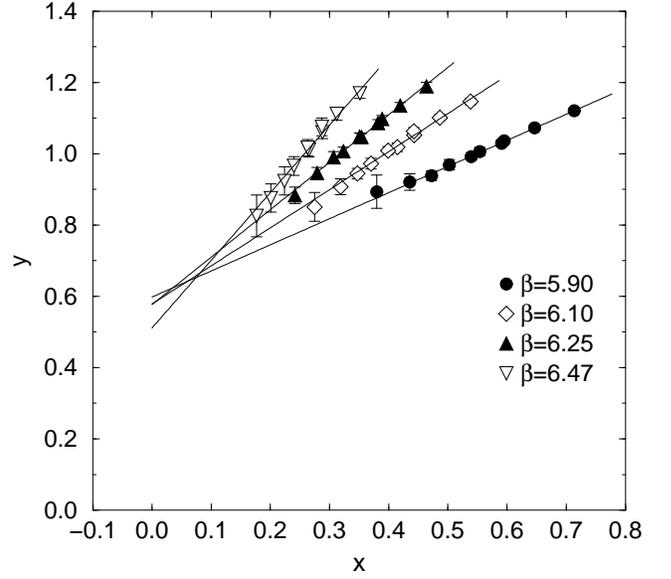}}
\caption{The same as Fig.~\protect\ref{fig:RhoRatioDeg},
but for degenerate octet baryons.}
\label{fig:OctRatioDeg}
\end{figure}

In Fig.~\ref{fig:OctRatioDeg} we show the ratio test as in 
Eq.~(\ref{eq:ratioVB}) for degenerate quark masses.
If $O(m_{PS}^2)$ terms in Eq.~(\ref{eq:ratioVB}), or equivalently 
the $O(m_{PS}^3)$ terms in the mass formula, are negligible, one would 
obtain $C_{1/2}\approx 0.6$ which is opposite in sign compared to 
Eq.~(\ref{eq:ChalfBo}).

\subsubsection{Chiral fit}
 
For octet baryons, it is natural to fit $\Sigma$- and $\Lambda$-like 
baryons simultaneously, because many coefficients of individual terms are 
related with each other. 
In these fits we need to include the $O(m_{PS}^3)$ terms 
to reproduce the negative curvature while maintaining consistency 
with Q$\chi$PT.

We find that the six-parameter fit as indicated by Eqs.~(\ref{eq:qchptS}) and 
(\ref{eq:qchptL}) shows several local minima in parameter space, 
and the minimization procedure converges to different minima 
depending on the jackknife ensemble. 
Some parameters exhibit irregular dependence 
on the lattice spacing, {\it e.g.},
$F= 0.150(20)$, 0.139(25), 0.103(43) and 0.247(49) 
for $\beta=5.90$, 6.10, 6.25, and 6.47.

Recalling that the octet-decuplet coupling $C$ is zero
within $2\sigma$ in the decuplet baryon fit, we set 
$C=0$ for octet baryons and obtain 
stable fits, 
{\it e.g.}, $F=$ 0.334(14), 0.326(14), 0.315(13), 0.299(31) 
for $\beta=$ 5.90, 6.10, 6.25 and 6.47. 

In view of the stability of parameters, we employ the simplified fit 
($C=0$). The fits are almost indistinguishable from 
those keeping $C\ne 0$ in the range of measured points.
See Fig.\ref{fig:chiralB590Deg}.

For the coefficient of the leading chiral singularity, we obtain 
$C_{1/2}= -0.118(4)$ as the weighted average over the four values of $\beta$.
This is smaller than a phenomenological estimate, 
$C_{1/2} = -(3\pi\delta/2) (D-3F)^2 \approx -0.27$
assuming $\delta=0.1$, $F=0.5$ and $D=0.75$.
The smallness of $C_{1/2}$ is not due to lattice cut-off effects,
because $C_{1/2}=-0.09(8)$ is obtained in the continuum limit
(Sec.\ref{sec:methodB}).

\section{Quark mass from local axial-vector current}
\label{appendix:mqlocal}

An alternative definition of the AWI quark mass is given by 
\bea
2 m_q^{AWI,local} a = \frac{Z_A^{local}}{Z_P} 2 m_q^{AWI,local(0)}, 
\label{eq:AWImqloc}
\\
2 m_q^{AWI,local(0)} = m_{PS} a \lim_{t\to\infty}
\frac{\langle A_4^{local}(t)P(0)\rangle}{\langle P(t)P(0)\rangle}, 
\label{eq:AWImqloc0}
\eea
where a time derivative in the numerator is substituted by $m_{PS}a$, and 
\be
A_\mu^{local}(n) = {\bar f}_n i\gamma_5\gamma_\mu g_n
\ee 
is the local axial-vector current. 

For $m_q^{AWI,local(0)}$, we first antisymmetrize the correlator
$\langle A_4^{local}(t)P(0)\rangle$ and then make a fit
\bea
&& m_{PS}a 
\frac{\langle A_4^{local}(t)P(0) \rangle}{\langle P(t) P(0) \rangle}
\approx \nonumber \\
&& 2m_q^{AWI,local(0)} \, \tanh(m_{PS}a(L_t/2-t)),
\label{eq:etaL}
\eea
where $m_{PS}a$ is already determined by the pseudoscalar propagator fit.
The results for $2m_q^{AWI,local(0)}$ are summarized in 
Table~\ref{tab:AWImqlocal}.
Chiral fits are made by a quadratic polynomial under the constraint
$\kappa_c^{AWI,local} = \kappa_c$.
For the renormalization coefficient for $m_q^{AWI,local(0)}$, we use 
Eq.~(\ref{eq:ZAlocal}).

\newpage
\newpage
\onecolumn

\begin{table*}
\caption{Simulation parameters. 
The lattice spacing $a$ is determined from $m_\rho$.
The last column ``acceptance'' is the mean acceptance rate in the 
pseudo-heat-bath update sweeps.}
\label{tab:param1}
\begin{center}
\begin{tabular}{lccccccc}
$\beta$ & $L_s^3 \times L_t$  & $a^{-1}$[GeV] & $a$[fm] & $L_sa$[fm] 
& \#conf. & iter./conf. & acceptance \\
\hline
5.90 & $32^3\,\, 56$  & 1.934(16) & 0.1020(8) & 3.26(3) & 800 &  200 & 0.85 \\
6.10 & $40^3\,\, 70$  & 2.540(22) & 0.0777(7) & 3.10(3) & 600 &  400 & 0.84 \\
6.25 & $48^3\,\, 84$  & 3.071(34) & 0.0642(7) & 3.08(3) & 420 & 1000 & 0.83 \\
6.47 & $64^3 112$     & 3.961(79) & 0.0498(10)& 3.18(6) & 150 & 2000 & 0.82 \\
\end{tabular}
\end{center}
\end{table*}

\begin{table*}[htbp]
\caption{Hopping parameters and corresponding $m_{PS}/m_{V}$ ratios.}
\label{tab:param2}
\begin{center}
\begin{tabular}{cccccc}
\multicolumn{6}{c}{$\beta=5.90$} \\ 
$\kappa$  & 0.15660 & 0.15740 & 0.15830 & 0.15890 & 0.15920 \\
$m_{PS}/m_V$ & 0.752(1) & 0.692(1) & 0.593(1) & 0.491(2) & 0.415(2) \\
\hline
\multicolumn{6}{c}{$\beta=6.10$} \\
$\kappa$  & 0.15280 & 0.15340 & 0.15400 & 0.15440 & 0.15460 \\
$m_{PS}/m_V$ & 0.751(1) & 0.684(1) & 0.581(2) & 0.474(2) & 0.394(3) \\  
\hline
\multicolumn{6}{c}{$\beta=6.25$} \\
$\kappa$  & 0.15075 & 0.15115 & 0.15165 & 0.15200 & 0.15220 \\
$m_{PS}/m_V$ & 0.760(1) & 0.707(2) & 0.609(2) & 0.502(2) & 0.411(3) \\
\hline
\multicolumn{6}{c}{$\beta=6.47$} \\
$\kappa$  & 0.14855 & 0.14885 & 0.14925 & 0.14945 & 0.14960 \\
$m_{PS}/m_V$ & 0.760(2) & 0.709(3) & 0.584(3) & 0.493(4) & 0.391(4) \\
\end{tabular}
\end{center}
\end{table*}

\begin{table*}
\caption{Measured execution time of each step in units of hour.}
\label{tab:simultime}
\begin{center}
\begin{tabular}{lcccc}
$\beta$                  & 5.90  & 6.10 & 6.25 & 6.47 \\
size                     & $32^3\times 56$ & $40^3\times 70$ 
                         & $48^3\times 84$ & $64^3\times 112$ \\ 
\#PU                     & 256 & 512 & 1024 & 2048 \\ 
\hline
configuration generation & 0.28(10\%)  & 0.65(14\%)  & 1.62(24\%)  & 4.5(29\%) \\
gauge fixing             & 0.13(\,5\%) & 0.22(\,5\%) & 0.33(\,5\%) & 1.8(12\%) \\
quark propagator         & 1.57(52\%)  & 2.72(57\%)  & 3.13(46\%)  & 6.6(42\%) \\
hadron propagator        & 1.01(33\%)  & 1.18(24\%)  & 1.67(24\%)  & 2.6(17\%) \\
\hline
total                    & 2.99        & 4.76        & 6.78        & 15.6 \\
\end{tabular}
\end{center}
\end{table*}

\begin{table*}
\caption{Quenched light hadron mass spectrum.
The first error is statistical and the second is systematic.
Deviation from experiment and its statistical significance 
are also given.}
\label{tab:MassFinal}
\begin{center}
\begin{tabular}{lc|lr|lr}
hadron    & experiment & 
\multicolumn{2}{c|}{$m_K$ input} &
\multicolumn{2}{c}{$m_\phi$ input} \\
          & [GeV]  & mass[GeV] & deviation
                   & mass[GeV] & deviation \\
\hline 
$K$       & 0.4977 &           &                       &
                     0.553(10)(08) &   11.2\% 5.6$\sigma$   \\
$K^*$     & 0.8961 & 0.858(09)(08) & $-$4.2\% 4.3$\sigma$  &
                     0.889(03)(06) & $-$0.8\% 2.3$\sigma$   \\
$\phi$    & 1.0194 & 0.957(13)(14) & $-$6.1\% 4.8$\sigma$  & 
                               &                        \\
$N$       & 0.9396 & 0.878(25)(14) & $-$6.6\% 2.5$\sigma$  &
                     0.878(25)(14) & $-$6.6\% 2.5$\sigma$   \\
$\Lambda$ & 1.1157 & 1.019(20)(09) & $-$8.6\% 4.7$\sigma$  &
                     1.060(13)(10) & $-$5.0\% 4.1$\sigma$   \\
$\Sigma$  & 1.1926 & 1.117(19)(11) & $-$6.4\% 4.1$\sigma$  &
                     1.176(11)(20) & $-$1.4\% 1.5$\sigma$   \\
$\Xi$     & 1.3149 & 1.201(17)(13) & $-$8.7\% 6.8$\sigma$  &
                     1.288(08)(09) & $-$2.0\% 3.5$\sigma$   \\
$\Delta$  & 1.2320 & 1.257(35)(10) &    2.0\% 0.7$\sigma$  &
                     1.257(35)(10) &    2.0\% 0.7$\sigma$   \\
$\Sigma^*$& 1.3837 & 1.359(29)(11) & $-$1.8\% 0.9$\sigma$  &
                     1.388(24)(11) &    0.3\% 0.2$\sigma$   \\
$\Xi^*$   & 1.5318 & 1.459(26)(10) & $-$4.7\% 2.8$\sigma$  &
                     1.517(16)(09) & $-$1.0\% 0.9$\sigma$   \\
$\Omega$  & 1.6725 & 1.561(24)(09) & $-$6.7\% 4.7$\sigma$  &
                     1.647(10)(15) & $-$1.5\% 2.6$\sigma$   \\
\end{tabular}
\end{center}
\end{table*}

\begin{table*}
\caption{Decay constants for light pseudoscalar and vector mesons.}
\label{tab:fFinal}
\begin{center}
\begin{tabular}{llcc}
              &                & experiment    & quenched QCD \\
\hline
$f_\pi$ [MeV] &                & $130.7(0.4)$ & $120.0(5.7)$ \\
$f_K$   [MeV] & ($m_K$-input)  & $159.8(1.5)$ & $138.8(4.4)$ \\
              &($m_\phi$-input)&               & $141.5(3.8)$ \\
$f_\pi/f_K-1$ & ($m_K$-input)  & $0.223(2)$    & $0.156(29)$   \\
\hline
$F_\rho$ [MeV]&                & $220(5)$   & $205.7(6.6)$ \\
$F_\phi$ [MeV]&($m_\phi$-input)& $239(3)$   & $229.4(5.7)$ \\
\end{tabular}
\end{center}
\end{table*}

\begin{table*}
\caption{Meson mass data in lattice units at the simulation points.
The fit range $[t_{min},t_{max}]$ and $\chi^2/N_{df}$ are also listed.
Errors of $\chi^2/N_{df}$ are estimated by the jackknife method.}
\label{tab:massmeson}
\begin{center}
\begin{tabular}{l|rrl|rrl}
\multicolumn{7}{c}{$\beta=5.90$} \\
       & \multicolumn{3}{c|}{Pseudo-Scalars} 
       & \multicolumn{3}{c}{Vectors} \\
       & range & $\chi^2/N_{df}$  & mass 
       & range & $\chi^2/N_{df}$  & mass \\
\hline
$s_1s_1$ &   10 28 & 0.55(36) & 0.38255(25)     
         &    7 28 & 1.03(46) & 0.50900( 60) \\ 
$s_2s_2$ &   10 28 & 0.49(34) & 0.33114(26)     
         &    7 26 & 1.11(51) & 0.47862( 71) \\ 
$u_1u_1$ &   10 28 & 0.49(34) & 0.26411(28)     
         &    7 22 & 0.34(32) & 0.44514( 96) \\ 
$u_2u_2$ &   10 28 & 0.53(36) & 0.20827(33)     
         &    7 18 & 0.35(38) & 0.42391(132) \\ 
$u_3u_3$ &   10 28 & 0.71(38) & 0.17145(54)     
         &    7 16 & 0.34(39) & 0.41311(175) \\
\end{tabular}
\end{center}
\end{table*}

\begin{table*}
\caption{The same as Table~\ref{tab:massmeson} for octet baryons.}
\label{tab:massoct}
\begin{center}
\begin{tabular}{l|rrl|rrl}
\multicolumn{7}{c}{$\beta=5.90$} \\
       & \multicolumn{3}{c|}{$\Sigma$-like} 
       & \multicolumn{3}{c}{$\Lambda$-like} \\
       & range & $\chi^2/N_{df}$  & mass 
       & range & $\chi^2/N_{df}$  & mass \\
\hline
$s_1s_1s_1$  &  5 22 & 0.47(35) & 0.7843(12) &&& \\ 
$s_2s_2s_2$  &  5 21 & 0.48(35) & 0.7267(13) &&& \\ 
$u_1u_1u_1$  &  5 18 & 0.41(36) & 0.6572(15) &&& \\ 
$u_2u_2u_2$  &  5 15 & 0.44(44) & 0.6048(18) &&& \\ 
$u_3u_3u_3$  &  5 12 & 0.28(46) & 0.5719(27) &&& \\
\end{tabular}
\end{center}
\end{table*}

\begin{table*}
\caption{The same as Table~\ref{tab:massmeson} for decuplet baryons.}
\label{tab:massdec}
\begin{center}
\begin{tabular}{l|rrl|rrl}
       & \multicolumn{3}{c|}{$\beta=5.90$} 
       & \multicolumn{3}{c}{$\beta=6.10$} \\
       & range & $\chi^2/N_{df}$  & mass 
       & range & $\chi^2/N_{df}$  & mass \\
\hline
$s_1s_1s_1$  &   5 18 & 0.41(37) & 0.8613(17)    
             &   7 23 & 0.83(49) & 0.6483(13) \\ 
$s_2s_2s_2$  &   5 16 & 0.38(39) & 0.8173(18)    
             &   7 21 & 0.71(48) & 0.6096(14) \\ 
$u_1u_1u_1$  &   5 14 & 0.59(55) & 0.7715(22)    
             &   7 18 & 1.12(69) & 0.5746(17) \\ 
$u_2u_2u_2$  &   5 12 & 0.96(81) & 0.7423(25)    
             &   7 15 & 1.12(80) & 0.5520(20) \\ 
$u_3u_3u_3$  &   5 11 & 0.99(92) & 0.7281(28)    
             &   7 14 & 1.11(86) & 0.5408(23) \\ 
$u_1u_1s_1$  &   5 15 & 0.40(43) & 0.8003(20)    
             &   7 19 & 0.80(56) & 0.5983(16) \\
\end{tabular}
\end{center}
\end{table*}

\begin{table*}
\caption{Bare AWI quark masses $2m_q^{AWI(0)}$ obtained 
at the simulation points. 
Here, $2m_q$ is a short cut notation for $m_{q_f} + m_{q_g}$. 
The fit range $[t_{min},t_{max}]$ is also given.}
\label{tab:AWImq}
\begin{center}
\begin{tabular}{l|rl|rl|rl|rl}
\ \ &\multicolumn{2}{c|}{$\beta=5.90$}&\multicolumn{2}{c|}{$\beta=6.10$}
&\multicolumn{2}{c|}{$\beta=6.25$}&\multicolumn{2}{c}{$\beta=6.47$} \\
$\kappa_1\kappa_2$ & range & $2m_q^{AWI(0)}$ & range & 
$2m_q^{AWI(0)}$ & range & $2m_q^{AWI(0)}$ & range & $2m_q^{AWI(0)}$ \\ 
\hline
$s_1s_1$ &  8 25 & 0.090662( 82) & 10 32 & 0.069701( 56) & 10 39 & 0.060678(48)  & 10 53 & 0.046245( 46) \\
$s_2s_2$ &  8 25 & 0.067556( 77) & 10 32 & 0.050123( 53) & 10 39 & 0.046676(47)  & 10 53 & 0.034949( 44) \\
$u_1u_1$ &  8 25 & 0.042373( 72) & 10 32 & 0.030977( 50) & 10 39 & 0.029459(44)  & 10 53 & 0.020087( 41) \\
$u_2u_2$ &  8 25 & 0.025840( 74) & 10 32 & 0.018363( 50) & 10 39 & 0.017590(42)  & 10 53 & 0.012752( 39) \\
$u_3u_3$ &  8 25 & 0.016953(201) & 10 32 & 0.011828( 68) & 10 39 & 0.010777(44)  & 10 53 & 0.007253( 40) \\
$s_1u_1$ &  8 25 & 0.066309( 78) & 10 32 & 0.050203( 54) & 10 39 & 0.044982(47)  & 10 53 & 0.033101( 45) \\
$s_1u_2$ &  8 25 & 0.057951( 78) & 10 32 & 0.043825( 54) & 10 39 & 0.038975(46)  & 10 53 & 0.029392( 44) \\
$s_1u_3$ &  8 25 & 0.053722( 82) & 10 32 & 0.040611( 56) & 10 39 & 0.035554(47)  & 10 53 & 0.026619( 45) \\
$s_2u_1$ &  8 25 & 0.054911( 75) & 10 32 & 0.040518( 52) & 10 39 & 0.038040(46)  & 10 53 & 0.027495( 43) \\
$s_2u_2$ &  8 25 & 0.046611( 75) & 10 32 & 0.034180( 52) & 10 39 & 0.032063(45)  & 10 53 & 0.023802( 43) \\
$s_2u_3$ &  8 25 & 0.042394( 79) & 10 32 & 0.030978( 54) & 10 39 & 0.028654(45)  & 10 53 & 0.021038( 43) \\
\end{tabular}
\end{center}
\end{table*}

\begin{table*}
\caption{Values of $x$, $y$ and $y_f$ used in the ratio tests 
for $m_{PS}^2$ and $f_{PS}$.}
\label{tab:PSratio}
\begin{center}
\begin{tabular}{lrrr}
\multicolumn{4}{c}{$\beta=5.90$} \\
$\kappa_1\kappa_2$ & $x$ & $y$ & $y_f$ \\
$s_1u_1$ & -0.09551( 24) & 0.98835( 38) & 1.00408( 59) \\
$s_1u_2$ & -0.25594( 87) & 0.9749 (12 ) & 1.0122 (17 ) \\
$s_1u_3$ & -0.4480 (59 ) & 0.9558 (62 ) & 1.0313 (80 ) \\
$s_2u_1$ & -0.03613( 10) & 0.99602( 17) & 1.00166( 24) \\
$s_2u_2$ & -0.15161( 60) & 0.98710( 84) & 1.0076 (12 ) \\
$s_2u_3$ & -0.3089 (50 ) & 0.9725 (57 ) & 1.0249 (75 ) \\
\hline
\multicolumn{4}{c}{$\beta=6.10$} \\
$\kappa_1\kappa_2$ & $x$ & $y$ & $y_f$ \\
$s_1u_1$ & -0.10843( 25) & 0.98764( 58) & 1.00369( 82) \\
$s_1u_2$ & -0.28809( 90) & 0.9718 (18 )	& 1.0128 (25 ) \\
$s_1u_3$ & -0.4988 (29 ) & 0.9553 (62 )	& 1.0346 (75 ) \\
$s_2u_1$ & -0.03845( 10) & 0.99589( 24)	& 1.00158( 31) \\
$s_2u_2$ & -0.16528( 61) & 0.9852 (13 )	& 1.0086 (17 ) \\
$s_2u_3$ & -0.3360 (23 ) & 0.9725 (58 )	& 1.0284 (66 ) \\
\hline
\multicolumn{4}{c}{$\beta=6.25$} \\
$\kappa_1\kappa_2$ & $x$ & $y$ & $y_f$ \\
$s_1u_1$ & -0.08628( 19) & 0.98986( 53) & 1.00232( 70) \\
$s_1u_2$ & -0.24925( 69) & 0.9712 (16 ) & 1.0094 (21 ) \\
$s_1u_3$ & -0.4746 (19 ) & 0.9456 (34 ) & 1.0243 (47 ) \\
$s_2u_1$ & -0.03518( 10) & 0.99583( 24) & 1.00116( 30) \\
$s_2u_2$ & -0.15627( 49) & 0.9817 (12 ) & 1.0068 (14 ) \\
$s_2u_3$ & -0.3459 (15 ) & 0.9599 (29 ) & 1.0201 (38 ) \\
\hline
\multicolumn{4}{c}{$\beta=6.47$} \\
$\kappa_1\kappa_2$ & $x$ & $y$ & $y_f$ \\
$s_1u_1$ & -0.11458( 32) & 0.9870 (15 ) & 1.0054 (17 ) \\
$s_1u_2$ & -0.26924( 95) & 0.9699 (37 ) & 1.0145 (43 ) \\
$s_1u_3$ & -0.5417 (27 ) & 0.9402 (81 ) & 1.0336 (99 ) \\
$s_2u_1$ & -0.05086( 16) & 0.99414( 75) & 1.00272( 83) \\
$s_2u_2$ & -0.16660( 67) & 0.9808 (26 ) & 1.0096 (30 ) \\
$s_2u_3$ & -0.3961 (23 ) & 0.9546 (69 ) & 1.0262 (82 ) \\
\end{tabular}
\end{center}
\end{table*}

\begin{table*}
\caption{Values of $\delta$ and $\alpha_\Phi$ from the fit 
(\protect\ref{eq:qchiptxy}).}
\label{tab:delalpha}
\begin{center}
\begin{tabular}{c|cc|cccc}
& \multicolumn{2}{c|}{One parameter fit} 
& \multicolumn{4}{c}{Two parameter fit} \\
& \multicolumn{2}{c|}{$y=1+\delta x$} 
& \multicolumn{4}{c}{$y=1+\delta x+\alpha_X z$} \\
$\beta$ & $\delta$ & $\chi^2/N_{df}$ & $\delta$ & $\alpha_X$ & $\alpha_\Phi$ & $\chi^2/N_{df}$ \\
\hline
5.90 & 0.106( 5) & 7.0 & 0.016(12) & 2.69(29) & 0.59( 6) & 0.61 \\ 
6.10 & 0.103( 6) & 2.2 & 0.042(21) & 1.88(55) & 0.40(12) & 0.12 \\ 
6.25 & 0.117( 7) & 0.1 & 0.112(15) & 0.14(36) & 0.03( 8) & 0.03 \\ 
6.47 & 0.113(13) & 0.0 & 0.113(33) & 0.01(88) & 0.00(19) & 0.02 \\ 
\end{tabular}
\end{center}
\end{table*}

\begin{table*}
\caption{Parameters from Q$\chi$PT chiral fit
(\protect\ref{eq:mps0}) with $\alpha_\Phi=0$
for pseudoscalar meson masses.}
\label{tab:PSQchPTPrm}
\begin{center}
\begin{tabular}{lcccc}
\multicolumn{5}{c}{degenerate fit} \\
& $\beta=5.90$  & $\beta=6.10$  & $\beta=6.25$  & $\beta=6.47$  \\ 
$\kappa_c$ & 0.1598315(68) & 0.1549903(55) & 0.1525355(69) & 0.1498049(79) \\
$A$ [lattice] & 1.1127(39) & 0.8784(42) & 0.7396(67) & 0.5862(85) \\
$A$ [GeV]  & 2.152(19) & 2.231(22) & 2.271(31) & 2.322(59) \\
$\delta$   & 0.1061(45) & 0.0815(68) & 0.0598(96) & 0.079(20)\\
$B$        & 1.012(29) & 1.096(47) & 1.088(55) & 1.42(15) \\
$\Lambda_\chi$ [GeV] & 0.761(7) & 0.789(8) & 0.803(11) & 0.821(21) \\
\hline
\multicolumn{5}{c}{non-degenerate fit with $\kappa_1=s_1$} \\
& $\beta=5.90$  & $\beta=6.10$  & $\beta=6.25$  & $\beta=6.47$  \\ 
$A$ [lattice] & 1.1161(49) & 0.8740(49) & 0.7339(62) & 0.5771(90) \\
$A$ [GeV]  & 2.158(20) & 2.220(23) & 2.254(30) & 2.286(58) \\
$\delta$   & 0.0943(57) & 0.0813(62) & 0.0774(88) & 0.090(24) \\
$B$        & 0.893(51) & 1.121(55) & 1.285(66) & 1.62(21) \\
\hline
\multicolumn{5}{c}{non-degenerate fit with $\kappa_1=s_2$} \\
& $\beta=5.90$  & $\beta=6.10$  & $\beta=6.25$  & $\beta=6.47$  \\ 
$A$ [lattice] & 1.1177(54) & 0.8731(55) & 0.7312(65) & 0.5731(96) \\
$A$ [GeV]  & 2.161(20) & 2.218(23) & 2.246(30) & 2.270(59) \\
$\delta$   & 0.0980(63) & 0.0827(64) & 0.0815(99) & 0.101(25) \\
$B$        & 0.906(66) & 1.143(65) & 1.360(79) & 1.78(24) \\
\end{tabular}
\end{center}
\end{table*}

\begin{table*}
\caption{Values of $\delta$ obtained from ratio tests for $m_{PS}$ and
$f_{PS}$ and Q$\chi$PT chiral fits. DG denotes degenerate fit and 
ND,$s_i$ non-degenerate fits with one of $\kappa$ being fixed to $s_i$.}
\label{tab:delta}
\begin{center}
\begin{tabular}{lccccc}
\multicolumn{2}{l}{} 
& $\beta=5.90$  & $\beta=6.10$  & $\beta=6.25$  & $\beta=6.47$  \\ 
\hline
\multicolumn{2}{l}{ratio} &&&& \\
\multicolumn{2}{l}{$m_{PS}$} 
& 0.106( 5)      & 0.103( 6)      & 0.117( 7)  & 0.113(13)  \\
\multicolumn{2}{l}{$f_{PS}$}
& 0.095(13)      & 0.091(17)      & 0.079(17)  & 0.111(32) \\
\hline
fit & $2A^2/\Lambda_\chi^2$ &&&& \\
DG & 4 & 0.122( 4)      & 0.091( 5)      & 0.065(11)  & 0.088( 9) \\
DG & 8 & 0.114( 5)      & 0.086( 8)      & 0.062(10)  & 0.083(23) \\
DG &16 & 0.106( 5)      & 0.082( 7)      & 0.060(10)  & 0.079(20) \\ 
DG &32 & 0.100( 4)      & 0.077( 6)      & 0.058( 9)  & 0.075(19) \\
ND,$s_1$ & 16 
& 0.094( 6)      & 0.081( 6)      & 0.077( 9)  & 0.090(24) \\
ND,$s_2$ & 16 
& 0.098( 6)      & 0.083( 6)      & 0.082(10)  & 0.101(25) \\
\end{tabular}
\end{center}
\end{table*}

\begin{table*}[htbp]
\caption{Parameters from the Q$\chi$PT fit 
(\protect\ref{eq:qchptVfinal}) for vector meson masses.}
\label{tab:VQchPTPrm}
\begin{center}
\begin{tabular}{lrrrr}
           & $\beta$=5.90 & $\beta$=6.10 & $\beta$=6.25 & $\beta$=6.47 \\ 
$\chi^2/N_{df}$  & 0.40  & 0.07  & 0.08 & 0.72  \\
$m_V^0$ (Lat) & 0.3969(49) & 0.3030(38) & 0.2502(40) & 0.1966(59) \\
$m_V^0$ (GeV)  & 0.7675(32)   & 0.7695(33)   & 0.7684(42)   & 0.7789(85) \\
$C_1$ (Lat) & 0.919(38) & 1.259(54) & 1.496(79) & 2.27(22) \\
$C_1$ (GeV$^{-1}$)  & 0.475(23) & 0.495(25) & 0.487(30) & 0.574(66)  \\
$C_{1/2}$ & $-$0.058(27) & $-$0.075(28) & $-$0.065(35) & $-$0.155(72) \\
\end{tabular}
\end{center}
\end{table*}

\begin{table*}[htbp]
\caption{Parameters from the Q$\chi$PT fit, 
(\protect\ref{eq:qchptSfinal}) and 
(\protect\ref{eq:qchptLfinal})
for octet baryon masses. 
$C_{1/2} = -(3\pi\delta/2) (D-3F)^2$ is the coefficient of 
the $m_{PS}$-linear term in the degenerate mass formula.}
\label{tab:OQchPTPrm}
\begin{center}
\begin{tabular}{lrrrr}
           & $\beta$=5.90 & $\beta$=6.10 & $\beta$=6.25 & $\beta$=6.47 \\ 
$\chi^2/df$        & 0.42         & 0.60         & 0.73         & 0.10 \\
$m_O^0$ (Lat)      & 0.5141(29)   & 0.3774(25)   & 0.3075(20)   & 0.2353(33)\\
$m_O^0$ (GeV)      & 0.9941(84)   & 0.9585(81)   & 0.945(10)    & 0.932(19) \\
$b_F$ (Lat)        & $-$0.534(27) & $-$0.696(36) & $-$0.823(37) &$-$0.97(12)\\
$b_F$ (GeV$^{-1}$) & $-$0.276(14) & $-$0.274(14) & $-$0.268(13) &$-$0.245(30)\\
$b_D$ (Lat)        & 0.034(13)    & 0.052(17)    & 0.071(20)    & 0.159(60) \\
$b_D$ (GeV$^{-1}$) & 0.018(7)     & 0.020(7)     & 0.023(6)     & 0.040(15) \\
$F$                & 0.334(14)    & 0.326(14)    & 0.315(13)    & 0.299(31) \\
$D$                & 0.484(13)    & 0.475(15)    & 0.450(13)    & 0.467(28) \\
$C_{1/2}$          & $-$0.126(15) & $-$0.120(15) & $-$0.116(13) &$-$0.087(30)\\
\end{tabular}
\end{center}
\end{table*}

\begin{table*}[htbp]
\caption{Parameters from the Q$\chi$PT fit (\protect\ref{eq:qchptDfinal}) 
for decuplet baryon masses. 
$C_{1/2} = -(5\pi\delta/6) H^2 $ is the coefficient of 
the $m_{PS}$-linear term in the degenerate mass formula.}
\label{tab:DQchPTPrm}
\begin{center}
\begin{tabular}{lrrrr}
           & $\beta$=5.90 & $\beta$=6.10 & $\beta$=6.25 & $\beta$=6.47 \\ 
$\chi^2/df$     & 0.03         & 0.03         & 0.03         & 0.04    \\
$m_D^0$ (Lat)   & 0.7144(64)   & 0.5301(52)   & 0.4309(48)   & 0.3390(87) \\
$m_D^0$ (GeV)   & 1.381(15)    & 1.346(14)    & 1.323(17)    & 1.343(36)  \\
$c$ (Lat)       & 0.483(16)    & 0.630(27)    & 0.735(31)    & 1.028(110) \\
$c$ (GeV$^{-1}$)& 0.250(9)     & 0.248(11)    & 0.239(11)    & 0.259(29)  \\ 
$H^2$           & 0.65(13)     & 0.55(15)     & 0.39(15)     & 0.70(40)   \\
$C^2$           & $-$0.49(26)  & $-$0.42(30)  & $-$0.75(28)  & 0.88(79)   \\
$C_{1/2}$       & $-$0.169(33) & $-$0.145(39) & $-$0.101(40) & $-$0.18(10)\\
\end{tabular}
\end{center}
\end{table*}

\begin{table*}
\caption{Hadron masses in lattice units from Q$\chi$PT fits.}
\label{tab:MassQchPTLat}
\begin{center}
\begin{tabular}{lcccc}
\multicolumn{5}{c}{$K$ input} \\
&$\beta$=5.90 & $\beta$=6.10 & $\beta$=6.25 & $\beta$=6.47 \\
$\kappa_c$       & 0.1598315( 68)& 0.1549903( 55)& 0.1525355( 69)& 0.1498049( 79) \\
$\kappa_{ud}$    & 0.1597449( 63)& 0.1549274( 47)& 0.1524834( 57)& 0.1497687( 65) \\
$\kappa_s$       & 0.157104 (49 )& 0.153103 (33 )& 0.151056 (33 )& 0.148730 (49 ) \\
$\pi$       & 0.06982( 58)  & 0.05315( 45)  & 0.04396( 48)  & 0.03408( 68)   \\
$\rho$      & 0.3974 (33 )  & 0.3025 (26 )  & 0.2502 (27 )  & 0.1940 (38 )   \\
$K$         & 0.2574 (22 )  & 0.1959 (17 )  & 0.1621 (18 )  & 0.1256 (25 )   \\
\end{tabular}
\end{center}
\end{table*}

\begin{table*}
\caption{Hadron masses in units of GeV from Q$\chi$PT fits.}
\label{tab:MassQchPTPhys}
\begin{center}
\begin{tabular}{lrrrrr}
\multicolumn{6}{c}{$K$ input} \\
&exp.&$\beta$=5.90 & $\beta$=6.10 & $\beta$=6.25 & $\beta$=6.47\\
$a^{-1}$    && 1.934(16) & 2.540(22) & 3.071(34) & 3.961(79)  \\

$K^*$       &0.8961& 0.8561( 29) & 0.8575( 30) & 0.8591( 40) & 0.8516( 71) \\
$\eta_s$    &      & 0.6786( 13) & 0.6860( 14) & 0.6879( 19) & 0.6893( 59) \\
$\phi$      &1.0194& 0.9472( 43) & 0.9512( 44) & 0.9538( 58) & 0.9445(102) \\
$N$         &0.9396& 1.0060( 87) & 0.9712( 83) & 0.9574(106) & 0.9472(194) \\
$\Delta$    &1.2320& 1.3722(128) & 1.3404(116) & 1.3229(142) & 1.3321(278) \\
\end{tabular}
\end{center}
\end{table*}

\begin{table*}
\caption{Coefficients $S$ in units of GeV of $O(a)$ terms in linear continuum
extrapolations of hadron masses. $\chi^2/N_{df}$ of the fits
are also shown.}
\label{tab:MassSlope}
\begin{center}
\begin{tabular}{llrrrrrrrrrrr}
input & \ & $K$ & $K^*$  & $\phi$ & $N$ & $\Lambda$ & $\Sigma$ & $\Xi$ & $\Delta$ & $\Sigma^*$ & $\Xi^*$ & $\Omega$ \\ 
\hline
$m_K$ & $S$             & & $-$0.004 & $-0.018$ & 0.278 & 0.208 & 0.173 & 0.144 & 0.173 & 0.118 & 0.083 & 0.062 \\
\     & $\chi^2/N_{df}$ & & 0.47 & 0.45 & 0.15 & 0.01 & 0.03 & 0.03 & 0.36 & 0.52 & 0.28 & 0.03 \\
\hline
$m_\phi$ & $S$          & 0.067 & 0.007 & & 0.278 & 0.214 & 0.169 & 0.149 & 0.173 & 0.120 & 0.092 & 0.079 \\
\     & $\chi^2/N_{df}$ & 0.35 & 0.49 & & 0.15 & 0.13 & 0.38 & 1.29 & 0.36 & 1.08 & 1.63 & 1.02 \\
\end{tabular}
\end{center}
\end{table*}

\begin{table*}
\caption{Comparison of statistical errors (Stat.), 
estimated $O(a^2)$ effects of continuum extrapolations and 
estimated systematic errors of chiral extrapolations 
(mass difference from Q$\chi$PT and polynomial chiral fits).
Deviations of the experimental value from our final mass prediction
are also given.
Errors or deviations quoted in percent are
relative to the central values of the predicted mass;
those in terms of $\sigma$ are normalized by statistical errors.}
\label{tab:deviations}
\begin{center}
\begin{tabular}{ll|rr|r}
\multicolumn{5}{c}{$K$ input} \\
\hline
& Stat. & $O(a^2)$ error & chiral fit error& deviation\\
$K^*$     &1.04\%& 0.00\% 0.00$\sigma$& 0.67\% 0.65$\sigma$&    4.42\%   4.26$\sigma$  \\
$\phi$    &1.35\%& 0.00\% 0.00$\sigma$& 1.35\% 1.00$\sigma$&    6.54\%   4.83$\sigma$  \\
$N$       &2.84\%& 1.12\% 0.39$\sigma$& 1.05\% 0.37$\sigma$&    7.01\%   2.47$\sigma$  \\
$\Lambda$ &2.00\%& 0.63\% 0.31$\sigma$& 0.24\% 0.12$\sigma$&    9.46\%   4.74$\sigma$  \\
$\Sigma$  &1.67\%& 0.43\% 0.26$\sigma$& $-$0.63\% 0.38$\sigma$& 6.80\%   4.07$\sigma$  \\
$\Xi$     &1.40\%& 0.30\% 0.21$\sigma$& 0.88\% 0.63$\sigma$&    9.49\%   6.78$\sigma$  \\
$\Delta$  &2.80\%& 0.43\% 0.15$\sigma$& $-$0.38\% 0.14$\sigma$& $-$2.00\%  0.72$\sigma$  \\
$\Sigma^*$&2.10\%& 0.20\% 0.10$\sigma$& $-$0.54\% 0.26$\sigma$& 1.84\%   0.88$\sigma$  \\
$\Xi^*$   &1.76\%& 0.10\% 0.06$\sigma$& 0.29\% 0.16$\sigma$&    4.97\%   2.83$\sigma$  \\
$\Omega$  &1.54\%& 0.06\% 0.04$\sigma$& $-$0.07\% 0.04$\sigma$& 7.17\%   4.65$\sigma$  \\
\hline
\\
\multicolumn{5}{c}{$\phi$ input} \\
\hline
& Stat. & $O(a^2)$ error & chiral fit error& deviation\\
$K$       &1.78\%& 0.06\% 0.04$\sigma$& $-$1.34\% 0.75$\sigma$&  $-$10.06\%  5.64$\sigma$  \\
$K^*$     &0.37\%& 0.00\% 0.00$\sigma$&  0.23\% 0.63$\sigma$&      0.84\%   2.27$\sigma$  \\
$\Lambda$ &1.27\%& 0.66\% 0.52$\sigma$& $-$0.21\% 0.17$\sigma$&  5.21\%   4.11$\sigma$  \\
$\Sigma$  &0.96\%& 0.41\% 0.43$\sigma$& $-$1.54\% 1.61$\sigma$&  1.41\%   1.48$\sigma$  \\
$\Xi$     &0.58\%& 0.32\% 0.55$\sigma$& $-$0.23\% 0.40$\sigma$&  2.06\%   3.52$\sigma$  \\
$\Sigma^*$&1.72\%& 0.21\% 0.12$\sigma$& $-$0.53\% 0.31$\sigma$&  $-$0.28\%  0.17$\sigma$  \\
$\Xi^*$   &1.05\%& 0.12\% 0.12$\sigma$&  0.09\% 0.08$\sigma$&      0.99\%   0.94$\sigma$  \\
$\Omega$  &0.60\%& 0.09\% 0.15$\sigma$& $-$0.64\% 1.07$\sigma$&  1.53\%   2.56$\sigma$  \\
\end{tabular}
\end{center}
\end{table*}

\begin{table*}
\caption{Hadron masses in units of GeV from polynomial fits.}
\label{tab:MassPolyPhys}
\begin{center}
\begin{tabular}{lrrrrr}
\multicolumn{6}{c}{$K$ input} \\
&exp.&$\beta$=5.90 & $\beta$=6.10 & $\beta$=6.25 & $\beta$=6.47\\
$a^{-1}$ &   & 1.970( 8) & 2.584(13) & 3.109(20) & 4.102(60) \\
$K^*$       &0.8961& 0.8726( 20) & 0.8722( 20) & 0.8678( 33) & 0.8672( 73) \\
$\eta_s$    &      & 0.6906( 12) & 0.6930( 12) & 0.6915( 17) & 0.6908( 59) \\
$\phi$      &1.0194& 0.9659( 19) & 0.9672( 22) & 0.9676( 25) & 0.9657( 65) \\
$N$         &0.9396& 1.0139( 64) & 0.9833( 73) & 0.9640( 97) & 0.9576(196) \\
$\Delta$    &1.2320& 1.3968( 76) & 1.3621( 83) & 1.3365( 92) & 1.3516(241) \\
\end{tabular}
\end{center}
\end{table*}

\begin{table*}
\caption{Hadron spectrum from linear continuum extrapolations 
of masses determined by polynomial chiral fits.}
\label{tab:MassPoly}
\begin{center}
\begin{tabular}{l|lr|lr}
\         & 
\multicolumn{2}{c|}{$m_K$ input} &
\multicolumn{2}{c}{$m_\phi$ input} \\
\         & mass(GeV) & $\chi^2/N_{df}$ 
          & mass(GeV) & $\chi^2/N_{df}$ \\
\hline 
$K$       &          &      &   
           0.546(06) & 0.07 \\  
$K^*$     &0.864(07) & 0.42 &   
           0.891(05) & 0.70 \\  
$\phi$    &0.970(06) & 0.08 &   
                     &      \\  
$N$       &0.887(22) & 0.18 &   
           0.887(22) & 0.18 \\  
$\Lambda$ &1.022(16) & 0.60 &   
           1.058(14) & 0.80 \\  
$\Sigma$  &1.110(14) & 0.62 &   
           1.158(12) & 0.74 \\  
$\Xi$     &1.212(13) & 0.89 &   
           1.285(11) & 1.73 \\  
$\Delta$  &1.252(24) & 1.07 &   
           1.252(24) & 1.07 \\  
$\Sigma^*$&1.351(20) & 1.14 &   
           1.380(17) & 1.34 \\  
$\Xi^*$   &1.463(18) & 1.05 &   
           1.518(14) & 1.71 \\  
$\Omega$  &1.560(17) & 0.43 &   
           1.637(11) & 1.14 \\  
\end{tabular}
\end{center}
\end{table*}

\begin{table*}
\caption{Values of $\kappa_c^{AWI}$ from the quark mass based on 
the axial-vector Ward identity and $\kappa_c$ from various chiral 
extrapolations of pseudoscalar meson masses.
The second column shows fit number in Table \protect\ref{tab:chi2extPS}.}
\label{tab:Kc}
\begin{center}
\begin{tabular}{llcccc}
&& $\beta=5.90$ & $\beta=6.10$ & $\beta=6.25$ & $\beta=6.47$ \\ 
$\kappa_c^{AWI}$(linear) & & 
0.1598254(38) & 0.1549765(21) & 0.1525172(13) & 0.1497968(11) \\
$\kappa_c^{AWI}$(quadratic) & & 
0.1598490(93) & 0.1549821(23) & 0.1525226(15) & 0.1498001(12) \\
$\kappa_c$(Q$\chi$PT)     & Fit 5 & 
0.1598315(68) & 0.1549903(55) & 0.1525355(69) & 0.1498049(79) \\
$\kappa_c({\rm quadratic})$ & Fit 2 &
0.1599531(73) & 0.1550430(50) & 0.1525650(53) & 0.1498244(78) \\
$\kappa_c({\rm cubic})$     & Fit 3 &
0.1598993(63) & 0.1550251(45) & 0.1525579(67) & 0.1498210(76) \\
\end{tabular}
\end{center}
\end{table*}

\begin{table*}
\caption{Parameters of chiral fits to quark masses based on 
the axial-vector Ward identity.}
\label{tab:mqextrap}
\begin{center}
\begin{tabular}{lrrrr}
&$\beta$=5.90 & $\beta$=6.10 & $\beta$=6.25 & $\beta$=6.47 \\
$A_0^{dg}$  &$-$4.3885(39)&$-$4.8871(38)&$-$5.1724(35)&$-$5.5069(47)\\
$A_1^{dg}$  &0.70142(62)&0.75739(59)&0.78888(53)&0.82492(71)\\
$\kappa_c^{AWI}$ & 0.1598299(31) & 0.1549768(18) & 0.1525160(13) & 0.1497959(11) \\
$B_1^{dg}$ &  0.6986(49) & 0.7386(70) &  0.753(12) &  0.799(21) \\
$B_2^{dg}$ &  0.022(28) &  0.163(58) &  0.37(12) &  0.39(30) \\
\hline
$A_0^{s_1}$ &$-$2.1705(16)&$-$2.4283(17)&$-$2.5771(15)&$-$2.7510(21)\\
$A_1^{s_1}$ &0.35410(26)&0.38170(26)&0.39765(23)&0.41553(31)\\
\hline
$A_0^{s_2}$ &$-$2.1554(22)&$-$2.4146(22)&$-$2.5654(19)&$-$2.7399(27)\\
$A_1^{s_2}$ &0.34989(35)&0.37809(34)&0.39481(28)&0.41304(40)\\
\end{tabular}
\end{center}
\end{table*}

\begin{table*}
\caption{Light quark masses in lattice units.}
\label{tab:MQLat}
\begin{center}
\begin{tabular}{lrrrr}
&$\beta$=5.90 & $\beta$=6.10 & $\beta$=6.25 & $\beta$=6.47 \\
$m_{ud}^{VWI}$    &0.001696(37)&0.001311(35)&0.001120(42)&0.000808(64)\\
$m_{ud}^{AWI}$    &0.001185(23) & 0.000969(21) & 0.000845(25) & 0.000646(42) \\
$m_s^{VWI}(K)$    & 0.0543(10) & 0.0398( 7) & 0.0321( 8) & 0.0241(12) \\
$m_s^{AWI}(K)$    & 0.0379( 7) & 0.0296( 5) & 0.0246( 6) & 0.0195( 9) \\
$m_s^{VWI}(\phi)$ & 0.0736(21) & 0.0523(15) & 0.0417(15) & 0.0318(20) \\
$m_s^{AWI}(\phi)$ & 0.0512(15) & 0.0389(11) & 0.0321(12) & 0.0258(16) \\
\end{tabular}
\end{center}
\end{table*}

\begin{table*}
\caption{Light quark masses in units of GeV in the ${\overline {\rm MS}}$
scheme at $\mu=2$ GeV.}
\label{tab:MQPhys}
\begin{center}
\begin{tabular}{lrrrr}
&$\beta$=5.90 & $\beta$=6.10 & $\beta$=6.25 & $\beta$=6.47 \\
$m_{ud}^{VWI}$    &0.00418( 6)& 0.00431( 9)& 0.00450(15)& 0.00424(28) \\
$m_{ud}^{AWI}$    &0.00302( 4)&0.00327( 5)&0.00348( 8)&0.00347(18)\\
$m_s^{VWI}(K)$    & 0.1338(13)&  0.1306(13)&  0.1289(17)&  0.1267(37) \\
$m_s^{AWI}(K)$    & 0.0965( 9)&0.1000( 8)&0.1014(11)&0.1049(26)\\
$m_s^{VWI}(\phi)$ & 0.1814(35)&  0.1716(32)&  0.1673(41)&  0.1669(67) \\
$m_s^{AWI}(\phi)$ & 0.1304(25)&0.1315(25)&0.1322(34)&0.1385(54)\\
\end{tabular}
\end{center}
\end{table*}

\begin{table*}
\caption{Parameters of continuum extrapolations of quark masses.
Parameters are defined by $m_q^{VWI} = m_0 + A^{VWI} a$ and
$m_q^{AWI} = m_0 + A^{AWI} a$.}
\label{tab:MQCont}
\begin{center}
\begin{tabular}{lrrrr}
&$m_0$ (MeV) & $A^{VWI}$ & $A^{AWI}$ & $\chi^2/N_{df}$ \\
$m_{ud}$ & 4.29(14) &  $-$0.12(31) &  $-$2.49(30) & 1.13 \\
$m_s(K)$ & 113.8(2.3) &  41.2(5.5) & $-$34.6(5.3) & 0.81 \\
$m_s(\phi)$ & 142.3(5.8) &  76(15) & $-$25(14) & 0.31 \\
\end{tabular}
\end{center}
\end{table*}

\begin{table*}
\caption{Plaquette values}
\label{tab:plaquette}
\begin{center}
\begin{tabular}{rrrr}
$\beta=5.90$  &  $\beta=6.10$  & $\beta=6.25$  & $\beta=6.47$ \\
0.58184394(109) & 0.60413086( 71) & 0.61810507( 55) &  0.63608939( 47) \\
\end{tabular}
\end{center}
\end{table*}

\begin{table*}
\caption{$\Lambda_{\overline{\rm MS}}$ in units of MeV.}
\label{tab:Lambda}
\begin{center}
\begin{tabular}{lrrrrr}
method & $\beta=5.90$  &  $\beta=6.10$  & $\beta=6.25$  & $\beta=6.47$  & $\beta=\infty$ \\
$\alpha_{MF}$ &  233.1(1.9) &  231.4(2.0) &  229.4(2.5) &  223.2(4.4) &  220.4(5.7) \\
$\alpha_V$    &  219.1(1.8) &  220.4(1.9) &  220.2(2.4) &  216.1(4.3) &  219.5(5.4) \\
$\alpha_P$ w/o BLM & 258.3(2.2) &  249.8(2.1) & 244.9(2.7) & 235.5(4.7) & 219.3(6.2) \\
$\alpha_p$ w/  BLM & 254.6(2.1) & 246.7(2.1) & 242.1(2.6) & 233.1(4.6) & 218.0(6.1) \\
$\alpha_p \to \Lambda_P$ &  258.2(2.2) & 249.8(2.1) & 244.9(2.7) & 235.5(4.7) & 219.2(6.2) \\
\end{tabular}
\end{center}
\end{table*}

\begin{table*}
\caption{Pseudoscalar meson decay constant, 
$Z_\kappa f_{PS}^{(0)}$, where $Z_\kappa$ 
is the $\kappa$-dependent part of the renormalization constant 
defined in Eq.(\protect\ref{eq:ZK}). 
Ranges for fits in Eq.(\protect\ref{eq:PPPS}) are also given.}
\label{tab:fp}
\begin{center}
\begin{tabular}{l|rl|rl}
       & \multicolumn{2}{c|}{$\beta=5.90$} 
       & \multicolumn{2}{c}{$\beta=6.10$} \\
       & range  & value 
       & range  & value \\
\hline
$s_1s_1$ & 12 28 & 0.10128(51)    
         & 12 35 & 0.07412(44) \\ 
$s_2s_2$ & 12 28 & 0.09379(53)    
         & 12 35 & 0.06834(47) \\ 
$u_1u_1$ & 12 28 & 0.08498(55)    
         & 12 35 & 0.06232(53) \\ 
$u_2u_2$ & 12 28 & 0.07875(59)    
         & 12 35 & 0.05816(61) \\ 
$u_3u_3$ & 12 28 & 0.07469(87)    
         & 12 35 & 0.05553(76) \\ 
$s_1u_1$ & 12 28 & 0.09297(53)    
         & 12 35 & 0.06809(48) \\ 
$s_1u_2$ & 12 28 & 0.08985(55)    
         & 12 35 & 0.06608(51) \\ 
$s_1u_3$ & 12 28 & 0.08833(56)    
         & 12 35 & 0.06526(55) \\ 
$s_2u_1$ & 12 28 & 0.08935(54)    
         & 12 35 & 0.06532(50) \\ 
$s_2u_2$ & 12 28 & 0.08627(55)    
         & 12 35 & 0.06332(53) \\ 
$s_2u_3$ & 12 28 & 0.08473(57)    
         & 12 35 & 0.06247(57) \\ 
\hline
\hline
       & \multicolumn{2}{c|}{$\beta=6.25$} 
       & \multicolumn{2}{c}{$\beta=6.47$} \\
       & range & value 
       & range & value \\
\hline
$s_1s_1$ & 17 42 & 0.05987(45)    
         & 20 56 & 0.04605(67) \\ 
$s_2s_2$ & 17 42 & 0.05606(47)    
         & 20 56 & 0.04324(72) \\ 
$u_1u_1$ & 17 42 & 0.05106(49)    
         & 20 56 & 0.03914(80) \\ 
$u_2u_2$ & 17 42 & 0.04734(53)    
         & 20 56 & 0.03690(84) \\ 
$u_3u_3$ & 17 42 & 0.04493(57)    
         & 20 56 & 0.03508(89) \\ 
$s_1u_1$ & 17 42 & 0.05536(47)    
         & 20 56 & 0.04257(73) \\ 
$s_1u_2$ & 17 42 & 0.05350(49)    
         & 20 56 & 0.04152(76) \\ 
$s_1u_3$ & 17 42 & 0.05251(50)    
         & 20 56 & 0.04086(80) \\ 
$s_2u_1$ & 17 42 & 0.05353(48)    
         & 20 56 & 0.04119(76) \\ 
$s_2u_2$ & 17 42 & 0.05170(50)    
         & 20 56 & 0.04014(78) \\ 
$s_2u_3$ & 17 42 & 0.05070(51)    
         & 20 56 & 0.03945(81) \\ 
\end{tabular}
\end{center}
\end{table*}

\begin{table*}
\caption{Pseudoscalar meson decay constant $Z_\kappa f_{PS}^{(0)}$.}
\label{tab:fPSLat}
\begin{center}
\begin{tabular}{llcccc}
&&$\beta$=5.90 & $\beta$=6.10 & $\beta$=6.25 & $\beta$=6.47 \\
\hline
$f_\pi$ &  &0.06882(92)& 0.05201(90)& 0.04172(67)& 0.03304(101) \\
$f_K$ & ($m_K$-input)  & 0.08343(61)& 0.06216(59)& 0.04970(55)& 0.03913(85) \\
      &($m_\phi$-input)& 0.08781(74)& 0.06486(62)& 0.05170(60)& 0.04074(90) \\
\end{tabular}
\end{center}
\end{table*}

\begin{table*}
\caption{Pseudoscalar meson decay constants in units of GeV.}
\label{tab:fPSPhys}
\begin{center}
\begin{tabular}{llcccc}
        &       & $\beta$=5.90 & $\beta$=6.10 & $\beta$=6.25 & $\beta$=6.47 \\
\hline
$f_\pi$ &                & 0.1220(20)& 0.1227(24)& 0.1198(23)& 0.1233(44) \\
$f_K$   & ($m_K$-input)  & 0.1479(14)& 0.1466(16)& 0.1427(19)& 0.1461(38) \\
        &($m_\phi$-input)& 0.1557(11)& 0.1530(14)& 0.1485(17)& 0.1521(34) \\
\end{tabular}
\end{center}
\end{table*}

\begin{table*}
\caption{Renormalization constant $Z_V$ determined by a non-perturbative 
method. Ranges of the fit shown at the bottom are common to all combinations of 
quark masses.}
\label{tab:ZV}
\begin{center}
\begin{tabular}{l|cccc}
         & $\beta=5.90$ & $\beta=6.10$ & $\beta=6.25$ & $\beta=6.47$  \\
\hline
$s_1s_1$ &0.53589( 20)& 0.59909( 20) & 0.63341( 23) & 0.67379( 23)    \\ 
$s_2s_2$ &0.53514( 24)& 0.59867( 27) & 0.63326( 31) & 0.67403( 30)    \\ 
$u_1u_1$ &0.53386( 34)& 0.59761( 42) & 0.63236( 54) & 0.67406( 49)    \\ 
$u_2u_2$ &0.53240( 52)& 0.59592( 70) & 0.63005(103) & 0.67367( 73)    \\ 
$u_3u_3$ &0.53098( 77)& 0.59415(111) & 0.62637(194) & 0.67272(126)    \\
$s_1u_1$ &0.53480( 25)& 0.59838( 28) & 0.63303( 33) & 0.67384( 33)    \\ 
$s_1u_2$ &0.53400( 29)& 0.59771( 34) & 0.63244( 43) & 0.67353( 39)    \\ 
$s_1u_3$ &0.53340( 33)& 0.59720( 41) & 0.63173( 54) & 0.67294( 48)    \\ 
$s_2u_1$ &0.53447( 28)& 0.59816( 33) & 0.63288( 40) & 0.67400( 38)    \\ 
$s_2u_2$ &0.53368( 33)& 0.59744( 41) & 0.63218( 52) & 0.67371( 45)    \\ 
$s_2u_3$ &0.53309( 38)& 0.59688( 49) & 0.63134( 66) & 0.67312( 55)    \\ 
\hline
range & 9 17 & 14 22 & 20 35 & 20 40 \\
\end{tabular}
\end{center}
\end{table*}

\begin{table*}
\caption{Renormalized vector meson decay constant $F_V a$.}
\label{tab:FVR}
\begin{center}
\begin{tabular}{l|cccc}
          & $\beta=5.90$ & $\beta=6.10$ & $\beta=6.25$ & $\beta=6.47$ \\
\hline
$s_1s_1$ &0.12390( 56)& 0.09476( 48)& 0.07647( 52)& 0.05843( 67)    \\ 
$s_2s_2$ &0.12211( 63)& 0.09277( 55)& 0.07496( 58)& 0.05626( 83)    \\ 
$u_1u_1$ &0.11987( 76)& 0.09037( 69)& 0.07330( 61)& 0.05606( 87)    \\ 
$u_2u_2$ &0.11864( 79)& 0.08960( 69)& 0.07211( 79)& 0.05571(120)    \\
$u_3u_3$ &0.11795( 99)& 0.08912( 88)& 0.07183( 90)& 0.05556(118)    \\
$s_1u_1$ &0.12147( 62)& 0.09233( 55)& 0.07441( 61)& 0.05695( 73)    \\ 
$s_1u_2$ &0.12054( 69)& 0.09136( 62)& 0.07397( 56)& 0.05664( 81)    \\ 
$s_1u_3$ &0.12025( 75)& 0.09106( 69)& 0.07362( 63)& 0.05675( 96)    \\ 
$s_2u_1$ &0.12081( 67)& 0.09151( 60)& 0.07380( 66)& 0.05662( 78)    \\ 
$s_2u_2$ &0.12005( 75)& 0.09061( 69)& 0.07345( 61)& 0.05642( 88)    \\ 
$s_2u_3$ &0.11978( 83)& 0.09029( 77)& 0.07317( 68)& 0.05651(105)    \\
\end{tabular}
\end{center}
\end{table*}

\begin{table*}
\caption{Vector meson decay constant $F_V$ at physical points 
in units of GeV.} 
\label{tab:FVPhys}
\begin{center}
\begin{tabular}{lcccc}
& $\beta=5.90$ & $\beta=6.10$ & $\beta=6.25$ & $\beta=6.47$ \\
\hline
$F_\rho$ & 0.2256(20) & 0.2227(21) & 0.2166(29) & 0.2163(60) \\
$F_\phi$ & 0.2416(19) & 0.2430(19) & 0.2360(24) & 0.2321(47) \\
\end{tabular}
\end{center}
\end{table*}

\begin{table*}
\caption{Pseudoscalar meson masses interpolated or slightly extrapolated to
$m_{PS}/m_V=0.75 (s_1)$, 0.7($s_2$), 0.6($u_1$), 0.5($u_2$) and 0.4($u_3$) and 
normalized by $m_V^{(0.75)}a$. Parameters and $\chi^2/df$ of their linear
continuum extrapolations $ma = m_0a + c\cdot m_V^{(0.75)}a$ are also given.}
\label{tab:PS075}
\begin{center}
\begin{tabular}{lrrrrrrr}
$\kappa_1\kappa_2$ & $\beta=5.90$ & $\beta=6.10$ & $\beta=6.25$ & $\beta=6.47$ & $m_0a$ & $c$ & $\chi^2/df$ \\
\hline
$s_1s_1$ &  0.7500( 0) &  0.7500( 0) &  0.7500( 0) &  0.7500( 0) &  0.7500( 0) &  0.000( 0) & 0.00 \\
$s_2s_2$ &  0.6635( 5) &  0.6623( 5) &  0.6624( 6) &  0.6606(16) &  0.6595(15) &  0.008( 4) & 0.38 \\
$u_1u_1$ &  0.5282( 8) &  0.5266( 9) &  0.5265(12) &  0.5253(25) &  0.5229(27) &  0.010( 6) & 0.09 \\
$u_2u_2$ &  0.4192(11) &  0.4179(12) &  0.4176(16) &  0.4179(31) &  0.4153(36) &  0.007( 9) & 0.07 \\
$u_3u_3$ &  0.3243(13) &  0.3235(14) &  0.3231(18) &  0.3247(35) &  0.3223(41) &  0.004(10) & 0.14 \\
$s_1u_1$ &  0.6482( 6) &  0.6460( 6) &  0.6462( 8) &  0.6439(29) &  0.6415(20) &  0.013( 5) & 0.66 \\
$s_1u_2$ &  0.6074( 7) &  0.6047( 7) &  0.6046( 9) &  0.6029(31) &  0.5989(22) &  0.016( 5) & 0.51 \\
$s_1u_3$ &  0.5790( 7) &  0.5757( 7) &  0.5752( 9) &  0.5740(33) &  0.5677(24) &  0.022( 6) & 0.43 \\
$s_2u_1$ &  0.5998( 8) &  0.5971( 8) &  0.5974(10) &  0.5941(34) &  0.5915(26) &  0.016( 6) & 0.65 \\
$s_2u_2$ &  0.5556( 8) &  0.5525( 8) &  0.5527(11) &  0.5500(35) &  0.5460(28) &  0.018( 7) & 0.58 \\
$s_2u_3$ &  0.5244( 9) &  0.5207( 8) &  0.5206(11) &  0.5185(36) &  0.5124(28) &  0.023( 7) & 0.59 \\
\end{tabular}
\end{center}
\end{table*}

\begin{table*}
\caption{The same as table \protect\ref{tab:PS075} for vector mesons.
Values of $m_V^{(0.75)}a$ at each $\beta$ are also shown.}
\label{tab:V075}
\begin{center}
\begin{tabular}{lrrrrrrr}
$\kappa_1\kappa_2$ & $\beta=5.90$ & $\beta=6.10$ & $\beta=6.25$ & $\beta=6.47$ & $m_0a$ & $c$ & $\chi^2/df$ \\
\hline
$s_1s_1$ &  1.0000( 0) &  1.0000( 0) &  1.0000( 0) &  1.0000( 0) &  1.0000( 0) &  0.000( 0) & 0.00 \\
$s_2s_2$ &  0.9478( 6) &  0.9462( 7) &  0.9463( 9) &  0.9437(22) &  0.9422(22) &  0.011( 5) & 0.38 \\
$u_1u_1$ &  0.8803(14) &  0.8777(15) &  0.8776(20) &  0.8755(42) &  0.8716(46) &  0.017(11) & 0.09 \\
$u_2u_2$ &  0.8383(23) &  0.8358(24) &  0.8353(31) &  0.8358(62) &  0.8306(72) &  0.015(17) & 0.07 \\
$u_3u_3$ &  0.8108(33) &  0.8087(34) &  0.8078(45) &  0.8118(87) &  0.8056(103) &  0.009(25) & 0.14 \\
$s_1u_1$ &  0.9400( 7) &  0.9386( 7) &  0.9386( 9) &  0.9372(21) &  0.9354(22) &  0.009( 5) & 0.13 \\
$s_1u_2$ &  0.9187(10) &  0.9173(10) &  0.9171(14) &  0.9167(28) &  0.9142(32) &  0.009( 7) & 0.05 \\
$s_1u_3$ &  0.9046(14) &  0.9033(14) &  0.9030(19) &  0.9037(36) &  0.9008(42) &  0.007(10) & 0.08 \\
$s_2u_1$ &  0.9140(10) &  0.9118(11) &  0.9118(14) &  0.9094(31) &  0.9067(32) &  0.014( 8) & 0.17 \\
$s_2u_2$ &  0.8928(13) &  0.8906(14) &  0.8905(18) &  0.8891(38) &  0.8856(42) &  0.014(10) & 0.07 \\
$s_2u_3$ &  0.8788(16) &  0.8767(17) &  0.8764(23) &  0.8763(45) &  0.8723(52) &  0.012(12) & 0.06 \\
\hline
$m_V^{(0.75)}a$ & 0.5086(12) &  0.3874(11) &  0.3212(10) &  0.2447(14) & & & \\  
\end{tabular}
\end{center}
\end{table*}

\begin{table*}
\caption{The same as table \protect\ref{tab:PS075} for $\Sigma$-like 
octet baryons.} 
\label{tab:O075}
\begin{center}
\begin{tabular}{lrrrrrrr}
$\kappa_1\kappa_2$ & $\beta=5.90$ & $\beta=6.10$ & $\beta=6.25$ & $\beta=6.47$ & $m_0a$ & $c$ & $\chi^2/df$ \\
\hline
$s_1s_1s_1$ &  1.5385(23) &  1.5175(23) &  1.5076(26) &  1.4983(45) & 1.4568(62) &  0.159(15) & 0.31 \\
$s_2s_2s_2$ &  1.4424(24) &  1.4160(26) &  1.4032(30) &  1.3938(55) & 1.3398(70) &  0.200(17) & 0.66 \\
$u_1u_1u_1$ &  1.2988(30) &  1.2672(33) &  1.2517(39) &  1.2471(78) & 1.1788(92) &  0.234(22) & 1.44 \\
$u_2u_2u_2$ &  1.1955(39) &  1.1618(43) &  1.1452(48) &  1.1458(102) &1.0701(117) &  0.243(28) & 1.73 \\
$u_3u_3u_3$ &  1.1195(49) &  1.0848(53) &  1.0679(56) &  1.0722(123) &0.9928(141) &  0.245(34) & 1.71 \\
$u_1u_1s_1$ &  1.3970(26) &  1.3704(29) &  1.3583(33) &  1.3521(62) & 1.2973(77) &  0.194(18) & 1.08 \\
$u_2u_2s_1$ &  1.3430(28) &  1.3151(33) &  1.3023(37) &  1.2991(71) & 1.2400(86) &  0.200(21) & 1.53 \\
$u_3u_3s_1$ &  1.3067(31) &  1.2782(37) &  1.2648(40) &  1.2638(77) & 1.2023(94) &  0.202(22) & 1.75 \\
$s_1s_1u_1$ &  1.4562(24) &  1.4320(26) &  1.4168(29) &  1.4112(52) & 1.3574(68) &  0.193(16) & 1.21 \\
$s_1s_1u_2$ &  1.4221(24) &  1.3971(27) &  1.3806(30) &  1.3766(55) & 1.3196(71) &  0.200(17) & 1.70 \\
$s_1s_1u_3$ &  1.3969(25) &  1.3717(27) &  1.3545(31) &  1.3515(57) & 1.2931(73) &  0.203(18) & 1.97 \\
$u_1u_1s_2$ &  1.3564(27) &  1.3276(31) &  1.3137(35) &  1.3080(68) & 1.2471(83) &  0.213(20) & 1.22 \\
$u_2u_2s_2$ &  1.2999(31) &  1.2697(35) &  1.2552(39) &  1.2524(78) & 1.1874(93) &  0.219(22) & 1.60 \\
$u_3u_3s_2$ &  1.2618(34) &  1.2308(39) &  1.2160(43) &  1.2153(85) & 1.1477(102) &  0.221(24) & 1.78 \\
$s_2s_2u_1$ &  1.3900(25) &  1.3622(28) &  1.3469(33) &  1.3404(62) & 1.2804(76) &  0.214(18) & 1.13 \\
$s_2s_2u_2$ &  1.3533(27) &  1.3247(29) &  1.3081(34) &  1.3033(65) & 1.2399(80) &  0.221(19) & 1.46 \\
$s_2s_2u_3$ &  1.3263(28) &  1.2975(30) &  1.2803(35) &  1.2764(68) & 1.2115(83) &  0.224(20) & 1.62 \\
\end{tabular}
\end{center}
\end{table*}

\begin{table*}
\caption{The same as table \protect\ref{tab:PS075} for $\Lambda$-like 
octet baryons.} 
\label{tab:L075}
\begin{center}
\begin{tabular}{lrrrrrrr}
$\kappa_1\kappa_2$ & $\beta=5.90$ & $\beta=6.10$ & $\beta=6.25$ & $\beta=6.47$ & $m_0a$ & $c$ & $\chi^2/df$ \\
\hline
$u_1u_1s_1$ &  1.3732(26) &  1.3451(28) &  1.3286(33) &  1.3229(62) & 1.2609(77) &  0.219(18) & 1.31 \\
$u_2u_2s_1$ &  1.3014(29) &  1.2718(31) &  1.2541(36) &  1.2509(72) & 1.1836(87) &  0.230(21) & 1.68 \\
$u_3u_3s_1$ &  1.2476(33) &  1.2174(34) &  1.1993(39) &  1.1978(80) & 1.1281(96) &  0.233(23) & 1.75 \\
$s_1s_1u_1$ &  1.4737(24) &  1.4504(25) &  1.4408(29) &  1.4313(53) & 1.3859(68) &  0.171(17) & 0.53 \\
$s_1s_1u_2$ &  1.4510(25) &  1.4270(27) &  1.4175(30) &  1.4085(56) & 1.3618(71) &  0.174(17) & 0.64 \\
$s_1s_1u_3$ &  1.4367(25) &  1.4123(29) &  1.4028(31) &  1.3941(59) & 1.3467(73) &  0.175(18) & 0.66 \\
$u_1u_1s_2$ &  1.3410(27) &  1.3113(30) &  1.2953(35) &  1.2897(68) & 1.2250(83) &  0.226(20) & 1.30 \\
$u_2u_2s_2$ &  1.2677(31) &  1.2365(34) &  1.2192(39) &  1.2164(80) & 1.1463(94) &  0.236(23) & 1.63 \\
$u_3u_3s_2$ &  1.2129(36) &  1.1811(38) &  1.1634(43) &  1.1624(90) & 1.0898(105) &  0.239(25) & 1.66 \\
$s_2s_2u_1$ &  1.4032(25) &  1.3759(28) &  1.3632(32) &  1.3550(61) & 1.2990(75) &  0.203(18) & 0.86 \\
$s_2s_2u_2$ &  1.3790(26) &  1.3509(30) &  1.3384(34) &  1.3309(64) & 1.2735(79) &  0.205(19) & 1.01 \\
$s_2s_2u_3$ &  1.3636(27) &  1.3352(31) &  1.3227(35) &  1.3156(66) & 1.2574(81) &  0.207(19) & 1.07 \\
\end{tabular}
\end{center}
\end{table*}

\begin{table*}
\caption{The same as table \protect\ref{tab:PS075} for decuplet baryons.} 
\label{tab:D075}
\begin{center}
\begin{tabular}{lrrrrrrr}
$\kappa_1\kappa_2$ & $\beta=5.90$ & $\beta=6.10$ & $\beta=6.25$ & $\beta=6.47$ & $m_0a$ & $c$ & $\chi^2/df$ \\
\hline
$s_1s_1s_1$ &  1.6918(29) &  1.6712(29) &  1.6639(35) &  1.6718(77) & 1.6244(87) &  0.129(21) & 3.02 \\
$s_2s_2s_2$ &  1.6164(33) &  1.5933(34) &  1.5851(41) &  1.5931(83) & 1.5415(98) &  0.143(23) & 2.94 \\
$u_1u_1u_1$ &  1.5205(40) &  1.4948(43) &  1.4845(50) &  1.4969(100) &1.4374(120) &  0.158(29) & 3.05 \\
$u_2u_2u_2$ &  1.4629(48) &  1.4354(50) &  1.4227(59) &  1.4402(122) &1.3725(143) &  0.172(34) & 3.04 \\
$u_3u_3u_3$ &  1.4271(58) &  1.3979(60) &  1.3827(69) &  1.4051(151) &1.3286(172) &  0.188(41) & 2.75 \\
$u_1u_1s_1$ &  1.5768(36) &  1.5529(37) &  1.5435(44) &  1.5553(89) & 1.4999(106) &  0.146(25) & 3.45 \\
$u_2u_2s_1$ &  1.5372(40) &  1.5123(42) &  1.5012(48) &  1.5176(99) & 1.4569(118) &  0.153(28) & 3.88 \\
$u_3u_3s_1$ &  1.5118(45) &  1.4859(48) &  1.4730(53) &  1.4943(111) &1.4278(131) &  0.159(32) & 4.22 \\
$s_1s_1u_1$ &  1.6339(32) &  1.6117(33) &  1.6033(39) &  1.6136(81) & 1.5622(94) &  0.136(23) & 3.53 \\
$s_1s_1u_2$ &  1.6135(33) &  1.5909(35) &  1.5816(40) &  1.5948(83) & 1.5408(98) &  0.138(24) & 4.08 \\
$s_1s_1u_3$ &  1.6000(35) &  1.5770(37) &  1.5667(42) &  1.5832(87) & 1.5265(103) &  0.139(25) & 4.69 \\
$u_1u_1s_2$ &  1.5522(37) &  1.5274(40) &  1.5177(47) &  1.5290(93) & 1.4722(111) &  0.152(27) & 3.14 \\
$u_2u_2s_2$ &  1.5129(42) &  1.4870(44) &  1.4757(51) &  1.4913(104) &1.4291(124) &  0.159(30) & 3.46 \\
$u_3u_3s_2$ &  1.4877(47) &  1.4609(50) &  1.4478(56) &  1.4680(117) &1.4001(138) &  0.167(33) & 3.71 \\
$s_2s_2u_1$ &  1.5841(35) &  1.5602(37) &  1.5513(44) &  1.5611(88) & 1.5069(104) &  0.147(25) & 3.11 \\
$s_2s_2u_2$ &  1.5641(37) &  1.5396(39) &  1.5299(45) &  1.5422(91) & 1.4855(109) &  0.150(26) & 3.46 \\
$s_2s_2u_3$ &  1.5508(39) &  1.5260(41) &  1.5153(47) &  1.5306(95) & 1.4711(114) &  0.152(27) & 3.88 \\
\end{tabular}
\end{center}
\end{table*}

\begin{table*}
\caption{Quark masses $m_qa$ in the ${\overline {\rm MS}}$ scheme at
$\mu=2$ GeV interpolated or slightly extrapolated to
$m_{PS}/m_V=0.75 (s_1)$, 0.7($s_2$), 0.6($u_1$), 0.5($u_2$) and 0.4($u_3$) and 
normalized by $m_V^{(0.75)}a$ together with parameters and $\chi^2/df$ of their 
linear continuum extrapolations $m_qa = m_{q,0}a + c\cdot m_V^{(0.75)}a$.
Lattice spacings at each $\beta$ determined from the physical value of
$m_V^{0.75}$ are also given.}
\label{tab:mq075}
\begin{center}
\begin{tabular}{lrrrrrrr}
$\kappa_1\kappa_2$ & $\beta=5.90$ & $\beta=6.10$ & $\beta=6.25$ & $\beta=6.47$ & $m_{q,0}a$ & $c$ & $\chi^2/df$ \\
\hline
$s_1$ &  0.1164( 3) &  0.1185( 4) &  0.1203( 5) &  0.1216(10) & 0.1265(11) & -0.020( 3) & 0.28 \\
$s_2$ &  0.0910( 3) &  0.0927( 4) &  0.0943( 5) &  0.0949( 9) & 0.0992(10) & -0.016( 2) & 0.51 \\
$u_1$ &  0.0568( 3) &  0.0581( 3) &  0.0593( 4) &  0.0598( 8) & 0.0630( 9) & -0.012( 2) & 0.40 \\
$u_2$ &  0.0347( 3) &  0.0357( 3) &  0.0365( 4) &  0.0370( 8) & 0.0393( 9) & -0.009( 2) & 0.15 \\
$u_3$ &  0.0199( 2) &  0.0205( 2) &  0.0209( 4) &  0.0215( 7) & 0.0226( 8) & -0.005( 2) & 0.05 \\
\hline
$a^{-1}$ (GeV) &  1.929(36) &  2.532(47) &  3.054(57) &  4.009(78) & & & \\
\end{tabular}
\end{center}
\end{table*}

\begin{table*}
\caption{Parameters of Q$\chi$PT fits made in the continuum limit.
Results are given in units of $m_V^{(0.75)}$ of the continuum theory 
and in physical units (GeV,GeV$^{-1}$). We also give parameters
obtained from an extrapolation of the parameters at finite $\beta$.}
\label{tab:ContChiralPrm}
\begin{center}
\begin{tabular}{lllrr|r} 
&$\chi^2/N_{df}$&Param.&in $m_V^{(0.75)}$ unit & in Phys. unit & extrap.(in Phys. unit)\\
PS      & 0.2 & $A$       &1.72(23)&1.69(23)       & 2.43(4)\\
        &     & $B$       &2.17(91)&               & 1.69(9)\\
        &     & $\delta$  &0.16(10)& \\
Vector  & 0.0 & $m_V^0$   &0.791(29)    &0.776(11) & 0.777(10)\\
        &     & $C_1$     &0.548(79)    &0.559(93) & 0.560(72)\\
        &     & $C_{1/2}$ &$-$0.133(98) &          & $-$0.136(82)\\
octet   & 0.0 & $m_O^0$   & 0.877(12)   & 0.861(20) & 0.863(24)\\
        &     & $b_F$     &$-$0.248(66) &$-$0.253(67) & $-$0.244(36)\\
        &     & $b_D$     & 0.040(44)   & 0.041(45)   & 0.040(18)\\
        &     & $F$       & 0.290(71)   &             & 0.280(36)\\
        &     & $D$       & 0.434(69)   &             & 0.416(35)\\
decuplet& 0.0 & $m_D^0$   &1.265(72) &1.241(74) & 1.251(43)\\
        &     & $c$       &0.220(77) &0.224(78) & 0.234(28)\\
        &     & $H^2$     &0.06(99)  &          & 0.16(40)\\
        &     & $C^2$     &0.3(3.0)  &          & $-$0.38(77)\\
\end{tabular}
\end{center}
\end{table*}

\begin{table*}
\caption{Hadron spectrum from Q$\chi$PT chiral fits in the continuum limit.
The scale $m_V^{(0.75)}$ of the continuum theory is also given. 
See text for details.
Deviation and its statistical significance are relative to our final result 
(data in \protect\ref{tab:MassFinal}).}
\label{tab:ContChiralMass}
\begin{center}
\begin{tabular}{l|lr|lr}
\         & 
\multicolumn{2}{c|}{$m_K$ input} &
\multicolumn{2}{c}{$m_\phi$ input} \\
\         & mass(GeV) & deviation   
          & mass(GeV) & deviation  \\
\hline 
$m_V^{(0.75)}$  & 0.981(18) & & 0.981(18) & \\ 
\hline
$K$       &            &                    
          &0.559( 6)&  1.0\%  0.6$\sigma$ \\
$K^*$     &0.856( 9)& $-$0.3\%  0.3$\sigma$
          &0.889( 4)&  0.0\%  0.0$\sigma$ \\
$\phi$    &0.951(12)& $-$0.6\%  0.4$\sigma$
          &            &                     \\
$N$       &0.876(23)& $-$0.2\%  0.1$\sigma$
          &0.876(23)& $-$0.2\%  0.1$\sigma$ \\
$\Lambda$ &1.011(17)& $-$0.8\%  0.4$\sigma$
          &1.057(12)& $-$0.3\%  0.3$\sigma$ \\
$\Sigma$  &1.109(16)& $-$0.7\%  0.4$\sigma$
          &1.173(10)& $-$0.2\%  0.2$\sigma$ \\
$\Xi$     &1.189(13)& $-$1.0\%  0.7$\sigma$
          &1.285( 8)& $-$0.3\%  0.5$\sigma$ \\
$\Delta$  &1.251(48)& $-$0.5\%  0.2$\sigma$
          &1.251(48)& $-$0.5\%  0.2$\sigma$ \\
$\Sigma^*$&1.353(32)& $-$0.4\%  0.2$\sigma$
          &1.385(31)& $-$0.2\%  0.1$\sigma$ \\
$\Xi^*$   &1.453(27)& $-$0.4\%  0.2$\sigma$
          &1.517(25)&  0.0\%  0.0$\sigma$ \\
$\Omega$  &1.551(20)& $-$0.6\% 0.4$\sigma$
          &1.647(14)&  0.0\%  0.0$\sigma$ \\
\end{tabular}
\end{center}
\end{table*}

\begin{table*}[htbp]
\caption{Smearing parameters for quark matrix 
inversion. Parameters $A$ and $B$ appear in the smearing function
$S(n)=A\exp(-B|n|)$. The number of iterations necessary 
for point (P) and smeared (S) sources are also given.}
\label{tab:param3}
\begin{center}
\begin{tabular}{cccccc}
\multicolumn{6}{c}{$\beta=5.90$ ($r^2_P=10^{-12}$, $r^2_S=10^{-7}$)} \\
$\kappa$  & 0.15660 & 0.15740 & 0.15830 & 0.15890 & 0.15920 \\
$A$          & 1.1427 & 1.1408 & 1.1387 &  1.1373 &  1.1366 \\
$B$          & 0.3290 & 0.3268 & 0.3242 &  0.3226 &  0.3217 \\
\#Iter(P)    & 220$\pm$8  & 307$\pm$13 & 495$\pm$26 & 780$\pm$170 & 1130$\pm$320 \\ 
\#Iter(S)    & 201$\pm$6  & 279$\pm$10 & 463$\pm$23 & 750$\pm$170 & 1080$\pm$290 \\ 
\hline
\multicolumn{6}{c}{$\beta=6.10$ ($r^2_P=10^{-12}$, $r^2_S=10^{-7}$)} \\
$\kappa$  & 0.15280 & 0.15340 & 0.15400 & 0.15440 & 0.15460 \\
$A$          & 1.0884 & 1.0863 & 1.0843 & 1.0830 & 1.0823 \\
$B$          & 0.2361 & 0.2339 & 0.2317 & 0.2303 & 0.2296 \\
\#Iter(P)    & 259$\pm$6  & 373$\pm$9  & 644$\pm$21 & 1170$\pm$60 & 1740$\pm$230 \\
\#Iter(S)    & 261$\pm$4  & 378$\pm$7  & 661$\pm$18 & 1220$\pm$60 & 1840$\pm$250 \\
\hline
\multicolumn{6}{c}{$\beta=6.25$ ($r^2_P=10^{-12}$, $r^2_S=10^{-7}$)} \\
$\kappa$  & 0.15075 & 0.15115 & 0.15165 & 0.15200 & 0.15220 \\
$A$          & 1.06304 & 1.06151 & 1.05961 & 1.05829 & 1.05754 \\
$B$          & 0.19336 & 0.19175 & 0.18974 & 0.18835 & 0.18755 \\
\#Iter(P)    & 275$\pm$6 & 359$\pm$8 & 586$\pm$15 & 1080$\pm$40 & 1900$\pm$120 \\
\#Iter(S)    & 298$\pm$3 & 387$\pm$5 & 633$\pm$11 & 1170$\pm$40 & 2130$\pm$130 \\ 
\hline
\multicolumn{6}{c}{$\beta=6.47$ ($r^2_P=10^{-14}$, $r^2_S=10^{-7}$)} \\
$\kappa$  & 0.14855 & 0.14885 & 0.14925 & 0.14945 & 0.14960 \\
$A$          & 1.02817 & 1.02688 & 1.02517 & 1.02431 & 1.02368 \\
$B$          & 0.13484 & 0.13351 & 0.13174 & 0.13085 & 0.13019 \\
\#Iter(P)    & 427$\pm$8 & 561$\pm$11 & 987$\pm$63 & 1600$\pm$50 & 3060$\pm$160 \\
\#Iter(S)    & 405$\pm$3 & 532$\pm$6  & 940$\pm$60 & 1520$\pm$40 & 2930$\pm$140 \\
\end{tabular}
\end{center}
\end{table*}

\begin{table*}
\caption{Values of $\chi^2_{full}$ for degenerate (DG) and
non-degenerate (ND) pseudoscalar meson mass fits.}
\label{tab:chi2fullPS}
\begin{center}
\begin{tabular}{c|cr|cr|cr}
& \multicolumn{2}{c|}{DG} &
  \multicolumn{2}{c|}{ND ($\kappa_1=s_1$)} &
  \multicolumn{2}{c}{ND ($\kappa_1=s_2$)} \\
$\beta$ & $N_{df}$ & $\chi^2_{full}$ 
        & $N_{df}$ & $\chi^2_{full}$ 
        & $N_{df}$ & $\chi^2_{full}$ \\
5.90 & 85 & 79(21) &  68 & 61(17)  & 68 & 61(18)  \\ 
6.10 &100 &119(26) &  80 & 92(22)  & 80 & 99(23)  \\
6.25 & 40 & 47(15) &  32 & 29(12)  & 32 & 33(13)  \\
6.47 & 70 &172(66) &  56 & 85(36)  & 56 & 86(35)  \\
\end{tabular}
\end{center}
\end{table*}

\begin{table*}
\caption{Chiral fit functions and $\chi^2_{ext}/N_{df}$ 
for pseudoscalar meson masses.
Fit 1 -- Fit 5 are for degenerate cases, while Fit 1n -- Fit 5n are
for non-degenerate cases with $m_s = (1/s_1 - 1/\kappa_c)/2$ for 
upper rows and $m_s = (1/s_2 - 1/\kappa_c)/2$ for lower rows.}
\label{tab:chi2extPS}
\begin{center}
\begin{tabular}{llccrrrr}
&&&& \multicolumn{4}{c}{$\chi^2_{ext}/N_{df}$} \\
      & fit function & parameters & $N_{df}$ 
      & $\beta=5.90$ & $\beta=6.10$ & $\beta=6.25$ & $\beta=6.47$ \\
\hline
Fit 1 & $m_{PS}^2 = 2Am$ & $\kappa_c,A$ & 3 
      & 1290(100) & 1117( 92) & 942( 82) & 630(170) \\
Fit 2 & $m_{PS}^2 = 2Am + 4Bm^2$ & $\kappa_c,A,B$ & 2 
      & 417( 43) & 172( 28) & 67( 28) & 59( 30) \\
Fit 3 & $m_{PS}^2 = 2Am + 4Bm^2 + 8Cm^3$ & $\kappa_c,A,B,C$ & 1 
      & 182( 43) & 125( 54) & 100( 55) & 83( 48) \\
Fit 4 & $m_{PS}^2 = 2Am[1-\delta(\ln(f_\Lambda m/A)+1)]$ & $\kappa_c,A,\delta$ & 2
      & 1018( 85) & 606( 61) & 376( 55) & 285( 92) \\
\lw{Fit 5} & $m_{PS}^2 = 2Am[1-\delta(\ln(f_\Lambda m/A)+1)]$ 
      & \lw{$\kappa_c,A,\delta,B$} & \lw{1}
      & \lw{116( 31)} & \lw{71( 35)} &  \lw{57( 31)} & \lw{26( 26)} \\
      & $\ \ \ \ \ \ \ + 4Bm^2$  &  &  &  &  &  \\
\hline
\lw{Fit 1n} & \lw{$m_{PS}^2 = A_s m_s + Am$} & \lw{$A_s,A$} & \lw{2} 
     & 630(110) & 455(55) & 430( 50) & 246( 67) \\
       &                           &         &
     & 381( 84) & 314(45) & 339( 41) & 233( 64) \\
\lw{Fit 2n} & $m_{PS}^2 = A(m_s+m) + B(m_s+m)^2 $ & \lw{$A,B,C$} & \lw{1}
     & 91( 26) & 20( 12) &  2(  3) &  0(  1) \\
       & $\ \ \ \ \ \ \ + C(m_s-m)^2$ & & 
     & 128( 35) & 36(17) & 4(  4) &  1(  4) \\
     & $m_{PS}^2 = A(m_s+m)[1-\delta(\ln(f_\Lambda m_s/A)$ & & 
     & 147( 40) & 46(19) & 12(  9) & 4(  9) \\
Fit 5n  & $\ \ \ \ \ \ \ + m/(m-m_s)\ln(m/m_s) ) ]$ & $A,\delta,B$ & 1
     &&&& \\
       & $\ \ \ \ \ \ \ + B(m_s+m)^2$ & &
     & 80(  50) &  62( 24) &  14( 10) &   8( 13) \\
\end{tabular}
\end{center}
\end{table*}

\begin{table*}
\caption{Bare AWI quark masses $2m_q^{AWI,local(0)}$ 
obtained at the simulation points. 
Here, $2m_q$ stands for $m_{q_f} + m_{q_g}$. 
The fit range $[t_{min},t_{max}]$ is also given.}
\label{tab:AWImqlocal}
\begin{center}
\begin{tabular}{l|rl|rl|rl|rl}
\ \ &\multicolumn{2}{c|}{$\beta=5.90$}&\multicolumn{2}{c|}{$\beta=6.10$}
&\multicolumn{2}{c|}{$\beta=6.25$}&\multicolumn{2}{c}{$\beta=6.47$} \\
\hline
$\kappa_1\kappa_2$ & range & value & range & value & range & value & range & value \\ 
$s_1s_1$ & 5 25 & 0.091309( 80) & 8 32 & 0.072038( 59) & 7 39 & 0.063377(74) & 10 53 & 0.048770( 80) \\ 
$s_2s_2$ & 5 25 & 0.067984( 74) & 8 32 & 0.051800( 53) & 7 39 & 0.048763(67) & 10 53 & 0.036887( 72) \\ 
$u_1u_1$ & 5 25 & 0.042573( 68) & 8 32 & 0.031987( 50) & 7 39 & 0.030771(57) & 10 53 & 0.021232( 60) \\ 
$u_2u_2$ & 5 25 & 0.025941( 69) & 8 32 & 0.018921( 56) & 7 39 & 0.018359(52) & 10 53 & 0.013501( 58) \\ 
$u_3u_3$ & 5 25 & 0.017144(142) & 8 32 & 0.012114(100) & 7 39 & 0.011225(60) & 10 53 & 0.007701( 62) \\ 
$s_1u_1$ & 5 25 & 0.066713( 75) & 8 32 & 0.051875( 54) & 7 39 & 0.046989(66) & 10 53 & 0.034941( 73) \\ 
$s_1u_2$ & 5 25 & 0.058270( 75) & 8 32 & 0.045274( 54) & 7 39 & 0.040706(64) & 10 53 & 0.031035( 73) \\ 
$s_1u_3$ & 5 25 & 0.054006( 78) & 8 32 & 0.041953( 57) & 7 39 & 0.037117(65) & 10 53 & 0.028115( 78) \\ 
$s_2u_1$ & 5 25 & 0.055217( 71) & 8 32 & 0.041860( 51) & 7 39 & 0.039740(62) & 10 53 & 0.029039( 67) \\ 
$s_2u_2$ & 5 25 & 0.046842( 71) & 8 32 & 0.035299( 52) & 7 39 & 0.033486(59) & 10 53 & 0.025147( 66) \\ 
$s_2u_3$ & 5 25 & 0.042594( 75) & 8 32 & 0.031982( 55) & 7 39 & 0.029911(60) & 10 53 & 0.022235( 70) \\ 
\end{tabular}
\end{center}
\end{table*}

\twocolumn

\newpage

\onecolumn

\twocolumn

\newpage

\end{document}